\documentclass[9pt]{extarticle}

\usepackage{microtype}
\usepackage[utf8]{inputenc}

\usepackage{eurosym}
\usepackage{url}
\usepackage{amssymb}
\usepackage{amsmath,amsfonts}
\usepackage{pifont}
\usepackage[square,numbers]{natbib}
\usepackage{colortbl}
\usepackage{multirow}
\usepackage{amsthm}
\usepackage{mathtools}
\usepackage{mathrsfs}
\usepackage{booktabs}
\usepackage{array}  % >{...}p{...} column prefixes used in tex/dataset.tex
\usepackage{tabularx} % auto-stretching X columns for the CVE narrative
\usepackage{textcomp}
\usepackage{manyfoot}
\usepackage[table,xcdraw]{xcolor}
\usepackage{fancyhdr}
\usepackage{wrapfig}

\definecolor{cai_primary}{HTML}{4C9A99}
\definecolor{cai_secondary}{HTML}{307FE2}
\definecolor{cai_accent}{HTML}{1D8348}
\definecolor{cai_dark}{HTML}{3F4444}
\definecolor{human_color}{HTML}{173C47}

\definecolor{graph_lightcyan}{HTML}{B8D8D8}
\definecolor{graph_gray}{HTML}{E8F0EF}
\definecolor{graph_navy}{HTML}{2D5A56}
\definecolor{graph_arrow}{HTML}{3D7A79}
\definecolor{graph_accent}{HTML}{6BBFB5}
\definecolor{graph_teal}{HTML}{4C9A99}
\definecolor{graph_human}{HTML}{F7FAFA}
\definecolor{cai_light}{HTML}{F5F5F5}

\definecolor{anthropic_color}{RGB}{204, 119, 102}
\definecolor{google_color}{RGB}{66, 133, 244}
\definecolor{openai_color}{RGB}{100, 100, 100}
\definecolor{mistral_color}{RGB}{255, 140, 0}

\definecolor{defender_color}{HTML}{1F618D}
\definecolor{static_color}{HTML}{2980B9}
\definecolor{dynamic_color}{HTML}{E67E22}
\definecolor{apt_agent_color}{HTML}{C0392B}

\definecolor{cc_color}{HTML}{C0392B}
\definecolor{codex_color}{HTML}{1F618D}
\definecolor{gcai_color}{HTML}{6BBFB5}
\definecolor{cai_orange}{HTML}{2E6260}
\definecolor{mistral_vibe}{HTML}{FF8C00}
\definecolor{provider_grey}{HTML}{6E7B7B}
\definecolor{provider_grey_light}{HTML}{A8B0B0}

\definecolor{group_cyber}{HTML}{FBE9E7}
\definecolor{group_agent}{HTML}{E3F2FD}
\definecolor{group_sft}{HTML}{E8F5E9}
\definecolor{group_chat}{HTML}{FFF8E1}
\definecolor{group_cyber_hdr}{HTML}{F4A99B}
\definecolor{group_agent_hdr}{HTML}{A0C8F0}
\definecolor{group_sft_hdr}{HTML}{A8D7AE}
\definecolor{group_chat_hdr}{HTML}{FFE08A}

\definecolor{hero_vol_a}{HTML}{1F4F4D}
\definecolor{hero_vol_b}{HTML}{3F8984}
\definecolor{hero_vol_c}{HTML}{56A8A0}
\definecolor{hero_vol_d}{HTML}{6BBFB5}
\definecolor{hero_vol_e}{HTML}{B8D8D8}
\definecolor{hero_role_off}{HTML}{C0392B}
\definecolor{hero_role_int}{HTML}{E67E22}
\definecolor{hero_role_biz}{HTML}{27AE60}
\definecolor{hero_role_def}{HTML}{1F618D}

\newcommand{\todo}[1]{} % review-pass notes suppressed; uncomment line below to re-enable
\DeclareRobustCommand{\aliasmini}{\texorpdfstring{\href{https://aliasrobotics.com/aliasLLMs.php}{\textcolor{cai_primary}{\texttt{alias2-mini}}}}{alias2-mini}}
\DeclareRobustCommand{\caitool}{\texorpdfstring{\href{https://github.com/aliasrobotics/cai}{\textcolor{cai_primary}{\texttt{CAI}}}}{CAI}}
\newcommand{\csidata}{\texorpdfstring{{\sffamily\bfseries CAI Dataset}}{CAI Dataset}}

\usepackage{graphicx}
\usepackage{subcaption}
\usepackage{float}
\usepackage{stfloats}
\usepackage[hyperfootnotes=false]{hyperref}
\hypersetup{
    colorlinks=true,
    urlcolor=cai_secondary,
    linkcolor=cai_secondary,
    filecolor=cai_accent,
    citecolor=cai_secondary,
}
\makeatletter
\g@addto@macro{\UrlBreaks}{\do\/\do\-\do\_\do\.\do\=\do\?\do\&}
\makeatother

\usepackage{tikz}
\usetikzlibrary{arrows.meta,positioning,shapes.geometric,calc,patterns,shadows,decorations.pathreplacing,fit,backgrounds}
\usepackage{pgfplots}
\pgfplotsset{compat=1.17,compat/show suggested version=false}
\usepgfplotslibrary{groupplots}

\usepackage{enumitem}
\usepackage[section]{placeins}
\usepackage[breakable,skins]{tcolorbox}
\usepackage{algorithm}
\usepackage{algorithmicx}
\usepackage{algpseudocode}
\usepackage{listings}

\usepackage{geometry}
\renewcommand{\headrulewidth}{0.4pt}
\renewcommand{\footrulewidth}{0.4pt}
\renewcommand{\headrule}{\hbox to\headwidth{\color{cai_primary}\leaders\hrule height \headrulewidth\hfill}}
\renewcommand{\footrule}{\hbox to\headwidth{\color{human_color}\leaders\hrule height \footrulewidth\hfill}}
\usepackage{caption,setspace}
\usepackage{titlesec}
\titleformat{\section}
  {\normalfont\Large\bfseries\color{cai_primary}}
  {\thesection}
  {1em}
  {}
  [\titlerule]
\titleformat{\subsection}
  {\normalfont\large\bfseries\color{human_color}}
  {\thesubsection}
  {1em}
  {}
\titleformat{\subsubsection}
  {\normalfont\normalsize\bfseries\color{cai_dark}}
  {\thesubsubsection}
  {1em}
  {}

\usepackage{authblk}

\renewcommand\Affilfont{\small\normalfont}
\definecolor{cai_affil_color}{HTML}{3F8984}

\makeatletter
\renewcommand\AB@affilsepx{\\\protect\Affilfont}
\let\orig@maketitle\maketitle
\renewcommand{\maketitle}{%
  \orig@maketitle%
  \vspace{-1.5em}%
  {\color{cai_primary!30}\hrule height 0.5pt}%
  \vspace{1em}%
}
\makeatother

\makeatletter
\renewenvironment{abstract}{%
  \small
  \noindent\ignorespaces
}{%
  \par
}
\makeatother

\newcommand{\statTotalFiles}{230,935}
\newcommand{\statTotalPrompts}{26,027,742}
\newcommand{\statUniqueIPs}{16,768}

\newcommand{\statUniqueModels}{4,187}
\newcommand{\statUniqueCountries}{123}
\newcommand{\statSpanDays}{428}
\newcommand{\statTotalFilesRaw}{230935}
\newcommand{\statTotalPromptsRaw}{26027742}
\newcommand{\statUniqueIPsRaw}{16768}

\newcommand{\statUniqueCountriesRaw}{123}
\newcommand{\statEarliestDate}{2025-03-22}
\newcommand{\statLatestDate}{2026-05-25}

\newcommand{\statCloudSizeTB}{18.07}

\newcommand{\statUniqueDomains}{23,147}
\newcommand{\statUniqueDomainsRaw}{23147}
\newcommand{\statTotalURLs}{683,606}

\newcommand{\statFilesScannedForTokens}{224,763}

\newcommand{\statTokenProvenance}{partial (97.3\% of files)}

\newcommand{\statOffensiveCount}{9,469,362}
\newcommand{\statDefensiveCount}{1,152,054}
\newcommand{\statBusinessCount}{7,144,999}
\newcommand{\statAttackerIntentCount}{5,220,028}
\newcommand{\statOffensiveCountRaw}{9469362}
\newcommand{\statDefensiveCountRaw}{1152054}
\newcommand{\statBusinessCountRaw}{7144999}
\newcommand{\statAttackerIntentCountRaw}{5220028}
\newcommand{\statOffensivePct}{36.4\%}
\newcommand{\statDefensivePct}{4.4\%}
\newcommand{\statBusinessPct}{27.5\%}
\newcommand{\statAttackerIntentPct}{20.1\%}

\newcommand{\statAvgPromptLen}{625.4}
\newcommand{\statMedianPromptLen}{75}
\newcommand{\statStdPromptLen}{41,252.2}

\newcommand{\statMaxPromptLen}{180,626,398}

\newcommand{\statSpainCentroidIPs}{12,730}

\newcommand{\statAttackCategoriesTable}{%
  Exploitation & 2,371,861 & 45.4\% \\
  Reconnaissance & 1,851,712 & 35.5\% \\
  General attack & 349,737 & 6.7\% \\
  Data exfiltration & 286,515 & 5.5\% \\
  Web attacks & 201,851 & 3.9\% \\
  Privilege escalation & 100,520 & 1.9\% \\
  Persistence & 45,091 & 0.9\% \\
  Social engineering & 12,741 & 0.2\% \\
}

\newcommand{\statTopBusinessCategoriesInline}{AI/ML: 2,130,065; API integration: 1,917,492; Data management: 1,620,997}

\newcommand{\statTorExitIPs}{59}
\newcommand{\statVpnIPs}{7}

\newcommand{\statHourlyCoords}{%
  (0,518531)
  (1,307962)
  (2,290591)
  (3,335711)
  (4,268111)
  (5,223091)
  (6,203421)
  (7,280571)
  (8,334870)
  (9,428203)
  (10,690430)
  (11,635096)
  (12,764382)
  (13,860662)
  (14,815638)
  (15,869395)
  (16,754958)
  (17,688539)
  (18,476512)
  (19,341418)
  (20,485212)
  (21,553316)
  (22,725207)
  (23,731145)
}
\newcommand{\statDailyCoords}{%
  (0,906445)
  (1,2123040)
  (2,4292663)
  (3,2675725)
  (4,1191360)
  (5,720110)
  (6,673629)
}

\newcommand{\statWeeklyTokenCoords}{%
  (1,13380)
  (2,468710)
  (3,6531190)
  (4,641283)
  (5,2542218)
  (6,2443499)
  (7,1517994)
  (8,3256002)
  (9,8001796)
  (10,6835314)
  (11,8908714)
  (12,8047344)
  (13,10280240)
  (14,8063872)
  (15,10119922)
  (16,9600677)
  (17,9521996)
  (18,8416125)
  (19,9538818)
  (20,16760006)
  (21,10703959)
  (22,16302650)
  (23,22914669)
  (24,23240962)
  (25,18870922)
  (26,22903282)
  (27,20313251)
  (28,29696497)
  (29,36967622)
  (30,31064931)
  (31,43539737)
  (32,56328238)
  (33,64423334)
  (34,51375887)
  (35,43460720)
  (36,50682943)
  (37,42588646)
  (38,46326948)
  (39,36485748)
  (40,49578571)
  (41,57435930)
  (42,60531044)
  (43,62836256)
  (44,61312156)
  (45,59565117)
  (46,36200745)
  (47,40188114)
  (48,72552457)
  (49,54261564)
  (50,52215373)
  (51,45473114)
  (52,39176793)
  (53,59552706)
  (54,47264241)
  (55,39837555)
  (56,35670969)
  (57,46715158)
  (58,43367798)
  (59,39682166)
  (60,60473540)
  (61,71985614)
  (62,40119430)
  (63,0)
}
\newcommand{\statWeeklyTokenCumCoords}{%
  (1,13380)
  (2,482090)
  (3,7013280)
  (4,7654563)
  (5,10196781)
  (6,12640280)
  (7,14158274)
  (8,17414276)
  (9,25416072)
  (10,32251386)
  (11,41160100)
  (12,49207444)
  (13,59487684)
  (14,67551556)
  (15,77671478)
  (16,87272155)
  (17,96794151)
  (18,105210276)
  (19,114749094)
  (20,131509100)
  (21,142213059)
  (22,158515709)
  (23,181430378)
  (24,204671340)
  (25,223542262)
  (26,246445544)
  (27,266758795)
  (28,296455292)
  (29,333422914)
  (30,364487845)
  (31,408027582)
  (32,464355820)
  (33,528779154)
  (34,580155041)
  (35,623615761)
  (36,674298704)
  (37,716887350)
  (38,763214298)
  (39,799700046)
  (40,849278617)
  (41,906714547)
  (42,967245591)
  (43,1030081847)
  (44,1091394003)
  (45,1150959120)
  (46,1187159865)
  (47,1227347979)
  (48,1299900436)
  (49,1354162000)
  (50,1406377373)
  (51,1451850487)
  (52,1491027280)
  (53,1550579986)
  (54,1597844227)
  (55,1637681782)
  (56,1673352751)
  (57,1720067909)
  (58,1763435707)
  (59,1803117873)
  (60,1863591413)
  (61,1935577027)
  (62,1975696457)
  (63,1975696457)
}
\newcommand{\statWeeksN}{63}

\newcommand{\statWeeklyTeamCumCoords}{%
  (1,0)
  (2,0)
  (3,20)
  (4,66)
  (5,119)
  (6,127)
  (7,130)
  (8,177)
  (9,244)
  (10,741)
  (11,900)
  (12,948)
  (13,1025)
  (14,1153)
  (15,1177)
  (16,1186)
  (17,1205)
  (18,1220)
  (19,1282)
  (20,1400)
  (21,1513)
  (22,1526)
  (23,1650)
  (24,1736)
  (25,1919)
  (26,1928)
  (27,2055)
  (28,2207)
  (29,3265)
  (30,3507)
  (31,3658)
  (32,3696)
  (33,3770)
  (34,3986)
  (35,4151)
  (36,4374)
  (37,4543)
  (38,4624)
  (39,4718)
  (40,5100)
  (41,5153)
  (42,5204)
  (43,5336)
  (44,5743)
  (45,6247)
  (46,6256)
  (47,6260)
  (48,6263)
  (49,6270)
  (50,6270)
  (51,6273)
  (52,6273)
  (53,6273)
  (54,6275)
  (55,6276)
  (56,6280)
  (57,6280)
  (58,6283)
  (59,6283)
  (60,6283)
  (61,6283)
  (62,6286)
  (63,6286)
}
\newcommand{\statWeeklyCommunityCumCoords}{%
  (1,5)
  (2,111)
  (3,300)
  (4,382)
  (5,557)
  (6,805)
  (7,1043)
  (8,1365)
  (9,2091)
  (10,2820)
  (11,3642)
  (12,4673)
  (13,6311)
  (14,7622)
  (15,9406)
  (16,11400)
  (17,13210)
  (18,15188)
  (19,17152)
  (20,19455)
  (21,21206)
  (22,24007)
  (23,28001)
  (24,30875)
  (25,33389)
  (26,35859)
  (27,38233)
  (28,41473)
  (29,44092)
  (30,46943)
  (31,50489)
  (32,54238)
  (33,58350)
  (34,61860)
  (35,65432)
  (36,68369)
  (37,71662)
  (38,75464)
  (39,78604)
  (40,82160)
  (41,84722)
  (42,87014)
  (43,89564)
  (44,92220)
  (45,95464)
  (46,98078)
  (47,101233)
  (48,105033)
  (49,108288)
  (50,112036)
  (51,115476)
  (52,118562)
  (53,122983)
  (54,126341)
  (55,129111)
  (56,132286)
  (57,135920)
  (58,140113)
  (59,143140)
  (60,146335)
  (61,150829)
  (62,155831)
  (63,156203)
}
\newcommand{\statTeamLogsTotal}{6,286}
\newcommand{\statCommunityLogsTotal}{156,203}

\newcommand{\statContinentEuropeCount}{12,284}
\newcommand{\statContinentEuropePct}{85.7\%}
\newcommand{\statContinentAsiaCount}{1,066}
\newcommand{\statContinentAsiaPct}{7.4\%}
\newcommand{\statContinentNorthAmericaCount}{603}
\newcommand{\statContinentNorthAmericaPct}{4.2\%}
\newcommand{\statContinentSouthAmericaCount}{165}
\newcommand{\statContinentSouthAmericaPct}{1.2\%}
\newcommand{\statContinentAfricaCount}{159}
\newcommand{\statContinentAfricaPct}{1.1\%}
\newcommand{\statContinentOceaniaCount}{52}
\newcommand{\statContinentOceaniaPct}{0.4\%}
\newcommand{\statContinentTotal}{14,329}
\newcommand{\statOrgHostingCount}{1,838}
\newcommand{\statOrgHostingPct}{11.4\%}
\newcommand{\statOrgIspCount}{6,116}
\newcommand{\statOrgIspPct}{37.8\%}
\newcommand{\statOrgOtherOrgCount}{8,232}
\newcommand{\statOrgOtherOrgPct}{50.9\%}
\newcommand{\statBubbleN}{1719}
\newcommand{\statBubblesEurope}{%
  (-3.34,47.76) [1]
  (2.57,48.95) [1]
  (14.37,40.94) [1]
  (-8.02,42.91) [1]
  (7.54,46.29) [1]
  (-5.97,37.38) [1]
  (-2.99,43.30) [1]
  (-6.14,36.42) [1]
  (2.35,48.94) [1]
  (-1.45,49.05) [1]
  (53.20,56.85) [1]
  (4.93,52.35) [1]
  (19.26,42.44) [1]
  (2.38,48.82) [1]
  (-16.25,28.47) [1]
  (17.59,48.38) [1]
  (19.40,50.14) [1]
  (8.96,46.01) [1]
  (-4.49,37.41) [1]
  (-3.01,40.31) [1]
  (-1.98,43.31) [1]
  (8.71,50.65) [1]
  (-9.34,38.70) [1]
  (10.31,52.20) [1]
  (0.15,45.65) [1]
  (-3.89,40.41) [1]
  (1.87,41.54) [1]
  (14.27,40.84) [1]
  (-2.70,42.83) [1]
  (-0.10,51.55) [1]
  (-1.50,47.21) [1]
  (2.17,41.41) [1]
  (-3.73,38.33) [1]
  (5.60,43.17) [1]
  (-8.77,42.60) [1]
  (-4.02,39.86) [1]
  (2.68,48.88) [1]
  (0.19,51.52) [1]
  (38.07,54.90) [1]
  (-3.02,43.32) [1]
  (1.69,46.82) [1]
  (-0.18,38.97) [1]
  (1.44,43.60) [1]
  (-0.39,44.81) [1]
  (9.47,45.48) [1]
  (5.86,51.83) [1]
  (-3.69,41.67) [1]
  (16.99,51.12) [1]
  (6.17,49.23) [1]
  (8.36,49.03) [1]
  (16.50,48.22) [1]
  (27.20,60.57) [1]
  (23.50,46.74) [1]
  (5.03,52.08) [1]
  (30.20,55.18) [1]
  (35.36,49.01) [1]
  (-4.11,40.94) [1]
  (-8.51,43.31) [1]
  (19.28,47.57) [1]
  (-8.14,41.63) [1]
  (-7.59,43.66) [1]
  (-3.00,43.05) [1]
  (4.88,52.38) [1]
  (2.02,41.42) [1]
  (-0.10,51.50) [1]
  (3.02,39.77) [1]
  (-2.95,42.69) [1]
  (-1.15,46.17) [1]
  (-1.70,38.23) [1]
  (2.40,49.01) [1]
  (-0.10,51.53) [1]
  (8.52,47.38) [1]
  (2.21,41.41) [1]
  (9.68,50.56) [1]
  (6.96,51.53) [1]
  (-1.63,42.81) [1]
  (-3.71,40.47) [1]
  (-3.72,40.41) [1]
  (-3.73,40.31) [1]
  (6.08,44.56) [1]
  (25.88,40.85) [1]
  (-2.56,51.54) [1]
  (-8.62,41.15) [1]
  (7.76,49.44) [1]
  (-4.41,40.78) [1]
  (13.58,42.85) [1]
  (24.86,44.85) [1]
  (7.47,51.51) [1]
  (6.61,46.53) [1]
  (30.75,46.49) [1]
  (20.44,39.36) [1]
  (10.06,53.58) [1]
  (19.95,50.01) [1]
  (9.97,53.61) [1]
  (23.70,38.03) [1]
  (2.21,41.44) [1]
  (-3.25,40.60) [1]
  (11.59,52.11) [1]
  (22.68,50.02) [1]
  (-2.26,43.30) [1]
  (-3.26,40.57) [1]
  (2.29,48.89) [1]
  (19.94,50.05) [1]
  (3.05,50.64) [1]
  (10.03,53.56) [1]
  (-16.60,28.02) [1]
  (-0.12,51.47) [1]
  (13.37,52.51) [1]
  (26.73,58.38) [1]
  (41.12,56.94) [1]
  (4.80,46.31) [1]
  (12.83,48.17) [1]
  (2.56,50.59) [1]
  (-1.68,48.12) [1]
  (13.06,45.64) [1]
  (-1.20,48.35) [1]
  (-4.11,43.39) [1]
  (-2.22,52.19) [1]
  (-8.19,43.35) [1]
  (10.40,63.43) [1]
  (17.66,44.17) [1]
  (1.24,41.28) [1]
  (-8.25,43.35) [1]
  (10.37,55.43) [1]
  (27.61,47.17) [1]
  (15.42,53.17) [1]
  (12.49,55.71) [1]
  (8.65,49.88) [1]
  (2.22,48.76) [1]
  (20.46,44.80) [1]
  (11.02,49.44) [1]
  (5.97,50.92) [1]
  (1.45,43.60) [1]
  (0.52,40.81) [1]
  (9.01,50.07) [1]
  (9.93,53.59) [1]
  (9.19,47.54) [1]
  (-2.41,43.03) [1]
  (-1.97,43.31) [1]
  (2.24,41.47) [1]
  (10.92,44.65) [1]
  (-1.28,38.08) [1]
  (9.97,53.57) [1]
  (50.35,53.37) [1]
  (-6.35,36.78) [1]
  (14.47,50.08) [1]
  (-0.14,51.54) [1]
  (-6.37,53.28) [1]
  (13.85,58.39) [1]
  (11.08,49.45) [1]
  (26.17,44.42) [1]
  (16.36,48.19) [1]
  (2.12,41.55) [1]
  (1.80,41.23) [1]
  (3.27,50.83) [1]
  (28.86,47.00) [1]
  (-5.86,38.96) [1]
  (-3.65,40.54) [1]
  (16.60,49.18) [1]
  (-3.78,37.78) [1]
  (-3.63,40.56) [1]
  (1.30,52.63) [1]
  (37.55,55.43) [1]
  (6.94,46.82) [1]
  (-3.64,40.54) [1]
  (21.63,47.53) [1]
  (6.43,53.14) [1]
  (-3.14,37.30) [1]
  (18.88,52.97) [1]
  (37.44,55.89) [1]
  (4.37,52.01) [1]
  (21.90,41.60) [1]
  (-0.02,51.51) [1]
  (37.59,55.73) [1]
  (18.12,59.25) [1]
  (16.30,39.31) [1]
  (24.42,61.00) [1]
  (28.62,44.30) [1]
  (-4.62,36.54) [1]
  (-0.99,37.61) [1]
  (11.54,46.43) [1]
  (-3.74,40.39) [1]
  (13.74,51.05) [1]
  (17.43,40.74) [1]
  (19.92,39.62) [1]
  (19.13,47.64) [1]
  (-0.52,39.17) [1]
  (17.33,51.03) [1]
  (7.33,53.58) [1]
  (-0.89,41.66) [1]
  (21.26,48.72) [1]
  (16.29,48.14) [1]
  (11.41,48.76) [1]
  (11.63,44.84) [1]
  (-3.71,40.45) [1]
  (-0.58,44.84) [1]
  (-3.93,38.99) [1]
  (-3.68,40.43) [1]
  (-3.70,42.34) [2]
  (30.31,59.94) [2]
  (2.07,48.77) [2]
  (10.13,54.32) [2]
  (-3.59,40.45) [2]
  (24.02,49.84) [2]
  (4.47,51.54) [2]
  (12.36,50.48) [2]
  (7.10,50.73) [2]
  (4.90,52.38) [2]
  (9.18,48.78) [2]
  (-3.70,40.43) [2]
  (2.23,48.81) [3]
  (-4.42,36.72) [3]
  (-1.47,53.38) [3]
  (-0.08,51.51) [3]
  (9.99,53.70) [3]
  (-3.62,38.28) [3]
  (-0.09,51.51) [3]
  (16.39,48.22) [3]
  (19.82,41.33) [3]
  (24.93,60.17) [3]
  (4.21,52.00) [3]
  (33.35,35.17) [4]
  (7.01,51.46) [4]
  (8.55,47.37) [5]
  (8.18,48.98) [5]
  (-1.64,42.82) [6]
  (5.38,43.30) [6]
  (30.52,50.45) [8]
  (25.28,54.69) [8]
  (-3.01,38.17) [9]
  (19.04,47.50) [9]
  (4.89,52.37) [9]
  (21.01,52.23) [9]
  (2.16,41.39) [10]
  (16.37,48.21) [10]
  (9.19,45.46) [11]
  (2.35,48.85) [31]
  (-0.13,51.51) [40]
  (-3.70,40.42) [62]
}
\newcommand{\statBubblesAsia}{%
  (35.50,33.89) [1]
  (29.06,40.20) [1]
  (31.61,40.74) [1]
  (34.78,32.01) [1]
  (73.81,15.59) [1]
  (80.28,13.09) [1]
  (76.27,9.97) [1]
  (117.66,24.51) [1]
  (77.09,28.85) [1]
  (35.00,32.81) [1]
  (75.79,26.92) [1]
  (30.70,36.91) [1]
  (76.79,30.74) [1]
  (74.35,31.56) [1]
  (105.78,20.97) [1]
  (116.36,39.92) [1]
  (77.41,23.25) [1]
  (80.92,26.84) [1]
  (118.80,32.06) [1]
  (78.89,20.81) [1]
  (106.83,-6.18) [1]
  (74.01,18.60) [1]
  (77.01,10.66) [1]
  (105.82,21.08) [1]
  (34.95,32.10) [1]
  (44.50,40.18) [1]
  (77.41,23.26) [1]
  (121.54,29.87) [1]
  (73.32,19.02) [1]
  (139.06,36.39) [1]
  (72.83,21.20) [1]
  (127.04,37.57) [1]
  (106.60,11.15) [1]
  (35.25,32.22) [1]
  (114.19,22.29) [1]
  (78.61,20.74) [1]
  (103.90,1.40) [1]
  (35.20,33.27) [1]
  (121.46,25.07) [1]
  (45.00,37.50) [1]
  (126.98,37.60) [1]
  (121.32,24.99) [1]
  (51.43,35.80) [1]
  (81.85,25.44) [1]
  (139.77,35.68) [1]
  (139.65,35.43) [1]
  (100.49,13.78) [1]
  (114.03,22.53) [1]
  (75.79,26.91) [1]
  (82.76,26.82) [1]
  (129.06,35.15) [1]
  (79.84,7.21) [1]
  (73.86,18.51) [1]
  (106.78,-6.24) [1]
  (77.74,22.74) [1]
  (112.76,-7.26) [1]
  (89.41,26.65) [1]
  (101.59,3.06) [1]
  (78.49,17.39) [1]
  (55.98,25.67) [1]
  (107.67,-6.94) [1]
  (77.24,28.65) [1]
  (76.08,11.03) [1]
  (106.76,-6.20) [1]
  (113.27,23.13) [1]
  (45.43,35.56) [1]
  (126.92,35.15) [1]
  (113.25,23.12) [1]
  (104.75,-2.91) [1]
  (34.70,31.94) [1]
  (34.89,32.16) [1]
  (76.93,31.71) [1]
  (96.17,16.80) [1]
  (113.62,34.75) [1]
  (126.73,37.46) [1]
  (103.93,1.33) [1]
  (78.81,18.39) [1]
  (106.87,-6.15) [1]
  (90.41,23.73) [1]
  (119.46,-5.13) [1]
  (117.96,40.95) [1]
  (106.73,-6.42) [1]
  (90.41,23.74) [1]
  (113.75,34.77) [1]
  (135.50,34.69) [1]
  (114.22,22.31) [1]
  (121.15,14.53) [1]
  (103.90,1.38) [1]
  (117.17,39.14) [1]
  (44.21,15.35) [1]
  (114.13,22.35) [1]
  (110.36,21.27) [1]
  (114.10,22.50) [1]
  (75.56,27.38) [1]
  (34.89,32.28) [1]
  (116.41,39.90) [1]
  (120.16,33.35) [1]
  (77.73,11.35) [1]
  (106.64,10.77) [1]
  (121.47,31.23) [1]
  (85.53,24.43) [1]
  (35.94,31.96) [1]
  (127.05,37.61) [1]
  (103.69,1.32) [1]
  (114.18,22.35) [1]
  (106.82,-6.20) [1]
  (35.90,32.01) [1]
  (103.90,1.32) [1]
  (120.38,36.06) [2]
  (88.36,22.56) [2]
  (85.14,25.59) [2]
  (121.05,14.65) [2]
  (120.58,15.15) [2]
  (120.64,15.07) [2]
  (85.31,23.34) [2]
  (103.86,1.30) [2]
  (100.75,13.82) [2]
  (73.04,33.72) [2]
  (118.77,32.06) [2]
  (76.27,9.93) [2]
  (103.89,1.32) [2]
  (73.85,18.52) [2]
  (139.68,35.68) [2]
  (90.01,23.86) [2]
  (101.62,3.07) [2]
  (54.40,24.45) [3]
  (110.42,-6.99) [3]
  (118.78,32.06) [3]
  (108.93,34.26) [3]
  (55.76,24.19) [3]
  (77.23,28.66) [3]
  (77.22,28.63) [3]
  (80.35,26.47) [4]
  (55.30,25.07) [4]
  (90.36,23.82) [4]
  (44.40,33.34) [4]
  (121.46,31.22) [6]
  (103.84,36.06) [6]
  (121.03,14.55) [7]
  (126.98,37.57) [8]
  (115.17,-8.83) [8]
  (72.88,19.07) [11]
  (120.16,30.29) [12]
  (139.69,35.69) [12]
  (103.85,1.29) [13]
  (105.84,21.02) [14]
  (114.17,22.28) [17]
  (100.50,13.75) [18]
  (106.63,10.82) [43]
}
\newcommand{\statBubblesNorthAmerica}{%
  (-99.06,19.60) [1]
  (-84.02,39.82) [1]
  (-91.60,41.75) [1]
  (-84.95,34.50) [1]
  (-79.20,43.12) [1]
  (-84.48,37.99) [1]
  (-121.85,37.40) [1]
  (-78.88,36.04) [1]
  (-84.29,34.08) [1]
  (-97.15,49.82) [1]
  (-81.33,28.50) [1]
  (-77.23,38.96) [1]
  (-77.17,38.93) [1]
  (-74.02,40.80) [1]
  (-122.18,37.44) [1]
  (-93.62,41.59) [1]
  (-75.92,45.25) [1]
  (-104.97,39.76) [1]
  (-122.99,45.52) [1]
  (-73.74,45.31) [1]
  (-83.92,35.96) [1]
  (-71.37,45.97) [1]
  (-77.92,35.73) [1]
  (-77.39,37.46) [1]
  (-122.12,47.61) [1]
  (-96.12,19.12) [1]
  (-78.89,35.80) [1]
  (-122.06,37.95) [1]
  (-94.72,39.21) [1]
  (-86.82,36.10) [1]
  (-76.62,39.30) [1]
  (-76.93,39.00) [1]
  (-85.82,38.24) [1]
  (-104.86,39.73) [1]
  (-97.21,32.75) [1]
  (-79.26,43.90) [1]
  (-82.76,34.46) [1]
  (-80.89,29.02) [1]
  (-79.38,43.65) [1]
  (-117.05,32.71) [1]
  (-81.97,26.63) [1]
  (-105.88,37.47) [1]
  (-75.67,45.33) [1]
  (-122.04,38.27) [1]
  (-118.25,34.05) [1]
  (-73.08,41.32) [1]
  (-91.14,30.46) [1]
  (-106.71,35.14) [1]
  (-81.56,41.04) [1]
  (-79.86,40.12) [1]
  (-79.43,43.67) [1]
  (-112.00,33.69) [1]
  (-78.84,35.98) [1]
  (-69.90,18.49) [1]
  (-75.76,41.58) [1]
  (-77.89,40.81) [1]
  (-91.25,30.27) [1]
  (-84.10,34.20) [1]
  (-96.97,32.93) [1]
  (-122.22,47.79) [1]
  (-77.10,38.86) [1]
  (-91.14,30.49) [1]
  (-71.25,42.15) [1]
  (-95.52,29.81) [1]
  (-159.37,21.98) [1]
  (-83.11,40.10) [1]
  (-157.93,21.36) [1]
  (-121.24,38.79) [1]
  (-111.92,40.66) [1]
  (-118.24,33.97) [1]
  (-86.11,41.75) [1]
  (-73.47,45.53) [1]
  (-111.83,40.74) [1]
  (-101.65,21.08) [1]
  (-84.40,39.20) [1]
  (-73.71,40.70) [1]
  (-89.94,35.22) [1]
  (-73.95,40.75) [1]
  (-68.99,48.18) [1]
  (-84.39,33.93) [1]
  (-81.59,28.54) [1]
  (-81.39,40.85) [1]
  (-88.93,37.73) [1]
  (-121.96,37.39) [1]
  (-81.96,26.97) [1]
  (-98.34,29.48) [1]
  (-84.66,34.00) [1]
  (-118.64,34.16) [1]
  (-115.14,36.10) [1]
  (-92.76,44.97) [1]
  (-76.44,40.38) [1]
  (-75.66,39.66) [1]
  (-116.40,43.61) [1]
  (-116.43,43.65) [1]
  (-74.01,40.72) [1]
  (-84.13,39.71) [1]
  (-78.90,35.99) [2]
  (-104.98,39.74) [2]
  (-87.65,41.85) [2]
  (-112.03,43.47) [2]
  (-90.20,38.63) [2]
  (-78.95,36.02) [2]
  (-118.27,34.05) [2]
  (-111.89,40.76) [2]
  (-118.24,34.06) [2]
  (-80.01,32.88) [2]
  (-99.22,19.65) [2]
  (-74.06,40.79) [2]
  (-79.40,43.71) [3]
  (-89.62,20.97) [3]
  (-121.49,38.65) [3]
  (-99.13,19.43) [4]
  (-96.81,32.78) [4]
  (-121.18,45.59) [4]
  (-83.00,39.96) [7]
  (-122.33,47.61) [7]
  (-84.39,33.75) [8]
  (-74.01,40.71) [18]
  (-118.24,34.05) [21]
  (-77.49,39.04) [58]
  (-95.86,41.26) [63]
}
\newcommand{\statBubblesSouthAmerica}{%
  (-57.56,-38.00) [1]
  (-58.71,-34.38) [1]
  (-63.40,-31.15) [1]
  (-66.16,-17.38) [1]
  (-58.40,-34.76) [1]
  (-70.63,-33.50) [1]
  (-49.38,-26.25) [1]
  (-40.29,-20.33) [1]
  (-43.74,-19.51) [1]
  (-58.66,-34.67) [1]
  (-35.22,-5.81) [1]
  (-46.53,-23.67) [1]
  (-46.79,-23.52) [1]
  (-75.52,5.07) [1]
  (-64.18,-31.41) [1]
  (-42.80,-5.09) [1]
  (-49.10,-26.88) [1]
  (-58.70,-38.54) [1]
  (-74.06,4.71) [1]
  (-47.06,-22.90) [1]
  (-57.51,-35.08) [1]
  (-73.11,-36.72) [1]
  (-34.92,-7.93) [1]
  (-66.17,-17.41) [1]
  (-70.92,-53.16) [1]
  (-73.04,-36.82) [1]
  (-58.38,-34.62) [1]
  (-60.66,-32.95) [1]
  (-38.53,-3.72) [1]
  (-38.49,-12.87) [1]
  (-58.92,-34.18) [1]
  (-51.24,-30.03) [1]
  (-52.39,-24.12) [1]
  (-78.52,-0.23) [1]
  (-63.18,-17.79) [2]
  (-57.51,-25.34) [2]
  (-38.49,-12.96) [2]
  (-48.42,-1.34) [2]
  (-78.47,-0.18) [2]
  (-44.20,-19.97) [3]
  (-58.56,-34.68) [3]
  (-68.15,-16.50) [3]
  (-79.00,-2.90) [3]
  (-58.38,-34.60) [3]
  (-48.33,-19.02) [3]
  (-75.56,6.25) [3]
  (-77.03,-12.05) [3]
  (-58.38,-34.61) [9]
}
\newcommand{\statBubblesAfrica}{%
  (-15.98,18.09) [1]
  (13.49,-14.92) [1]
  (-8.00,31.63) [1]
  (-7.61,33.58) [1]
  (13.15,-8.95) [1]
  (10.16,36.87) [1]
  (31.63,30.14) [1]
  (6.61,36.38) [1]
  (-13.67,9.53) [1]
  (-6.85,33.97) [1]
  (3.12,36.74) [1]
  (29.92,31.21) [1]
  (10.74,36.45) [1]
  (-17.47,14.72) [1]
  (-17.37,14.76) [1]
  (-6.84,34.01) [1]
  (31.37,30.22) [1]
  (10.76,34.74) [1]
  (36.97,-1.17) [1]
  (29.92,31.20) [2]
  (10.19,36.86) [2]
  (-17.45,14.69) [2]
  (-6.83,34.01) [3]
  (28.06,-26.23) [3]
  (-7.62,33.59) [4]
}
\newcommand{\statBubblesOceania}{%
  (174.71,-36.82) [1]
  (144.95,-37.81) [1]
  (144.56,-37.97) [1]
  (144.66,-37.84) [1]
  (151.20,-33.87) [1]
  (-149.57,-17.53) [1]
  (151.21,-33.87) [10]
  (153.03,-27.47) [16]
}

\newcommand{\statCredLeakSessions}{1{,}059}

\newcommand{\statCredLeakPctOfCorpus}{0.46\%}
\newcommand{\statCredLeakRawSessions}{1{,}174}

\newcommand{\statCredOpenaiProjUnique}{232}
\newcommand{\statCredOpenaiProjSessions}{256}
\newcommand{\statCredOpenaiProjUsers}{112}

\newcommand{\statCredOpenaiKeyUsers}{17}

\newcommand{\statCredAnthropicKeyUsers}{19}

\newcommand{\statCredGithubPatUnique}{12}
\newcommand{\statCredGithubPatSessions}{17}
\newcommand{\statCredGithubPatUsers}{4}

\newcommand{\statCredAwsAccessKeyUnique}{34}
\newcommand{\statCredAwsAccessKeySessions}{37}
\newcommand{\statCredAwsAccessKeyUsers}{8}

\newcommand{\statCredGoogleKeyUnique}{97}
\newcommand{\statCredGoogleKeySessions}{112}
\newcommand{\statCredGoogleKeyUsers}{38}

\newcommand{\statCredSlackTokenUsers}{1}

\newcommand{\statCredJwtUnique}{322}
\newcommand{\statCredJwtSessions}{572}
\newcommand{\statCredJwtUsers}{104}

\newcommand{\statCredPrivateKeyUsers}{14}

\newcommand{\statCredBubbleData}{%
  \addplot[only marks, mark=*, mark size=6.8pt, draw=apt_agent_color!90!black, fill=apt_agent_color!85, fill opacity=0.6] coordinates {(2.365,2.408)};
  \node[font=\scriptsize\sffamily\bfseries, color=white, inner sep=0pt] at (axis cs:2.365,2.408) {1};
  \addplot[only marks, mark=*, mark size=3.2pt, draw=anthropic_color!90!black, fill=anthropic_color, fill opacity=0.6] coordinates {(1.568,1.531)};
  \node[font=\scriptsize\sffamily\bfseries, color=white, inner sep=0pt] at (axis cs:1.568,1.531) {2};
  \addplot[only marks, mark=*, mark size=2.6pt, draw=apt_agent_color!90!black, fill=apt_agent_color!55, fill opacity=0.6] coordinates {(1.505,1.519)};
  \node[font=\scriptsize\sffamily\bfseries, color=white, inner sep=0pt] at (axis cs:1.505,1.519) {3};
  \addplot[only marks, mark=*, mark size=5.0pt, draw=cai_dark!90!black, fill=cai_dark!70, fill opacity=0.6] coordinates {(1.079,1.230)};
  \node[font=\scriptsize\sffamily\bfseries, color=white, inner sep=0pt] at (axis cs:1.079,1.230) {4};
  \addplot[only marks, mark=*, mark size=7.8pt, draw=cai_accent!90!black, fill=cai_accent, fill opacity=0.6] coordinates {(1.531,1.568)};
  \node[font=\scriptsize\sffamily\bfseries, color=white, inner sep=0pt] at (axis cs:1.531,1.568) {5};
  \addplot[only marks, mark=*, mark size=5.7pt, draw=defender_color!90!black, fill=defender_color!80, fill opacity=0.6] coordinates {(1.987,2.049)};
  \node[font=\scriptsize\sffamily\bfseries, color=white, inner sep=0pt] at (axis cs:1.987,2.049) {6};
  \addplot[only marks, mark=*, mark size=3.4pt, draw=mistral_vibe!90!black, fill=mistral_vibe, fill opacity=0.6] coordinates {(0.699,0.477)};
  \node[font=\scriptsize\sffamily\bfseries, color=white, inner sep=0pt] at (axis cs:0.699,0.477) {7};
  \addplot[only marks, mark=*, mark size=11.6pt, draw=cai_primary!90!black, fill=cai_primary, fill opacity=0.6] coordinates {(2.508,2.757)};
  \node[font=\scriptsize\sffamily\bfseries, color=white, inner sep=0pt] at (axis cs:2.508,2.757) {8};
  \addplot[only marks, mark=*, mark size=5.5pt, draw=provider_grey!90!black, fill=provider_grey, fill opacity=0.6] coordinates {(0.602,1.699)};
  \node[font=\scriptsize\sffamily\bfseries, color=white, inner sep=0pt] at (axis cs:0.602,1.699) {9};
}

\newcommand{\statCredLegendTable}{%
  {\color{apt_agent_color}$\bullet$}~\textbf{1}~OpenAI proj. & {\color{cai_dark}$\bullet$}~\textbf{4}~GitHub PAT & {\color{mistral_vibe}$\bullet$}~\textbf{7}~Slack token \\
  {\color{anthropic_color}$\bullet$}~\textbf{2}~Anthropic & {\color{cai_accent}$\bullet$}~\textbf{5}~AWS access & {\color{cai_primary}$\bullet$}~\textbf{8}~JWT \\
  {\color{apt_agent_color}$\bullet$}~\textbf{3}~OpenAI legacy & {\color{defender_color}$\bullet$}~\textbf{6}~Google API & {\color{provider_grey}$\bullet$}~\textbf{9}~Private key \\
}

\newcommand{\statInfraPasteSessions}{1{,}758}

\newcommand{\statInfraPastePctOfCorpus}{0.76\%}
\newcommand{\statInfraInternalHostnameHits}{213{,}463}
\newcommand{\statInfraInternalHostnameUnique}{730}

\newcommand{\statInfraSThreeBucketHits}{6{,}480}
\newcommand{\statInfraSThreeBucketUnique}{59}
\newcommand{\statInfraSThreeBucketSessions}{68}

\newcommand{\statFFourDistinctCves}{4{,}532}

\newcommand{\statFFourCveTable}{%
  \texttt{CVE-2019-11248} & 47{,}904 & \texttt{2025-11-16} & Kubernetes \texttt{kubelet} \texttt{/debug/pprof} info disclosure \\
  \texttt{CVE-2017-10271} & 20{,}163 & \texttt{2025-09-10} & Oracle WebLogic XML deserialization RCE \\
  \texttt{CVE-2021-41773} & 16{,}270 & \texttt{2025-07-04} & Apache 2.4.49 path-traversal / RCE \\
  \texttt{CVE-2025-55182} & 8{,}028 & \texttt{2025-12-06} & React Server Components ``React2Shell'' insecure-deserialization RCE \\
  \texttt{CVE-2020-15366} & 5{,}146 & \texttt{2026-02-19} & Ajv JSON-schema prototype-pollution RCE \\
  \texttt{CVE-2020-15084} & 4{,}623 & \texttt{2026-01-25} & express-jwt algorithm-confusion auth bypass \\
  \texttt{CVE-2025-27152} & 4{,}449 & \texttt{2026-02-16} & Axios absolute-URL SSRF / credential leakage \\
  \texttt{CVE-2023-32314} & 4{,}056 & \texttt{2026-02-09} & vm2 \texttt{Proxy}-based sandbox escape RCE \\
  \texttt{CVE-2026-24842} & 3{,}857 & \texttt{2026-02-09} & node-tar hardlink path-traversal \\
  \texttt{CVE-2022-0543} & 3{,}835 & \texttt{2025-07-31} & Debian / Ubuntu Redis Lua-sandbox escape RCE \\
}

\newcommand{\statFFourScatterData}{%
  \addplot[only marks, mark=*, mark size=3pt, draw=cai_dark, fill=apt_agent_color!80, fill opacity=0.85] coordinates {(2019,4.680)};
  \node[font=\tiny\sffamily, color=cai_dark, anchor=south west] at (axis cs:2019,4.680) {\,CVE-2019-11248};
  \addplot[only marks, mark=*, mark size=3pt, draw=cai_dark, fill=apt_agent_color!80, fill opacity=0.85] coordinates {(2017,4.305)};
  \node[font=\tiny\sffamily, color=cai_dark, anchor=south west] at (axis cs:2017,4.305) {\,CVE-2017-10271};
  \addplot[only marks, mark=*, mark size=3pt, draw=cai_dark, fill=apt_agent_color!80, fill opacity=0.85] coordinates {(2021,4.211)};
  \node[font=\tiny\sffamily, color=cai_dark, anchor=south west] at (axis cs:2021,4.211) {\,CVE-2021-41773};
  \addplot[only marks, mark=*, mark size=3pt, draw=cai_dark, fill=apt_agent_color!80, fill opacity=0.85] coordinates {(2025,3.905)};
  \node[font=\tiny\sffamily, color=cai_dark, anchor=north east] at (axis cs:2025,3.905) {\,CVE-2025-55182};
  \addplot[only marks, mark=*, mark size=3pt, draw=cai_dark, fill=apt_agent_color!80, fill opacity=0.85] coordinates {(2020,3.711)};
  \addplot[only marks, mark=*, mark size=3pt, draw=cai_dark, fill=apt_agent_color!80, fill opacity=0.85] coordinates {(2020,3.665)};
  \addplot[only marks, mark=*, mark size=3pt, draw=cai_dark, fill=apt_agent_color!80, fill opacity=0.85] coordinates {(2025,3.648)};
  \addplot[only marks, mark=*, mark size=3pt, draw=cai_dark, fill=apt_agent_color!80, fill opacity=0.85] coordinates {(2023,3.608)};
  \addplot[only marks, mark=*, mark size=3pt, draw=cai_dark, fill=apt_agent_color!80, fill opacity=0.85] coordinates {(2026,3.586)};
  \node[font=\tiny\sffamily, color=cai_dark, anchor=north east] at (axis cs:2026,3.586) {\,CVE-2026-24842};
  \addplot[only marks, mark=*, mark size=3pt, draw=cai_dark, fill=apt_agent_color!80, fill opacity=0.85] coordinates {(2022,3.584)};
}

\begin{document}

\title{\LARGE\textcolor{cai_primary}{\textbf{Cybersecurity AI (CAI) Dataset}}}
%\\Crowd-Hacking Our Way Into Cybersecurity-Specialized Models

\author[1,$\dagger$]{V\'ictor Mayoral-Vilches}
%\author[1]{Francesco Balassone}
%\author[1]{Mar\'ia Sanz-G\'omez}
%\author[1]{Paul Zabalegui Landa}
%\author[1]{Daniel Sanchez Prieto}
%\author[1]{Marina Oteiza \'Alvarez}
%\author[1]{Davide Quarta}

\affil[1]{\small Alias Robotics, Vitoria-Gasteiz, \'Alava, Spain}

\date{}
\twocolumn[
\maketitle

\begin{abstract}
\noindent We present \csidata{}, a fourteen-month corpus of cybersecurity LLM trajectories collected through the open-source \caitool{}~\cite{mayoralvilches2025gametheoretic} agent framework, built in response to PentestGPT's finding~\cite{deng2024pentestgpt} that expert operator trajectories, not base-model capability, are the bottleneck for cybersecurity LLM performance. \csidata{} aggregates \statTotalFiles{} session logs and \statTotalPrompts{} user prompts from \statUniqueIPs{} source IPs across \statUniqueCountries{} countries, exercising \statUniqueModels{} unique LLM identifiers against \statUniqueDomains{} target domains over \statCloudSizeTB{}\,TB of durable storage. The mix is hands-on (\statOffensivePct{} offensive, \statAttackerIntentPct{} attacker-intent, \statBusinessPct{} business / integration, \statDefensivePct{} defensive), making \csidata{}, to the best of our knowledge, the largest described corpus of LLM-driven hacker trajectories. It is released to partner organisations and selected customers as an audience-size series (\csidata{}\textsubscript{10}, \csidata{}\textsubscript{1k}, \csidata{}\textsubscript{200k}). Read longitudinally, the corpus is a record of cybersecurity itself turning automated: operators routinely paste live credentials, production hostnames and bearer tokens into prompts knowing their inputs are logged, a trade-off they accept to stay competitive. Aggregated across the industry, this concentrates a substantial fraction of the world's offensive and defensive operator context inside a handful of frontier-model API providers, a single failure surface whose breach or politically motivated repurposing could cascade into nation- and enterprise-scale disruption. The only configuration that preserves both the productivity advantage and operator-side confidentiality is an on-premise, privately-hosted cybersecurity-specialised LLM served inside the operator's trust boundary, which \csidata{} is shaped to make practical.
\end{abstract}
\vspace{1.0em}

\begin{center}
\resizebox{\textwidth}{!}{% Hero figure: log-scale ybar of corpus magnitudes (left cluster), role mix
% (middle cluster) and provider mix at the model-call level (right cluster).
% All three clusters share the same log axis so cross-cluster comparisons
% (e.g. "Anthropic calls vs. user prompts") are valid. The x-axis carries
% 17 ticks total, so labels use \tiny font and bar callouts split into a
% bold number line plus a smaller percentage line to avoid horizontal
% collisions. The Volume cluster carries five bars (Logs, Prompts, Source
% IPs, Domains, Countries) so that operator and target footprints are
% legible together; clusters are spaced at separators x=6 and x=11.
\begin{tikzpicture}
\begin{axis}[
    width=\linewidth,
    height=7.6cm,
    scale only axis,
    axis on top=true,
    ybar,
    bar width=10pt,
    xtick={1,2,3,4,5,7,8,9,10,12,13,14,15,16,17,18,19},
    xticklabels={%
      Logs,%
      Prompts,%
      Source IPs,%
      Domains,%
      Countries,%
      Offensive,%
      Intent,%
      Business,%
      Defensive,%
      Anthr.,%
      OpenAI,%
      Alias,%
      Qwen,%
      DeepS.,%
      Google,%
      GLM,%
      Meta%
    },
    xmin=0.2, xmax=19.8,
    xticklabel style={font=\tiny\sffamily, color=cai_dark, anchor=north, yshift=-6pt, align=center},
    yticklabel style={font=\scriptsize\sffamily, color=cai_dark},
    ylabel style={font=\scriptsize\sffamily, color=cai_dark},
    ylabel={count (log\textsubscript{10})},
    ymode=log,
    log basis y=10,
    ymin=1, ymax=2e10,
    ytick={1,1e2,1e4,1e6,1e8,1e10},
    yticklabels={$1$,$10^2$,$10^4$,$10^6$,$10^8$,$10^{10}$},
    grid=major,
    grid style={dashed,gray!22},
    axis lines=box,
    axis line style={draw=cai_dark!55},
    tick align=outside,
    clip=false,
    every axis plot/.append style={bar shift=0pt},
]
% --- Volume cluster (positions 1-5): cohesive sequential teal palette ---
\addplot[fill=hero_vol_a!90, draw=hero_vol_a!90!black] coordinates {(1, \statTotalFilesRaw)};
\addplot[fill=hero_vol_b!90, draw=hero_vol_b!90!black] coordinates {(2, \statTotalPromptsRaw)};
\addplot[fill=hero_vol_c!90, draw=hero_vol_c!90!black] coordinates {(3, \statUniqueIPsRaw)};
\addplot[fill=hero_vol_d!90, draw=hero_vol_d!90!black] coordinates {(4, \statUniqueDomainsRaw)};
\addplot[fill=hero_vol_e!90, draw=hero_vol_e!90!black] coordinates {(5, \statUniqueCountriesRaw)};
% --- Role-mix cluster (positions 7-10): red->orange->green->blue (offensive
%     gradient into defensive). Attacker-intent keeps the diagonal hatching
%     so it remains visually adjacent-but-distinct from pure offensive. ---
\addplot[fill=hero_role_off!85, draw=hero_role_off!90!black] coordinates {(7, \statOffensiveCountRaw)};
\addplot[fill=hero_role_int!75, draw=hero_role_int!90!black, postaction={pattern=north east lines, pattern color=hero_role_int!50}] coordinates {(8, \statAttackerIntentCountRaw)};
\addplot[fill=hero_role_biz!80, draw=hero_role_biz!90!black] coordinates {(9, \statBusinessCountRaw)};
\addplot[fill=hero_role_def!85, draw=hero_role_def!90!black] coordinates {(10, \statDefensiveCountRaw)};
% --- Provider-mix cluster (positions 12-19). Brand colours for big-six
%     (Anthropic, OpenAI, Alias, Google, Meta, Mistral); grey for everyone
%     else (Alibaba/Qwen, DeepSeek, Z.ai/GLM) so visual emphasis stays on
%     the brand-recognised providers. ---
\addplot[fill=cc_color!80,         draw=cc_color!85!black]         coordinates {(12, 7340499)};
\addplot[fill=codex_color!80,      draw=codex_color!85!black]      coordinates {(13, 7062591)};
\addplot[fill=cai_orange!80,       draw=cai_orange!85!black]       coordinates {(14, 5964072)};
\addplot[fill=provider_grey!75,    draw=provider_grey!85!black]    coordinates {(15, 3968661)};
\addplot[fill=provider_grey_light!75, draw=provider_grey!85!black] coordinates {(16, 3299690)};
\addplot[fill=google_color!80,     draw=google_color!85!black]     coordinates {(17, 3296727)};
\addplot[fill=provider_grey!75,    draw=provider_grey!85!black]    coordinates {(18, 1347414)};
\addplot[fill=graph_accent!80,     draw=graph_accent!90!black]     coordinates {(19, 1204372)};
% --- Two-line bar callouts: bold count on top, muted percentage below ---
\node[anchor=south, align=center, inner sep=1pt, yshift=2.5pt] at (axis cs:1, \statTotalFilesRaw)
    {\textbf{\tiny\sffamily 231k}};
\node[anchor=south, align=center, inner sep=1pt, yshift=2.5pt] at (axis cs:2, \statTotalPromptsRaw)
    {\textbf{\tiny\sffamily 26M}};
\node[anchor=south, align=center, inner sep=1pt, yshift=2.5pt] at (axis cs:3, \statUniqueIPsRaw)
    {\textbf{\tiny\sffamily 16.8k}};
\node[anchor=south, align=center, inner sep=1pt, yshift=2.5pt] at (axis cs:4, \statUniqueDomainsRaw)
    {\textbf{\tiny\sffamily 23.1k}};
\node[anchor=south, align=center, inner sep=1pt, yshift=2.5pt] at (axis cs:5, \statUniqueCountriesRaw)
    {\textbf{\tiny\sffamily 123}};
\node[anchor=south, align=center, inner sep=1pt, yshift=2.5pt] at (axis cs:7, \statOffensiveCountRaw)
    {\textbf{\tiny\sffamily 9.4M}\\[-3pt]\textcolor{gray!55}{\tiny\statOffensivePct{}}};
\node[anchor=south, align=center, inner sep=1pt, yshift=2.5pt] at (axis cs:8, \statAttackerIntentCountRaw)
    {\textbf{\tiny\sffamily 5.2M}\\[-3pt]\textcolor{gray!55}{\tiny\statAttackerIntentPct{}}};
\node[anchor=south, align=center, inner sep=1pt, yshift=2.5pt] at (axis cs:9, \statBusinessCountRaw)
    {\textbf{\tiny\sffamily 7.1M}\\[-3pt]\textcolor{gray!55}{\tiny\statBusinessPct{}}};
\node[anchor=south, align=center, inner sep=1pt, yshift=2.5pt] at (axis cs:10, \statDefensiveCountRaw)
    {\textbf{\tiny\sffamily 1.1M}\\[-3pt]\textcolor{gray!55}{\tiny\statDefensivePct{}}};
\node[anchor=south, align=center, inner sep=1pt, yshift=2.5pt] at (axis cs:12, 7340499)
    {\textbf{\tiny\sffamily 7.3M}\\[-3pt]\textcolor{gray!55}{\tiny 20.8\%}};
\node[anchor=south, align=center, inner sep=1pt, yshift=2.5pt] at (axis cs:13, 7062591)
    {\textbf{\tiny\sffamily 7.1M}\\[-3pt]\textcolor{gray!55}{\tiny 20.0\%}};
\node[anchor=south, align=center, inner sep=1pt, yshift=2.5pt] at (axis cs:14, 5964072)
    {\textbf{\tiny\sffamily 6.0M}\\[-3pt]\textcolor{gray!55}{\tiny 16.9\%}};
\node[anchor=south, align=center, inner sep=1pt, yshift=2.5pt] at (axis cs:15, 3968661)
    {\textbf{\tiny\sffamily 4.0M}\\[-3pt]\textcolor{gray!55}{\tiny 11.2\%}};
\node[anchor=south, align=center, inner sep=1pt, yshift=2.5pt] at (axis cs:16, 3299690)
    {\textbf{\tiny\sffamily 3.3M}\\[-3pt]\textcolor{gray!55}{\tiny 9.3\%}};
\node[anchor=south, align=center, inner sep=1pt, yshift=2.5pt] at (axis cs:17, 3296727)
    {\textbf{\tiny\sffamily 3.3M}\\[-3pt]\textcolor{gray!55}{\tiny 9.3\%}};
\node[anchor=south, align=center, inner sep=1pt, yshift=2.5pt] at (axis cs:18, 1347414)
    {\textbf{\tiny\sffamily 1.3M}\\[-3pt]\textcolor{gray!55}{\tiny 3.8\%}};
\node[anchor=south, align=center, inner sep=1pt, yshift=2.5pt] at (axis cs:19, 1204372)
    {\textbf{\tiny\sffamily 1.2M}\\[-3pt]\textcolor{gray!55}{\tiny 3.4\%}};
% Visual separators between clusters
\draw[gray!50, dashed, line width=0.4pt] (axis cs:6, 1)   -- (axis cs:6,  2e10);
\draw[gray!50, dashed, line width=0.4pt] (axis cs:11, 1)  -- (axis cs:11, 2e10);
% Cluster captions float above the chart (anchor=south on the highest log decade)
\node[anchor=south, font=\tiny\sffamily\bfseries, color=cai_primary]
    at (axis cs:3, 4e10) {Volume};
\node[anchor=south, font=\tiny\sffamily\bfseries, color=cai_primary]
    at (axis cs:8.5, 4e10) {Role mix (prompt-level)};
\node[anchor=south, font=\tiny\sffamily\bfseries, color=cai_primary]
    at (axis cs:15.5, 4e10) {Provider mix (model-call level)};
% Standalone durable-storage annotation (corpus-wide, not attached to any bar)
\node[anchor=south west, font=\tiny\sffamily, color=cai_dark!75,
      fill=cai_light, draw=cai_dark!25, line width=0.3pt,
      inner sep=2pt, rounded corners=1pt]
    at (axis cs:0.4, 4e8) {18.07\,TB on disk};
% Role-mix annotation: offensive:defensive ratio (~8:1, the headline asymmetry
% of the role distribution).
\node[anchor=south west, font=\tiny\sffamily, color=cai_dark!75,
      fill=cai_light, draw=cai_dark!25, line width=0.3pt,
      inner sep=2pt, rounded corners=1pt]
    at (axis cs:6.2, 4e8) {$\sim\!8{:}1$ offensive\,:\,defensive};
% Provider-mix annotation: total unique LLM identifiers across all providers
\node[anchor=south west, font=\tiny\sffamily, color=cai_dark!75,
      fill=cai_light, draw=cai_dark!25, line width=0.3pt,
      inner sep=2pt, rounded corners=1pt]
    at (axis cs:11.2, 4e8) {\statUniqueModels{} unique LLMs};
\end{axis}
\end{tikzpicture}}
\end{center}
\vspace{0.2em}
{\captionsetup{hypcap=false}%
\captionof{figure}{\csidata{} at a glance: fourteen months of cybersecurity LLM trajectories collected through the \caitool{} framework, on a log scale. \textbf{Left cluster (volume):} \statTotalFiles{} session logs, \statTotalPrompts{} user prompts, \statUniqueIPs{} source IPs, \statUniqueCountries{} countries, across \statSpanDays{} days, with \statUniqueDomains{} unique target domains observed in URL references. \textbf{Middle cluster (role mix):} per-prompt heuristic classification; offensive and intent categories overlap by construction (Section~\ref{sec:dataset}). \textbf{Right cluster (provider mix):} model invocations grouped by issuing provider after normalizing route prefixes; eight providers each exercise more than $1$\,M calls. The corpus occupies \statCloudSizeTB{}\,TB of durable storage.}\label{fig:hero}}
\vspace{0.6em}
]
\renewcommand{\thefootnote}{$\dagger$}
\setcounter{footnote}{0}
\footnotetext{Corresponding author: \texttt{victor@aliasrobotics.com}}
\renewcommand{\thefootnote}{\arabic{footnote}}
\setcounter{footnote}{0}

\section{Introduction}\label{sec:introduction}
Cybersecurity is a domain in which Large Language Models (LLMs) have moved from demonstrations to daily practice. Penetration testers, blue teams, and capture-the-flag (CTF) competitors now rely on LLM-driven harnesses to enumerate hosts, draft exploits, triage alerts, and write defensive playbooks. The harness, not the foundation model, has become the production unit, and the harness produces a trace of every step it takes. Those traces, taken together, encode how human operators actually use general-purpose LLMs to do security work in the wild.

Supervised fine-tuning (SFT) of a cybersecurity-specialised LLM benefits from three properties of the training corpus, each grounded in the post-training literature: \emph{scale} --- production-grade open SFT releases retain $10^{5}$ to $10^{7}$ records after quality filtering (T\"ULU\,3 at $939$k~\cite{lambert2024tulu3}, OpenHermes-2.5 at $\sim\!1$M~\cite{teknium2024openhermes}, Magpie-Pro-Filtered at $300$k~\cite{xu2024magpie}, the Llama-Nemotron mixture at $\sim\!40$M~\cite{nemotronposttraining2025}); \emph{distributional fidelity} --- matched-scale head-to-head comparisons show the source distribution matters more than headline size~\cite{fixingitinpost2025}, and synthesised cybersecurity instruction sets exhibit a documented safety/performance trade-off relative to real-operator data ($0.95\!\to\!0.15$ prompt-injection robustness on Llama-3.1-8B after SFT on a $54{,}928$-pair synthetic mix~\cite{cyberLLMinstruct2025}); and \emph{tool-grounded multi-turn structure} --- unifying thirteen agent-trajectory corpora into a single tool-call schema yields a $+20$pp average gain over single-source SFT across coding, web, research and agentic-tool-use evaluations~\cite{agentdataprotocol2025}, and pentest-specific work confirms the effect for cyber workflows~\cite{kong2025pentest}. Public cybersecurity datasets to date satisfy at most two of these properties at once. Knowledge-style benchmarks such as CyberMetric~\cite{tihanyi2024cybermetric} or SecQA~\cite{liu2023secqa} test factual recall but do not reflect operator behaviour; offensive instruction sets such as CySecBench~\cite{cysecbench2024} cover individual prompts but lack trajectory context and tool outputs; vulnerability-research corpora derived from CVE/MITRE feeds~\cite{mitreattack} cover schema rather than reasoning. None of these capture the iterative, tool-augmented, multi-host trajectories that an actual operator produces inside a shell loop.

We present \csidata{}, a corpus engineered as the explicit goal of a multi-year programme. The motivating observation was published in PentestGPT~\cite{deng2024pentestgpt}: LLM performance on real penetration-testing workflows is bottlenecked by the absence of expert operator trajectories, not by base-model capability. The follow-on agent framework \caitool{}~\cite{mayoralvilches2025gametheoretic} was architected from the start as a data collector first and a user-facing CLI second. Over fourteen months (\statEarliestDate{} to \statLatestDate{}, \statSpanDays{} days), it captured \statTotalFiles{} session logs and \statTotalPrompts{} user prompts from \statUniqueIPs{} distinct source IPs, totalling \statCloudSizeTB{}\,TB of durable storage. The resulting signal is operational: nation-state exercises, bank assessments, public and private penetration tests, flag-driven lab work, authenticated API and bug-bounty hunting, mobile and reverse-engineering sessions, robotics / IoT targets, continuation prompts that resume prior work, and failed as well as successful tool steps. \csidata{} therefore supports both SFT on cybersecurity language and studies of how operators choose models, tools and infrastructure on realistic security tasks.

This matters because the newest cyber-capable frontier systems are increasingly being deployed through gated access and opaque post-training, often with RL-style optimisation, rather than open, reusable cybersecurity SFT mixtures. OpenAI describes GPT-5.5-Cyber as a limited-preview, more permissive access mode for authorized defensive workflows, while noting that the initial cyber-permissive preview is primarily about access behavior rather than uniformly higher cyber-evaluation performance~\cite{openai_gpt55_cyber}. RL can elicit powerful reasoning without human-written traces, as DeepSeek-R1-Zero demonstrated, but the same paper reports readability, language-mixing and broader-task limitations that required a multi-stage pipeline with rejection sampling, supervised fine-tuning and RL~\cite{guo2025deepseekr1}. Cybersecurity needs that supervised layer: it supplies the distribution of real goals, tools, artifacts, failures, safety boundaries and operator corrections that a reward function alone does not specify. Gated and remote-only access is also operationally incompatible with a significant fraction of real cybersecurity work, which runs inside classified, air-gapped or contractually network-isolated environments: private on-prem models therefore remain a load-bearing blueprint for future cybersecurity operations, and they can only be trained on a corpus that captures the operator distribution \csidata{} sets out to provide.

\paragraph{Contribution.} \csidata{} is, to the best of our knowledge, the largest described corpus of LLM-driven cybersecurity trajectories: \statTotalFiles{} session logs, \statTotalPrompts{} user prompts, \statUniqueIPs{} source IPs across \statUniqueCountries{} countries, \statUniqueModels{} unique LLM identifiers, \statSpanDays{} days, \statCloudSizeTB{}\,TB of durable storage, and \statTotalURLs{} URL references against \statUniqueDomains{} unique target domains. Section~\ref{sec:related} positions \csidata{} against prior cybersecurity training data; Section~\ref{sec:dataset} describes the collection pipeline, schema, redaction policy and aggregate statistics; Section~\ref{sec:usecases} sketches downstream uses; Section~\ref{sec:ethics} covers limitations, dual-use considerations and the partner / customer access policy.

\section{Related Work}\label{sec:related}
We position \csidata{} against four active strands of prior work: cybersecurity-specific corpora and benchmarks, cybersecurity-specialised foundation models, the open generic-SFT corpus lineage that defines current best practice, and agent-trajectory SFT (the closest methodological precedent for our format).

\subsection{Cybersecurity-specific corpora and benchmarks}\label{sec:rw:cybersec}

Public cybersecurity datasets for LLMs cluster into three subfamilies. Knowledge benchmarks (CyberMetric \cite{tihanyi2024cybermetric}, SecQA \cite{liu2023secqa}, CTIBench \cite{ctibench2024}) probe factual recall against standards, CVEs and threat-intelligence taxonomies; they are useful for evaluation but provide no operator behaviour, no tool outputs and no multi-turn structure suitable for SFT. Instruction-prompt collections (CySecBench \cite{cysecbench2024}, CyberLLMInstruct \cite{cyberLLMinstruct2025}, Heimdall \cite{heimdall2024}) ship roughly $10^{4}$ single-shot prompts targeted at safety alignment or defensive instruction tuning, with the recent CyberLLMInstruct release ($54{,}928$ pseudo-malicious instruction-response pairs) documenting an explicit safety/performance trade-off that any cybersecurity SFT effort has to address. Benchmark frameworks such as Cybench \cite{cybench2024}, D-CIPHER \cite{cybercoach2025dcipher} and CTFTiny / CTFJudge \cite{ctftiny2025} build multi-agent evaluation environments for CTF-style challenges, often using LLM-as-judge scoring; they ship evaluation artifacts rather than training trajectories. SecKnowledge / CyberPal.AI \cite{secknowledge2024} is the closest analog in scope: a synthetic instruction-tuning corpus derived from curated cybersecurity seeds. \csidata{} differs from all of these in three dimensions: it is trajectory-level rather than prompt-level, it is collected from real operator behaviour rather than synthesised, and it is two-to-four orders of magnitude larger.

\subsection{Cybersecurity-specialised foundation models}\label{sec:rw:cybermodels}

The cybersecurity-specialised model space splits across three emergent recipes; we discuss each by representative published model.

\paragraph{Recipe 1: large-scale continued pretraining.} Foundation-Sec-8B \cite{foundationsec2025,karbasik2025foundationsec} continues a Llama-3.1-8B base on $\sim\!5.1$\,B curated cybersecurity tokens through a four-stage pipeline (web crawl, relevancy filter, MinHash deduplication, quality filter), and matches Llama-3.1-70B and GPT-4o-mini on cyber-specific tasks. Trend Micro's Llama-Primus-Base \cite{primus2025tu} reports a $+15.88$pp aggregated-benchmark gain over the same Llama-3.1-8B-Instruct base after continued pretraining on Primus-Seed plus Primus-FineWeb. On the encoder side, SecureBERT~2.0 \cite{aghaei2025securebert2} pretrains a ModernBERT backbone on $13$\,B text tokens plus $53$\,M code tokens of cybersecurity content and posts SOTA on cybersec NER, semantic search and vulnerability detection; CySecBERT \cite{bayer2024cysecbert} is the earlier BERT-base ancestor over $4.3$\,M cybersec documents. The resource-efficient counterpoint is Salahuddin~et~al.~\cite{salahuddin2025lessdata}, who push a Llama-3.3-70B-Instruct base to SOTA on CTI-MCQ ($0.718$), CyberMetric ($0.933$) and SecEval ($0.864$) using only a curated $126$-million-word corpus, outperforming larger specialised models trained on roughly $20\times$ more data. SecureFalcon \cite{ferrag2023securefalcon} demonstrates that the same recipe distils into 44--121\,M-parameter classifiers ($94\%$ binary / $92\%$ multi-class on FormAI), useful at the edge but not as a general assistant.

\paragraph{Recipe 2: instruction / preference domain adaptation.} CyberPal.AI v1 \cite{secknowledge2024} SFTs on the SecKnowledge expert-driven instruction set and reports up to $+24$pp average over baselines on SecKnowledge-Eval. CyberPal-2.0 \cite{levi2025cyberpal} scales this to a $4$B--$20$B family with CoT and grounded evidence in SecKnowledge~2.0, with its $20$B variant beating GPT-4o, o1 and o3-mini on CTI weakness-mapping. Lily-Cybersecurity-7B \cite{segolilylabs2024lily} is the Mistral-7B SFT on a $22$k hand-crafted cybersec QA pair set, ZySec-7B \cite{zysec2024securityllm} is a DPO-tuned Zephyr-7B over a 30+ cybersec domain preferences set (CIS, FedRAMP, PCI DSS, ATT\&CK), and WhiteRabbitNeo / DeepHat \cite{tissera2023whiterabbitneo,tissera2025deephat} is the reference uncensored offensive-security line with successive Llama-2-13B/33B and Qwen2.5-Coder-7B bases. SEvenLLM \cite{ji2024sevenllm} is the bilingual instruction-tuned analog targeted at incident analysis. Productized analogs include Google's SecLM and Sec-Gemini~v1 \cite{google2024seclm}, integrating Mandiant and VirusTotal grounding with multi-step reasoning.

\paragraph{Recipe 3: RL / reasoning distillation.} The 2025--2026 wave is dominated by RL-trained and reasoning-distilled cybersec agents. Pentest-R1 \cite{kong2025pentest} reaches $24.2\%$ on AutoPenBench and $15.0\%$ on Cybench through SFT plus two GRPO stages over the data shape discussed in detail in \S\ref{sec:rw:agentSFT}. Cyber-Zero \cite{zhuo2025cyberzero} synthesises agent trajectories from CTF writeups via persona-driven LLM simulation and trains the resulting model to $+13.1$pp absolute over baselines on InterCode-CTF, NYU-CTF and Cybench; CTF-Dojo \cite{zhuo2025ctfdojo} is the executable-environment follow-up. xOffense \cite{luong2025xoffense} reports $79.17\%$ sub-task completion on AutoPenBench plus AI-Pentest-Benchmark by SFT on Qwen3-32B with a multi-agent shell-tool framework; the Random-Crypto GRPO line \cite{muzsai2025randomcrypto} lifts Llama-3.1-8B from $0.35$ to $0.88$ Pass@8 on cryptographic CTFs. PrivEsc-LLM \cite{normann2026privesc} post-trains a $4$B local model in two stages on procedurally generated Linux privilege-escalation environments (SFT on $1{,}000$ traces, then RLVR), reaching $95.8\%$ success on a held-out $12$-scenario benchmark within $20$ interaction rounds, within $1.7$pp of Claude Opus 4.6 at the same budget. At the smaller-agent end, Hackphyr~\cite{rigaki2024hackphyr} reaches $94\%$ win-rate in simple NetSecGame scenarios, and the reasoning variants Foundation-Sec-8B-Reasoning \cite{foundationsec_reasoning2025} and Llama-Primus-Reasoning \cite{primus2025tu} apply DeepSeek-R1-style reasoning distillation directly to a cybersec base. Huang~\cite{huang2025cybersecsft} sits across recipes~2 and 3, comparing SFT vs.\ LoRA vs.\ QLoRA on cybersecurity tasks; Yu~et~al.~\cite{yu2025cybersmall} builds a small cybersec-expert LM via multi-source corpus plus CTI-Bench chain-of-thought distillation.

\paragraph{Where \csidata{} fits.} Continued-pretraining corpora describe \emph{what} cybersecurity texts look like; instruction / preference sets describe \emph{which} curated cybersec questions a model should answer; RL pipelines optimise against narrow verifiable reward surfaces (flag capture, exploit success). None of the three recipes by itself supplies the empirical distribution of \emph{what operators actually do} with LLMs over multi-step tool-augmented engagements: the triage, failed commands, tool-output interpretation, patch validation, scope-boundary discussion, benign business context and operator repair that appear in real sessions. The current frontier trend exposes the same gap from the opposite direction. GPT-5.5-Cyber shows that gated, post-trained or access-tuned frontier models can become highly capable at vulnerability discovery and authorized security workflows~\cite{openai_gpt55_cyber}, but such releases do not by themselves create a reusable public distribution of cybersecurity behaviour, and Foundation-Sec explicitly reports that the scarcity of specialised cybersecurity training data limits adoption~\cite{foundationsec2025}. CyberLLMInstruct also shows that cybersecurity fine-tuning changes both capability and safety, requiring explicit mitigation rather than reward-only optimism~\cite{cyberLLMinstruct2025}. A dataset at \csidata{} scale supplies the missing substrate underneath all three recipes: continued-pretraining mixes can ingest the operator-prose subset, instruction tuning can be re-weighted by the role-mix and target-indicator metadata, and reasoning / RL pipelines can use the trajectory bodies as both expert demonstration and reward grounding.

\subsection{Open SFT corpora and post-training mixtures}\label{sec:rw:openSFT}

The reference open generalist SFT release is T\"ULU~3 \cite{lambert2024tulu3}, whose SFT mix retains $939{,}000$ records after a four-axis GPT-4o LLM-as-judge cascade (helpfulness, instruction-following, honesty, truthfulness) and is followed by DPO and RLVR stages. OpenHermes-2.5 \cite{teknium2024openhermes} ships $1{,}001{,}551$ multi-source conversations; SmolTalk \cite{smoltalk2024} ships approximately $1.1$\,M records targeted at small-LM training. A direct head-to-head comparison \cite{fixingitinpost2025} shows that the two corpora yield measurably different students under matched training. Magpie \cite{xu2024magpie} demonstrates that $4$\,million self-synthesised instructions filtered to $300{,}000$ by quality and difficulty can match much larger mixes. NVIDIA's Llama-Nemotron \cite{llamanemotron2025} consolidates a five-stage recipe (NAS, recovery pretrain, SFT, RL, RLHF) and publicly releases its $40$\,M$+$-sample post-training mixture \cite{nemotronposttraining2025}; the mix is currently the largest open post-training corpus and explicitly trains the model to switch between reasoning-on and reasoning-off modes. Hermes~3 \cite{hermes3_2024} consolidated the ChatML format with Nous-style \texttt{<tool\_call>} XML envelopes for tool-augmented agentic SFT, and the Qwen3 post-training release \cite{qwen3_2025} adopts the same convention as its default tool-call template. The NVIDIA Front-Loading-Reasoning study \cite{frontloadingreasoning2025} provides additional evidence that reasoning data in pretraining stabilises subsequent SFT and reduces catastrophic forgetting, a finding directly relevant to any \csidata{}-driven SFT pipeline.

Table~\ref{tab:dataset_compare} places the training datasets referenced across this Related Work section on a single scale axis alongside \csidata{}. Pure evaluation benchmarks (SecQA, Cybench, CTIBench, CyberMetric, CySecBench, SWE-bench) are discussed in the prose above but excluded from the table: they ship task descriptors for evaluation rather than trajectories or instructions for training, and folding them into a training-dataset comparison would conflate two different artefact classes. Among cybersecurity training datasets, \csidata{} is approximately $65\times$ larger than the next-largest entry (SecKnowledge, $\sim\!400$k) and three-to-four orders of magnitude larger than the cybersecurity-agent SFT corpora (CTF-Dojo, PrivEsc-LLM, Pentest-R1, Cyber-Zero). The agent-trajectory rows (ToolACE, ToolMind, ADP) are the closest methodological comparators by record \emph{shape} (tool-augmented multi-turn) but are one-to-three orders of magnitude smaller than \csidata{} measured in user-prompt-equivalents.

\begin{table}[!tbp]
\centering
\footnotesize
\setlength{\tabcolsep}{3pt}
\renewcommand{\arraystretch}{1.04}
\resizebox{\columnwidth}{!}{%
\begin{tabular}{@{}lrlrl@{}}
\toprule
\textbf{Dataset} & \textbf{Records} & \textbf{Unit} & \textbf{Prompts (eq.)} & \textbf{Tools} \\
\midrule
\rowcolor{group_cyber_hdr}\multicolumn{5}{@{}l}{\textit{Cybersecurity (\S\ref{sec:rw:cybersec})}} \\
\rowcolor{group_cyber}CTF-Dojo \cite{zhuo2025ctfdojo}                    & $486$                 & trajectory & $\sim\!2{,}430$\textsuperscript{$\ddagger$}      & yes \\
\rowcolor{group_cyber}PrivEsc-LLM \cite{normann2026privesc}              & $1{,}000$             & trajectory & $\sim\!5{,}000$\textsuperscript{$\ddagger$}      & yes \\
\rowcolor{group_cyber}Random-Crypto \cite{muzsai2025randomcrypto}        & $5{,}000$             & task       & $5{,}000$             & yes \\
\rowcolor{group_cyber}Pentest-R1 \cite{kong2025pentest}                  & $\sim\!14{,}000$      & TCO tuple  & $\sim\!14{,}000$      & yes \\
\rowcolor{group_cyber}Heimdall \cite{heimdall2024}                       & $\sim\!21{,}000$      & dialogue   & $\sim\!21{,}000$\textsuperscript{$\dagger$}      & no \\
\rowcolor{group_cyber}Lily-Cybersecurity \cite{segolilylabs2024lily}    & $22{,}000$            & QA pair    & $22{,}000$            & no \\
\rowcolor{group_cyber}Cyber-Zero \cite{zhuo2025cyberzero}                & $9{,}464$             & trajectory & $\sim\!47{,}320$\textsuperscript{$\ddagger$}     & yes \\
\rowcolor{group_cyber}CyberLLMInstruct \cite{cyberLLMinstruct2025}       & $54{,}928$            & instr/resp & $54{,}928$            & no \\
\rowcolor{group_cyber}SecKnowledge \cite{secknowledge2024}               & $\sim\!400{,}000$     & instr/resp & $\sim\!400{,}000$     & no \\
\rowcolor{cai_light}\textbf{\csidata{}} (this work)                     & \textbf{$\statTotalFiles{}$} & \textbf{trajectory} & \textbf{$\statTotalPrompts{}$}\textsuperscript{$\ddagger$} & \textbf{yes} \\
\midrule
\rowcolor{group_agent_hdr}\multicolumn{5}{@{}l}{\textit{Agent-trajectory \& tool-use (\S\ref{sec:rw:agentSFT})}} \\
\rowcolor{group_agent}ToolACE \cite{toolace2024}                         & $\sim\!11{,}000$      & trajectory & $\sim\!55{,}000$\textsuperscript{$\ddagger$}     & yes \\
\rowcolor{group_agent}ToolMind \cite{toolmind2025}                       & $\sim\!360{,}000$     & trajectory & $\sim\!1{,}800{,}000$\textsuperscript{$\ddagger$} & yes \\
\rowcolor{group_agent}ADP \cite{agentdataprotocol2025}                   & $1{,}300{,}000$       & trajectory & $\sim\!13{,}100{,}000$\textsuperscript{$\ddagger$} & yes \\
\midrule
\rowcolor{group_sft_hdr}\multicolumn{5}{@{}l}{\textit{Open generalist SFT (\S\ref{sec:rw:openSFT})}} \\
\rowcolor{group_sft}Magpie filtered \cite{xu2024magpie}                & $300{,}000$           & prompt     & $300{,}000$           & no \\
\rowcolor{group_sft}DeepSeek-R1 SFT \cite{guo2025deepseekr1}           & $800{,}000$           & reasoning  & $800{,}000$           & no \\
\rowcolor{group_sft}T\"ULU\,3 SFT \cite{lambert2024tulu3}              & $939{,}000$           & prompt     & $939{,}000$           & no \\
\rowcolor{group_sft}OpenHermes-2.5 \cite{teknium2024openhermes}        & $1{,}001{,}551$       & dialogue   & $\sim\!1{,}001{,}551$\textsuperscript{$\dagger$}  & no \\
\rowcolor{group_sft}SmolTalk \cite{smoltalk2024}                       & $\sim\!1{,}100{,}000$ & prompt     & $\sim\!1{,}100{,}000$ & some \\
\rowcolor{group_sft}AM-R1-Distilled \cite{amDeepSeekR1Distilled2025}   & $1{,}400{,}000$       & reasoning  & $1{,}400{,}000$       & no \\
\rowcolor{group_sft}Llama-Nemotron \cite{nemotronposttraining2025}     & $\sim\!40{,}000{,}000$         & prompt     & $\sim\!40{,}000{,}000$         & some \\
\midrule
\rowcolor{group_chat_hdr}\multicolumn{5}{@{}l}{\textit{Crowd-sourced chat}} \\
\rowcolor{group_chat}LMSYS-Chat-1M \cite{lmsyschat2024}                 & $1{,}000{,}000$       & dialogue   & $\sim\!2{,}000{,}000$\textsuperscript{$\ddagger$} & no \\
\rowcolor{group_chat}WildChat \cite{wildchat2024}                       & $\sim\!1{,}000{,}000$ & dialogue   & $\sim\!2{,}520{,}000$\textsuperscript{$\ddagger$} & no \\
\bottomrule
\end{tabular}}
\caption{Training-dataset comparison for the corpora referenced across this section, normalised to a user-prompt-equivalent axis. Pure evaluation benchmarks (SecQA~\cite{liu2023secqa}, Cybench~\cite{cybench2024}, CTIBench~\cite{ctibench2024}, CyberMetric~\cite{tihanyi2024cybermetric}, CySecBench~\cite{cysecbench2024}, SWE-bench~\cite{jimenez2024swebench}) are discussed in the prose but excluded because they ship task descriptors for evaluation rather than trajectories or instructions for training. \textbf{Records}: headline release count in the dataset's native unit. \textbf{Prompts (eq.)}: approximate user-prompt-equivalents. Conversion rules: instruction/response pairs, TCO tuples and single-turn dialogues map $1{:}1$ to a user prompt; multi-turn trajectories or dialogues are multiplied by the average per-record turn count reported in the source paper. \textsuperscript{$\dagger$}\,$1{:}1$ approximation: source unit is structurally close to a single user prompt (single-turn QA dialogue, predominantly single-turn instruction record). \textsuperscript{$\ddagger$}\,Approximation via reported per-record turn average: LMSYS-Chat-1M $\times 2.0$~\cite{lmsyschat2024}, WildChat $\times 2.52$~\cite{wildchat2024}, ADP $\times 10.1$~\cite{agentdataprotocol2025}, ToolACE / ToolMind $\times 5$ (multi-turn function-calling default in the same family). Cybersecurity-agent trajectories without a published turn average (CTF-Dojo, PrivEsc-LLM, Cyber-Zero) use the same $\times 5$ conservative default; CTF and privesc engagements are typically longer than function-calling tasks (e.g.\ Cyber-Zero caps trajectories at $40$ paired turns and PrivEsc-LLM uses round budgets up to $R{=}60$), so these rows likely \emph{understate} rather than overstate their prompt-equivalent counts. Random-Crypto reports $5{,}000$ procedurally generated challenge specifications used for GRPO training; the row counts each challenge $1{:}1$ to a prompt-equivalent, even though GRPO produces many rollouts per challenge during training (the actual training-trajectory count is generator-dependent and not published). For \csidata{} the trajectories-to-prompts ratio is observed directly: \statTotalFiles{} trajectories carry \statTotalPrompts{} user prompts, averaging $\sim\!113$ user prompts per trajectory (an order of magnitude denser than ADP's $10.1$-turn average); we report this conservative user-prompt-equivalent figure rather than a higher agent-step count.}\label{tab:dataset_compare}
\end{table}

\subsection{Agent-trajectory SFT and tool-use corpora}\label{sec:rw:agentSFT}

The closest methodological precedent for \csidata{} is the recent body of work on agent-trajectory SFT. SWE-bench \cite{jimenez2024swebench} established the trajectory-corpus paradigm for software engineering. ToolACE \cite{toolace2024} introduced a large-scale synthetic function-calling SFT dataset that explicitly trains for format diversity (every tool call is rendered as JSON, YAML, XML or Markdown over a $26{,}507$-API pool); its Llama-3.1-8B-Instruct fine-tune (ToolACE-8B) is a standard tool-use baseline. ToolACE-R \cite{liu2025toolacer} extends this with iterative refinement and reports cross-template generalisation on BFCL and API-Bank. ToolMind \cite{toolmind2025} pushes the same direction to multi-turn reasoning-enhanced tool-use trajectories. The Agent Data Protocol (ADP) \cite{agentdataprotocol2025} is the landmark on agent-trajectory standardisation: it unifies thirteen heterogeneous trajectory corpora into one record schema (Action union of API-, Code- and Message-actions; Observation union of Text and Web), reports gains of roughly $+20$pp average over single-source SFT (e.g.\ SWE-Bench 7B from $0.4\%$ to $20.2\%$, AgentBench-OS from $3.5\%$ to $27.1\%$) and provides the converter pipeline that re-renders ADP records into the target tokenizer's chat template. CodeAct \cite{wang2024codeact} provides the orthogonal lever: collapsing heterogeneous tool calls into a single Python action space yields up to $+20\%$ success-rate over JSON tool-calling on API-Bank, decoupling skill from envelope. The PIPE study \cite{pipe2025trajectorySFT} is the central critical finding: across $16$ environments from AgentBench and Agent-Gym, trajectory-SFT \emph{substantially amplifies interface shortcutting}; trained agents degrade sharply under minimal interface rewrites, while non-trajectory-trained models remain largely stable.

\paragraph{Harness-overfit risk and mitigations for \csidata{}.} The PIPE finding is a direct caution for \csidata{} consumers: a model SFT'd na\"ively on trajectories from a single harness can overfit to that harness's specific tool-call syntax and degrade against any other harness. Two factors blunt this risk for the corpus we describe. First, recent collection is already multi-scaffold: Appendix~\ref{app:schema_evolution} (schema versions v3 and v4) documents that the v4 line shape carries a top-level \texttt{scaffold} field, and 2025--2026 logs are produced by the CSI publishing scaffold, the \caitool{} CLI, and additional harnesses routing through the same proxy. The latest data is therefore not \caitool{}-specific. Second, the published mitigations are mature enough to be applied at the consumer side. Our recommended recipe, drawing on ADP \cite{agentdataprotocol2025}, ToolACE / ToolACE-R \cite{toolace2024,liu2025toolacer} and CodeAct \cite{wang2024codeact}, has three steps: (i) at conversion time, project every trajectory into an ADP-compatible abstract action record and preserve the originating \texttt{scaffold} as provenance metadata, so the literal envelope is never baked into the assistant turn; (ii) at training time, re-render each record into multiple envelopes per example with uniform sampling (OpenAI \texttt{tool\_calls} JSON, Hermes-3 \texttt{<tool\_call>} XML, Qwen3-Coder \texttt{<function=...>} XML, CodeAct Python) and mix in $20$--$30\%$ non-cybersecurity ADP sources (SWE-Gym, CodeActInstruct, Mind2Web) to break corpus-specific priors; (iii) at evaluation time, apply PIPE-style interface rewrites on a held-out slice (alias tool names, swap envelope family, reorder kwargs) and report Interface Reliance alongside task success, with a target of $<5\%$ delta versus the native-harness result before claiming portability. This recipe makes \csidata{} a substrate for harness-portable cybersecurity models rather than a single-scaffold corpus.

\paragraph{Pentest-R1 and the cost of pre-decomposed formats.} The closest cybersecurity-specific point in this family is Pentest-R1 \cite{kong2025pentest}. Its base model is \texttt{DeepSeek-R1-0528-Qwen3-8B}, whose reported pre-finetune baseline is $3.0\%$ on AutoPenBench and $7.5\%$ on Cybench (unguided, pass@3). The full Pentest-R1 model reaches $24.2\%$ on AutoPenBench and $15.0\%$ on Cybench through a sequential pipeline: a Stage~1 SFT pass over $\approx\!14\mathrm{K}$ pre-decomposed Thought-Command-Observation tuples (constructed from $500{+}$ expert walkthroughs), then GRPO offline RL on the same tuples, then online RL in InterCode-CTF. The reported ablations are instructive about which technique contributes what. SFT alone on the TCO tuples leaves the AutoPenBench / Cybench scores at the base's $3.0\%$ / $7.5\%$, i.e.\ no measurable solve gain. GRPO offline-RL alone moves the model to $9.1\%$ / $12.5\%$. The full pipeline (SFT $+$ GRPO offline $+$ GRPO online) reaches $24.2\%$ / $15.0\%$. The two takeaways for a corpus-design paper: \emph{(i)} SFT is not the most performant regime on this data shape, yet it is a load-bearing prerequisite: removing it from the full pipeline drops AutoPenBench from $24.2\%$ to $9.1\%$, so SFT contributes $+15.1$pp on AutoPenBench and $+2.5$pp on Cybench when layered before GRPO; \emph{(ii)} Pentest-R1's records fix the schema as $\{\,\textsf{initial\_prompt},\;\textsf{steps}\!=\![\{\textsf{thought},\textsf{command},\textsf{result}\}]\,\}$, lifting the \emph{thought} and \emph{command} fields out of the chat body. This pre-decomposition tightly couples the data shape to one tool-call surface; modern instruction-tuned bases reward strict adherence to the native \texttt{<tool\_call>$\{$name,arguments$\}$</tool\_call>} XML envelope, and across multi-step CTF interactions the template mismatch compounds and costs solves. \csidata{} addresses both problems at the data-shape level: the released-slice body preserves the original \texttt{messages[]} list with native tool-call envelopes, so the trainer chooses the decomposition rather than inheriting one, and the corpus is roughly three-to-four orders of magnitude larger across logs and prompts than the Pentest-R1 walkthrough set. 

\section{Dataset}\label{sec:dataset}
\subsection{Collection and schema}\label{sec:dataset:rawschema}

\begin{figure*}[!tbp]
\centering
\resizebox{0.96\textwidth}{!}{% Token accumulation over time (weekly buckets). Derived from per-file
% caches via tools/build_paper_numbers.py; mirrors the
% create_token_accumulation_chart in logs.py with the CSI palette.
\begin{tikzpicture}
\begin{axis}[
    name=growthA,
    width=0.42\linewidth,
    height=5.0cm,
    scale only axis,
    ybar, bar width=2.5pt,
    title={(a)~Weekly token volume},
    title style={font=\small\bfseries\sffamily, color=cai_primary, yshift=-2pt},
    xtick={1,21,42,62},
    xticklabels={Mar'25,Aug'25,Dec'25,May'26},
    xticklabel style={font=\tiny\sffamily, color=cai_dark, rotate=30, anchor=north east},
    xmin=0, xmax=\statWeeksN,
    xlabel style={font=\small\sffamily, color=cai_dark},
    xlabel={Week},
    ylabel style={font=\small\sffamily, color=cai_dark},
    ylabel={Tokens (M)},
    yticklabel style={font=\small\sffamily, color=cai_dark},
    ytick={0,20000000,40000000,60000000,80000000},
    yticklabels={0,20,40,60,80},
    ymin=0, ymax=80000000,
    grid=major,
    grid style={dashed,gray!22},
    axis lines=box,
    axis line style={draw=cai_dark!55},
    tick align=outside,
    scaled y ticks=false,
]
\addplot+[fill=cai_primary!75, draw=cai_primary!85!black, bar shift=0pt]
    coordinates {\statWeeklyTokenCoords};
\end{axis}

\begin{axis}[
    name=growthB,
    at={($(growthA.east)+(2.4cm,0)$)},
    anchor=west,
    width=0.42\linewidth,
    height=5.0cm,
    scale only axis,
    title={(b)~Cumulative token corpus},
    title style={font=\small\bfseries\sffamily, color=cai_primary, yshift=-2pt},
    xtick={1,21,42,62},
    xticklabels={Mar'25,Aug'25,Dec'25,May'26},
    xticklabel style={font=\tiny\sffamily, color=cai_dark, rotate=30, anchor=north east},
    xmin=0, xmax=\statWeeksN,
    xlabel style={font=\small\sffamily, color=cai_dark},
    xlabel={Week},
    ylabel style={font=\small\sffamily, color=cai_dark},
    ylabel={Cumulative tokens (B)},
    yticklabel style={font=\small\sffamily, color=cai_dark},
    ytick={0,500000000,1000000000,1500000000,2000000000},
    yticklabels={0,0.5,1.0,1.5,2.0},
    ymin=0, ymax=2100000000,
    grid=major,
    grid style={dashed,gray!22},
    axis lines=box,
    axis line style={draw=cai_dark!55},
    tick align=outside,
    scaled y ticks=false,
]
\addplot+[draw=apt_agent_color, line width=1.4pt, mark=none, smooth]
    coordinates {\statWeeklyTokenCumCoords};
\addplot+[draw=cai_primary!75, line width=0.7pt, dashed, mark=none, smooth, forget plot]
    coordinates {(0,0) (\statWeeksN,2043313728)};
\end{axis}
\end{tikzpicture}}
\caption{Corpus growth over the 14-month collection window aggregated to weekly buckets. (a)~Weekly token volume; (b)~cumulative volume with a dashed straight-line reference between the first and the last week. Token values are a growth-shape proxy from \statFilesScannedForTokens{} scanned files (\statTokenProvenance{}); headline scale claims use the observed log, prompt and storage counts.}\label{fig:growth}
\end{figure*}
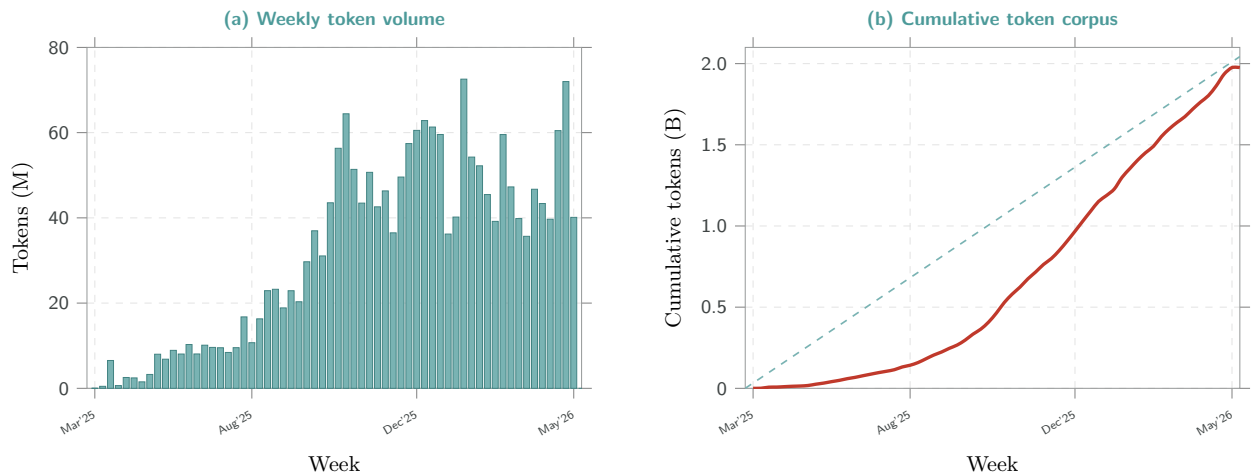

\caitool{}~\cite{mayoralvilches2025gametheoretic} is the first open-source agent framework for cybersecurity automation, distributed both as a framework and as a CLI agent with telemetry as a first-class subsystem of the scaffold. Every session emits a per-session JSONL log capturing model request, model response, tool call, tool output and assistant message, and ships it to our servers; the disclosure and opt-out are documented in the project README. Telemetry is enabled by default because the bottleneck identified in PentestGPT~\cite{deng2024pentestgpt} is the absence of expert operator trajectories. The collection model is behavioural: we record what real operators do with their own prompts on their own targets. No prompt is injected, no synthetic data is generated. Trajectories collected through 2025--2026 span multiple scaffold generations rather than a single harness (Appendix~\ref{app:schema_evolution}).

\caitool{} is distributed free of charge to researchers; in lieu of payment for research use, we ask the research community to contribute usage data that helps identify areas for improvement, understand how the framework is being used in practice, and prioritise new features. That contribution is precisely what makes \csidata{} possible, and the corpus is, in turn, fed back into the same community as detection-accuracy improvements, published findings and the dataset slices defined later in this section. \caitool{} is additionally an output of an EU-funded research programme; use of the framework is correspondingly framed as a contribution to that public-research endeavour, which is a second load-bearing argument for the default-on collection model alongside the technical scarcity of expert operator trajectories. The legal basis for the data collection itself is documented in Section~\ref{sec:ethics:disclose}.

A session log is one JSONL file whose lines are an ordered stream of records: \texttt{session\_start} envelopes, user and assistant messages, and full upstream chat-completion request/response pairs carrying the request model, the \texttt{messages} array, and the provider response. Tool results and subsequent user messages remain in the same file, preserving the act--observe--revise loop. The corpus spans four on-disk schema versions, enumerated in Appendix~\ref{app:schema_evolution}.

\paragraph{Release slices.} Slices follow an audience-size shorthand and are intended as a continuous series: \csidata{}\textsubscript{10}, \csidata{}\textsubscript{1k}, and \csidata{}\textsubscript{200k}. Further \csidata{}\textsubscript{N} slices at higher record counts may follow as more data accumulates. Access is restricted to partner organisations and customers. The redaction recipe applied at the consumer side is documented in Section~\ref{sec:dataset:leaks} alongside the empirical leak-surface measurements that motivate it.

\subsection{Volume and growth}\label{sec:stats:volume}

At the time of writing, the corpus comprises \statTotalFiles{} session logs and \statTotalPrompts{} user prompts produced over \statSpanDays{} days (\statEarliestDate{} to \statLatestDate{}), totalling \statCloudSizeTB{}\,TB of durable storage. One session corresponds to one JSONL log; the corpus is a collection of \statTotalFiles{} trajectories rather than a single concatenated stream, so downstream consumers can iterate, shard or sample at the session boundary. \statUniqueIPs{} distinct source IPs exercise \statUniqueModels{} unique LLM identifiers. Prompt length is heavy-tailed (measured in characters throughout): mean \statAvgPromptLen{}, median \statMedianPromptLen{}, standard deviation \statStdPromptLen{}; the longest prompt observed is \statMaxPromptLen{} characters (a single bulk-paste of multi-stage planning context). The order-of-magnitude gap between the median and the mean is the operator-workflow signature: most prompts are short shell-loop directives, but a long tail of planning briefs and continuation contexts drives the mean up.

\begin{figure}[!htbp]
\centering
\resizebox{\columnwidth}{!}{% Team vs community cumulative log count over time. Solid segments are
% observed; the dotted segment after week 63 is a six-month linear
% extrapolation fitted on the last 10 weeks of observed data (slope
% ~3.3k logs/week for the community series; the team series is
% effectively flat over the same window).
\begin{tikzpicture}
\begin{axis}[
    width=0.86\linewidth,
    height=5.4cm,
    scale only axis,
    title={Cumulative log count by source population},
    title style={font=\small\bfseries\sffamily, color=cai_primary, yshift=-2pt},
    xtick={1,21,42,63,89},
    xticklabels={Mar'25,Aug'25,Dec'25,May'26,Nov'26},
    xticklabel style={font=\tiny\sffamily, color=cai_dark, rotate=30, anchor=north east},
    xmin=0, xmax=92,
    xlabel style={font=\small\sffamily, color=cai_dark},
    xlabel={Week},
    ylabel style={font=\small\sffamily, color=cai_dark},
    ylabel={Cumulative logs},
    yticklabel style={font=\small\sffamily, color=cai_dark, /pgf/number format/.cd, 1000 sep={,}, fixed},
    ymin=0, ymax=260000,
    grid=major,
    grid style={dashed,gray!22},
    axis lines=box,
    axis line style={draw=cai_dark!55},
    tick align=outside,
    scaled y ticks=false,
    legend style={
        font=\tiny\sffamily, color=cai_dark,
        legend pos=north west,
        draw=gray!30, fill=white, fill opacity=0.92,
        cells={anchor=west},
    },
]
% Observed community curve (solid)
\addplot+[draw=graph_accent, line width=1.6pt, mark=none, smooth]
    coordinates {\statWeeklyCommunityCumCoords};
\addlegendentry{Community observed (\statCommunityLogsTotal{} logs)}
% Observed team curve (solid)
\addplot+[draw=apt_agent_color, line width=1.6pt, mark=none, smooth]
    coordinates {\statWeeklyTeamCumCoords};
\addlegendentry{Team observed (\statTeamLogsTotal{} logs)}
% Community projection (dotted): linear extrapolation of last 10 weeks,
% slope ~3.3k logs/week, six-month horizon ending around week 89.
\addplot+[draw=graph_accent, line width=1.4pt, mark=none, densely dotted, forget plot]
    coordinates {(63,156203) (70,179430) (76,199338) (82,219246) (89,242471)};
\addlegendentry{Linear projection ($+26$ weeks)}
% Team projection (dotted): effectively flat (~1 log/week net contribution).
\addplot+[draw=apt_agent_color, line width=1.4pt, mark=none, densely dotted, forget plot]
    coordinates {(63,6286) (89,6317)};
% Vertical reference at the observed/projected boundary
\draw[gray!55, dashed, line width=0.4pt] (axis cs:63,0) -- (axis cs:63,260000);
\node[anchor=south, font=\tiny\sffamily, color=cai_dark!75,
      fill=cai_light, draw=cai_dark!25, line width=0.3pt,
      inner sep=2pt, rounded corners=1pt]
    at (axis cs:77, 30000) {projected};
\end{axis}
\end{tikzpicture}}
\caption{Cumulative log count by source population, with a six-month linear projection (dotted) fitted on the last $10$ weeks of observed data. ``Team'' refers to the publisher's internal contributors; ``Community'' is the union of all externally-attributable named-user sessions. Community dominates team by approximately $24\times$ (\statCommunityLogsTotal{} versus \statTeamLogsTotal{}); the projection extends community to $\sim\!242$k logs and team to a near-flat $\sim\!6.3$k logs at week~$89$ (Nov~2026). The two observed curves sum to $162{,}489$ logs and do not match the corpus headline of \statTotalFiles{} session logs: the remaining $\sim\!68$k logs run under \texttt{root} or unidentified filename-usernames (typically internal lab containers, CI runners, automated harnesses and pristine VMs) and are excluded from the named-user attribution rather than redistributed across the two curves.}\label{fig:teamcomm}
\end{figure}
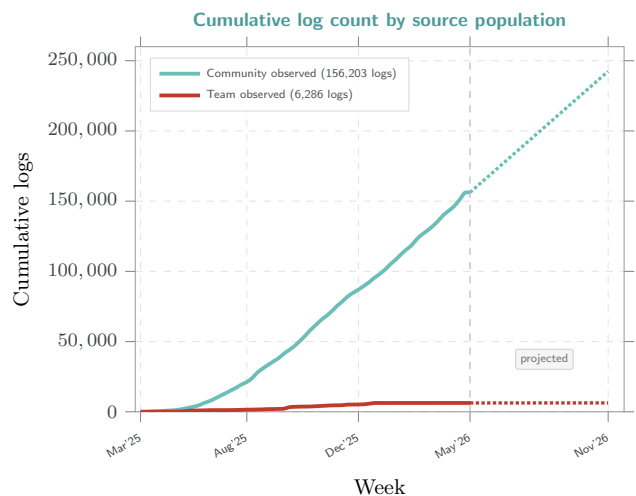

\paragraph{Growth rate and retraining cadence.} Over the collection window, the corpus has accumulated at an average rate of roughly $540$ session logs and $60$k user prompts per day, with Figure~\ref{fig:growth} confirming a near-linear cumulative trajectory and a sustained weekly cadence. At that rate a daily increment represents $\sim\!0.23\%$ of the current corpus, a weekly increment $\sim\!1.6\%$, and a monthly increment $\sim\!6.9\%$. For an SFT pipeline consuming \csidata{} as a rolling training corpus, this suggests \emph{weekly} retraining as the natural cadence (the new-data delta crosses one percent, large enough for measurable distributional shift), with daily checkpoints reserved for online-learning or rapid-feedback settings and monthly retraining sufficient when consumers prioritise stability over freshness. Figure~\ref{fig:teamcomm} additionally separates the publisher's team contribution from the community contribution and shows the community curve dominating the team curve by approximately $24\times$ in log count among the $162{,}489$ named-user sessions, with a six-month dotted projection extending community to $\sim\!242$k logs while the team curve stays flat (the residual $\sim\!68$k root/unidentified-username sessions are not redistributed across the two curves; see the figure caption).

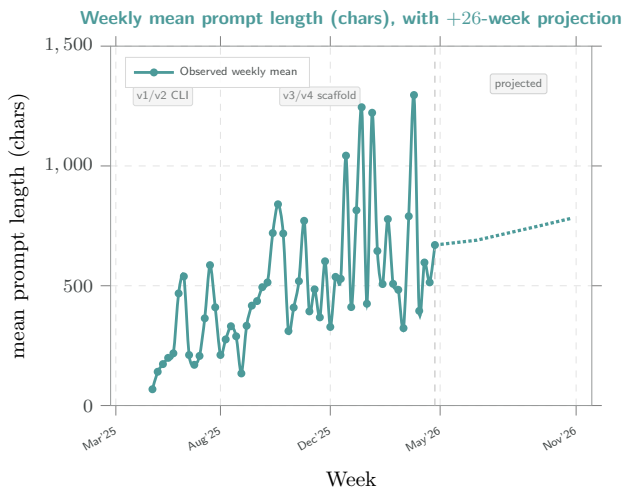
\begin{figure}[!htbp]
\centering
\resizebox{\columnwidth}{!}{% Weekly mean user-prompt length (characters) over the collection
% window, with a linear projection (dotted) extending the last-12-week
% trend an additional six months. The series tracks scaffold maturation:
% the early-window minimum (~70 chars) reflects v1 CAI CLI shell-loop
% prompts; the late-window mean (~700 chars) is dominated by v3/v4 CSI
% scaffold sessions whose users append multi-step engagement context.
\begin{tikzpicture}
\begin{axis}[
    width=0.86\linewidth,
    height=5.4cm,
    scale only axis,
    title={Weekly mean prompt length (chars), with $+26$-week projection},
    title style={font=\small\bfseries\sffamily, color=cai_primary, yshift=-2pt},
    xtick={1,21,42,63,89},
    xticklabels={Mar'25,Aug'25,Dec'25,May'26,Nov'26},
    xticklabel style={font=\tiny\sffamily, color=cai_dark, rotate=30, anchor=north east},
    xmin=0, xmax=92,
    xlabel style={font=\small\sffamily, color=cai_dark},
    xlabel={Week},
    ylabel style={font=\small\sffamily, color=cai_dark},
    ylabel={mean prompt length (chars)},
    yticklabel style={font=\small\sffamily, color=cai_dark, /pgf/number format/.cd, 1000 sep={,}, fixed},
    ymin=0, ymax=1500,
    grid=major,
    grid style={dashed,gray!22},
    axis lines=box,
    axis line style={draw=cai_dark!55},
    tick align=outside,
    scaled y ticks=false,
    legend style={
        font=\tiny\sffamily, color=cai_dark,
        legend pos=north west,
        draw=gray!30, fill=white, fill opacity=0.92,
        cells={anchor=west},
    },
]
% Observed weekly mean (solid line + markers)
\addplot+[draw=cai_primary, line width=1.2pt, mark=*, mark size=1.1pt,
          mark options={draw=cai_primary, fill=cai_primary}, smooth]
    coordinates {
        (8,68) (9,141) (10,173) (11,199) (12,218) (13,468) (14,539)
        (15,211) (16,170) (17,207) (18,364) (19,586) (20,410) (21,211)
        (22,276) (23,331) (24,289) (25,134) (26,333) (27,417) (28,436)
        (29,494) (30,514) (31,720) (32,840) (33,718) (34,311) (35,409)
        (36,519) (37,771) (38,393) (39,485) (40,368) (41,602) (42,328)
        (43,537) (44,529) (45,1043) (46,411) (47,815) (48,1245) (49,425)
        (50,1222) (51,645) (52,507) (53,778) (54,508) (55,483) (56,323)
        (57,790) (58,1296) (59,395) (60,597) (61,514) (62,670)
    };
\addlegendentry{Observed weekly mean}
% Linear projection (last-12-week fit): slope ~5 chars/week, extends to
% week 88 (~Nov 2026), well within the range observed in recent weeks.
\addplot+[draw=cai_primary, line width=1.4pt, mark=none, densely dotted, forget plot]
    coordinates {(62,670) (70,690) (76,720) (82,750) (88,781)};
\addlegendentry{Linear projection ($+26$ weeks)}
% Vertical reference at the observed/projected boundary
\draw[gray!55, dashed, line width=0.4pt] (axis cs:62,0) -- (axis cs:62,1500);
% Scaffold-version annotations (approximate transition weeks based on the
% v1->v2->v3->v4 timeline documented in Appendix A; transitions are gradual,
% not sharp, so the annotations are illustrative.)
\node[anchor=south, font=\tiny\sffamily, color=cai_dark!70,
      fill=cai_light, draw=cai_dark!20, line width=0.3pt,
      inner sep=1.5pt, rounded corners=1pt]
    at (axis cs:10, 1250) {v1/v2 CLI};
\node[anchor=south, font=\tiny\sffamily, color=cai_dark!70,
      fill=cai_light, draw=cai_dark!20, line width=0.3pt,
      inner sep=1.5pt, rounded corners=1pt]
    at (axis cs:40, 1250) {v3/v4 scaffold};
\node[anchor=south, font=\tiny\sffamily, color=cai_dark!75,
      fill=cai_light, draw=cai_dark!25, line width=0.3pt,
      inner sep=2pt, rounded corners=1pt]
    at (axis cs:78, 1300) {projected};
\end{axis}
\end{tikzpicture}}
\caption{Weekly mean user-prompt length (characters) over the $14$-month collection window, with a six-month linear projection (dotted) fitted on the last $12$ weeks of observed data. The early-window floor near $70$~chars reflects v1/v2 \caitool{} CLI shell-loop prompts; the late-window range of $400$--$1{,}300$~chars is dominated by v3/v4 CSI-scaffold sessions whose users routinely append multi-step engagement context, ntlmrelayx pivot state and tmux-resumption briefs. The projection extends the last-$12$-week slope (\,$\sim\!5$\,chars/week) to week~$88$ ($\sim$Nov~$2026$); the variance is high week-to-week and the projection should be read as a trend line rather than a point forecast.}\label{fig:prompt_evolution}
\end{figure}

\paragraph{Prompt-length evolution.} Beyond corpus size, the \emph{shape} of individual prompts has shifted as automation needs and scaffold sophistication have grown. Two trends in the wider literature predict longer prompts over time: agentic frameworks have shifted operator workflow from single-line shell-loop directives to multi-step planning briefs with embedded context~\cite{agentdataprotocol2025,wang2024codeact}, and crowd-sourced LLM corpora exhibit a measurable growth in mean user-prompt length as users habituate to the model~\cite{wildchat2024}. \csidata{} reproduces both trends: Figure~\ref{fig:prompt_evolution} plots the weekly mean prompt length across the collection window and shows the mean drift from $\sim\!150$ characters in the v1/v2 \caitool{} CLI era (Mar--Aug 2025) to a $400$--$1{,}300$-character range during the v3/v4 CSI publishing-scaffold era (Sep 2025 onward). Fitting a linear trend on the last $12$ weeks gives a slope of $\sim\!5$ characters of mean length added per week; extending that trend dottedly six months forward projects a weekly mean of $\sim\!780$ characters by November $2026$, which would put \csidata{} prompts in the same order of magnitude as the average WildChat prompt ($295$ tokens at the user side~\cite{wildchat2024}). The variance is high week-to-week, dominated by occasional bulk-paste sessions, but the floor (median, not shown) has stayed in the $50$--$140$-character band across the entire window, confirming that the upward drift is in the tail rather than the body of the distribution.

\begin{table*}[!t]
\centering
\small
\setlength{\tabcolsep}{6pt}
\renewcommand{\arraystretch}{1.05}
\begin{minipage}[t]{0.48\textwidth}
\centering
\begin{tabular*}{\textwidth}{@{\extracolsep{\fill}}lr@{}}
\toprule
\textbf{Security tool} & \textbf{Mentions} \\
\midrule
  \texttt{nc}         & $6{,}954{,}169$ \\
  \texttt{nmap}       & $1{,}691{,}235$ \\
  \texttt{curl}       & $\phantom{0,}738{,}295$ \\
  \texttt{nikto}      & $\phantom{0,}230{,}989$ \\
  \texttt{burp}       & $\phantom{0,}203{,}149$ \\
  \texttt{sqlmap}     & $\phantom{0,}160{,}285$ \\
  \texttt{gobuster}   & $\phantom{0,}144{,}055$ \\
  \texttt{metasploit} & $\phantom{00,}99{,}997$ \\
\bottomrule
\end{tabular*}
\subcaption{Top eight security tools mentioned. The next seven entries (\texttt{dirb}, \texttt{shodan}, \texttt{hydra}, \texttt{subfinder}, \texttt{wpscan}, \texttt{netcat}, \texttt{john}) each sit below $100$k mentions.}
\end{minipage}\hfill%
\begin{minipage}[t]{0.48\textwidth}
\centering
\begin{tabular*}{\textwidth}{@{\extracolsep{\fill}}lrr@{}}
\toprule
\textbf{Attack category} & \textbf{Prompts} & \textbf{Share} \\
\midrule
\statAttackCategoriesTable
\bottomrule
\end{tabular*}
\subcaption{Attack-category distribution among offensive-intent prompts.}
\end{minipage}
\caption{Tool and attack-pattern breakdowns.}\label{tab:tools_and_attacks}
\end{table*}

\subsection{Roles, tools and attack categories}\label{sec:stats:roles}

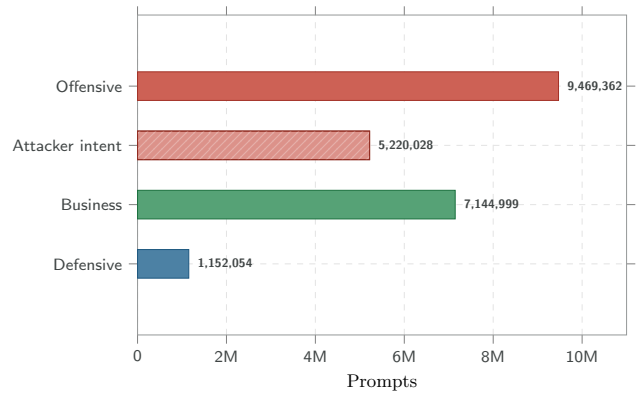
\begin{figure}[!htbp]
\centering
\resizebox{\columnwidth}{!}{% Role classification breakdown. xbar with numeric y coords + manual tick
% labels (symbolic y coords + y dir=reverse triggered a pgfplots transform
% bug that emitted "x coord trafo unsupported"). Display order top-to-bottom
% is therefore controlled by the y values: largest at top (y=4) etc.
\begin{tikzpicture}
\begin{axis}[
    width=\linewidth,
    height=5.5cm,
    scale only axis,
    xbar,
    bar width=14pt,
    bar shift=0pt,
    xmin=0, xmax=11000000,
    xtick={0,2000000,4000000,6000000,8000000,10000000},
    xticklabels={0,2M,4M,6M,8M,10M},
    ymin=0, ymax=5,
    ytick={1,2,3,4},
    yticklabels={Defensive,Business,{Attacker intent},Offensive},
    enlarge y limits=0.04,
    xticklabel style={font=\small\sffamily, color=cai_dark},
    yticklabel style={font=\small\sffamily, color=cai_dark},
    xlabel style={font=\small\sffamily, color=cai_dark},
    xlabel={Prompts},
    grid=major,
    grid style={dashed,gray!22},
    axis lines=box,
    axis line style={draw=cai_dark!55},
    tick align=outside,
    scaled x ticks=false,
]
\addplot[fill=apt_agent_color!80, draw=apt_agent_color!85!black]
    coordinates {(\statOffensiveCountRaw,4)};
\addplot[fill=apt_agent_color!55, draw=apt_agent_color!75!black,
         postaction={pattern=north east lines, pattern color=apt_agent_color!40}]
    coordinates {(\statAttackerIntentCountRaw,3)};
\addplot[fill=cai_accent!75, draw=cai_accent!85!black]
    coordinates {(\statBusinessCountRaw,2)};
\addplot[fill=defender_color!80, draw=defender_color!90!black]
    coordinates {(\statDefensiveCountRaw,1)};
% Value annotations placed manually so we keep them readable without the
% nodes-near-coords machinery that crashed pgfplots on this axis layout.
\node[anchor=west, font=\scriptsize\bfseries\sffamily, color=cai_dark] at (axis cs:\statOffensiveCountRaw,4) {\,\statOffensiveCount{}};
\node[anchor=west, font=\scriptsize\bfseries\sffamily, color=cai_dark] at (axis cs:\statAttackerIntentCountRaw,3) {\,\statAttackerIntentCount{}};
\node[anchor=west, font=\scriptsize\bfseries\sffamily, color=cai_dark] at (axis cs:\statBusinessCountRaw,2) {\,\statBusinessCount{}};
\node[anchor=west, font=\scriptsize\bfseries\sffamily, color=cai_dark] at (axis cs:\statDefensiveCountRaw,1) {\,\statDefensiveCount{}};
\end{axis}
\end{tikzpicture}}
\caption{Role classification of user prompts. Offensive and attacker-intent prompts together cover the majority of the corpus; the hatched fill on \emph{attacker intent} marks its overlap with offensive.}\label{fig:roles}
\end{figure}

Each prompt is classified with four heuristic predicates. The distribution: \statOffensivePct{} match offensive patterns (\statOffensiveCount{}), \statAttackerIntentPct{} show explicit attacker intent (\statAttackerIntentCount{}), \statBusinessPct{} carry business or integration value (\statBusinessCount{}), \statDefensivePct{} are defensive (\statDefensiveCount{}); the four categories overlap by construction. Figure~\ref{fig:roles} shows the breakdown. The top-tool mix is dominated by network-level tools (\texttt{nc}, \texttt{nmap}, \texttt{curl}) and web-application scanners (\texttt{nikto}, \texttt{burp}); reconnaissance and exploitation dominate the MITRE-aligned attack-category distribution (Table~\ref{tab:tools_and_attacks}). Table~\ref{tab:operational_signals} reports corpus-level evidence for verifiable CTF goals, multi-turn continuation, production API / bug-bounty traffic, mobile and reverse-engineering work, robotics / IoT targets, and network-origin intelligence; the three largest business-value categories are \statTopBusinessCategoriesInline{}.

\begin{table*}[!tbp]
\centering
\footnotesize\sffamily
\setlength{\tabcolsep}{6pt}
\renewcommand{\arraystretch}{1.22}
\begin{tabular*}{\textwidth}{@{}
  >{\raggedright\arraybackslash\bfseries}p{0.20\textwidth}
  >{\raggedright\arraybackslash}p{0.30\textwidth}
  >{\raggedright\arraybackslash}p{0.44\textwidth}
@{}}
\toprule
\textbf{Signal} & \textbf{Top markers in the corpus} & \textbf{Why it matters} \\
\midrule
\rowcolor{cai_light}
Verifiable lab outcomes &
  \texttt{flag}\hfill\textbf{$20.8$\,M}\newline
  ``CTF challenge''\hfill\textbf{$2.7$\,M} &
  Outcome-bearing raw material for \texttt{verified\_outcome} extraction, DPO pairs and RLVR rewards. \\
\addlinespace[1pt]
Act--observe--revise continuity &
  ``continue working''\hfill\textbf{$2.6$\,M}\newline
  ``based on previous''\hfill\textbf{$2.5$\,M}\newline
  \texttt{error}\hfill\textbf{$2.4$\,M}\newline
  \texttt{success}\hfill\textbf{$544$\,k} &
  Captures iterative repair, not just isolated instructions; this is the behaviour needed for tool-use and failure-recovery training. \\
\addlinespace[1pt]
\rowcolor{cai_light}
Production API \& bug-bounty traffic &
  \texttt{api}\hfill\textbf{$4.3$\,M}\newline
  \texttt{endpoint}\hfill\textbf{$3.1$\,M}\newline
  ``bug bounty''\hfill\textbf{$504$\,k}\newline
  \texttt{authorization}\hfill\textbf{$495$\,k} &
  Authenticated HTTP traffic, headers and endpoint paths make the redaction layer central rather than cosmetic. \\
\addlinespace[1pt]
Mobile \& reverse-engineering work &
  \texttt{reversing}\hfill\textbf{$1.7$\,M}\newline
  \texttt{mobile}\hfill\textbf{$654$\,k}\newline
  \texttt{Android}\hfill\textbf{$361$\,k}\newline
  \texttt{APK}\hfill\textbf{$201$\,k} &
  Task mix includes APK analysis, dynamic loading, SSL pinning and malware-style reversing, not only web scanning. \\
\addlinespace[1pt]
\rowcolor{cai_light}
Robotics / IoT \& embedded targets &
  \texttt{robot}\hfill\textbf{$406$\,k}\newline
  \texttt{IoT}\hfill\textbf{$69$\,k}\newline
  \texttt{Roomba}\hfill\textbf{$3$\,k} &
  A smaller but distinctive slice covering physical-device and embedded security workflows that generic cyber corpora rarely expose. \\
\addlinespace[1pt]
Network-origin intelligence &
  Tor-exit IPs\hfill\textbf{$59$}\newline
  VPN IPs\hfill\textbf{$7$}\newline
  Suspicious IPs\hfill\textbf{$59$} &
  Logs support passive studies of operator infrastructure under aggregate-only release; no raw IP disclosure. \\
\bottomrule
\end{tabular*}
\caption{Operational-realism signals. Each row lists the top keyword / phrase markers with their corpus-wide mention counts (right-aligned per row, bold) and a brief note on the SFT-pipeline value. The marker counts are aggregate occurrence counters across all user prompts, not de-duplicated trajectories.}\label{tab:operational_signals}
\end{table*}

\subsection{Temporal and geographic distribution}\label{sec:stats:geo}

\begin{figure*}[!tbp]
\centering
\resizebox{0.92\textwidth}{!}{% Temporal distribution, two side-by-side axes. The horizontal sep is
% large enough that panel (b)'s ylabel does not intrude into panel (a),
% and each axis carries its own scale-only-axis sizing.
\begin{tikzpicture}
\begin{axis}[
    name=tempA,
    width=0.36\linewidth,
    height=4.8cm,
    scale only axis,
    ybar, bar width=4.5pt,
    title={(a)~Hour of day (UTC)},
    title style={font=\small\bfseries\sffamily, color=cai_primary, yshift=-2pt},
    xtick={0,4,8,12,16,20},
    xmin=-0.6, xmax=23.6,
    xlabel={Hour}, xlabel style={font=\small\sffamily, color=cai_dark},
    ylabel={Prompts}, ylabel style={font=\small\sffamily, color=cai_dark},
    xticklabel style={font=\small\sffamily, color=cai_dark},
    yticklabel style={font=\small\sffamily, color=cai_dark},
    yticklabel={\pgfmathprintnumber[fixed,precision=0,1000 sep={,}]{\tick}},
    ymin=0,
    grid=major,
    grid style={dashed,gray!22},
    axis lines=box,
    axis line style={draw=cai_dark!55},
    tick align=outside,
    scaled y ticks=false,
    enlarge x limits=0.04,
]
\addplot+[fill=cai_primary!75, draw=cai_primary!85!black, bar shift=0pt]
    coordinates {\statHourlyCoords};
\end{axis}

\begin{axis}[
    name=tempB,
    at={($(tempA.east)+(2.6cm,0)$)},
    anchor=west,
    width=0.36\linewidth,
    height=4.8cm,
    scale only axis,
    ybar, bar width=14pt,
    title={(b)~Day of week},
    title style={font=\small\bfseries\sffamily, color=cai_primary, yshift=-2pt},
    xtick={0,1,2,3,4,5,6},
    xticklabels={Mon,Tue,Wed,Thu,Fri,Sat,Sun},
    xmin=-0.6, xmax=6.6,
    xlabel={Day}, xlabel style={font=\small\sffamily, color=cai_dark},
    ylabel={Prompts}, ylabel style={font=\small\sffamily, color=cai_dark},
    xticklabel style={font=\small\sffamily, color=cai_dark},
    yticklabel style={font=\small\sffamily, color=cai_dark},
    yticklabel={\pgfmathprintnumber[fixed,precision=0,1000 sep={,}]{\tick}},
    ymin=0,
    grid=major,
    grid style={dashed,gray!22},
    axis lines=box,
    axis line style={draw=cai_dark!55},
    tick align=outside,
    scaled y ticks=false,
    enlarge x limits=0.06,
]
\addplot+[fill=gcai_color!80, draw=gcai_color!90!black, bar shift=0pt]
    coordinates {\statDailyCoords};
\end{axis}
\end{tikzpicture}}
\caption{Temporal distribution of user prompts. (a)~Hourly activity (UTC); (b)~weekday activity.}\label{fig:temporal}
\end{figure*}
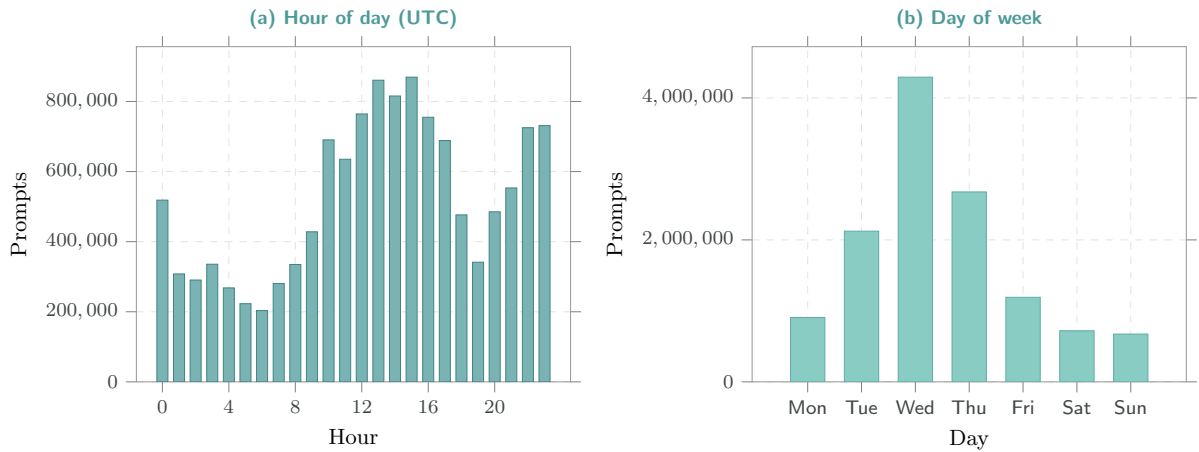

\begin{figure*}[!tbp]
\centering
\resizebox{\textwidth}{!}{% World bubble map of distinct source IPs.
%
% Each bubble is one (lat, lon) cluster at 2-decimal precision. Bubble area
% scales with log10(count); bubble colour encodes continent. Land
% silhouettes are generated from Natural Earth 1:110m land polygons and
% simplified for compact TikZ rendering. The Vitoria-Gasteiz
% centroid artifact (12,730 IPs at 42.85, -2.67) is excluded from the map
% and called out separately so it does not dominate the visual scale.
%
% Bubble coordinates are emitted per continent by tools/build_paper_numbers.py
% (\statBubblesEurope, \statBubblesAsia, ...). pgfplots reads them via the
% scatter / point meta machinery.
%
% Coordinate system: pgfplots axis in geographic units (lon ∈ [-180, 180],
% lat ∈ [-90, 90]). We render at width=\textwidth and height ≈ width/2 for
% the natural equirectangular aspect ratio.
\definecolor{cont_europe}{HTML}{4C9A99}
\definecolor{cont_asia}{HTML}{C0392B}
\definecolor{cont_nam}{HTML}{1F618D}
\definecolor{cont_sam}{HTML}{E67E22}
\definecolor{cont_africa}{HTML}{B58900}
\definecolor{cont_oceania}{HTML}{8E44AD}
\definecolor{cont_land}{HTML}{DCE6E4}
\definecolor{cont_land_edge}{HTML}{9CB8B5}
\definecolor{ocean_color}{HTML}{F7FAFA}

\pgfplotsset{
    worldland/.style={
        fill=cont_land,
        draw=cont_land_edge,
        line width=0.24pt,
        line join=round,
    }
}

\begin{tikzpicture}
\begin{axis}[
    width=\textwidth,
    height=0.5\textwidth,
    scale only axis,
    axis equal image=false,
    xmin=-180, xmax=180,
    ymin=-58, ymax=82,
    xtick={-180,-120,-60,0,60,120,180},
    ytick={-60,-30,0,30,60},
    xticklabel style={font=\tiny\sffamily, color=cai_dark},
    yticklabel style={font=\tiny\sffamily, color=cai_dark},
    xlabel style={font=\tiny\sffamily, color=cai_dark},
    ylabel style={font=\tiny\sffamily, color=cai_dark},
    xlabel={Longitude},
    ylabel={Latitude},
    grid=major,
    grid style={dashed,gray!18},
    axis lines=box,
    axis line style={draw=cai_dark!55},
    tick align=outside,
    clip=true,
    enlargelimits=false,
    legend image code/.code={
        \draw [#1, draw=none] (0cm,-0.05cm) rectangle (0.20cm,0.08cm);
    },
    legend style={
        font=\tiny\sffamily, color=cai_dark,
        legend pos=south west,
        draw=gray!30, fill=white, fill opacity=0.94,
        cells={anchor=west},
        row sep=0.5pt, inner sep=2pt,
    },
    axis background/.style={fill=ocean_color},
]
% Natural Earth land silhouettes, drawn behind the bubbles in plot
% coordinates so they align with the GeoIP buckets.
\input{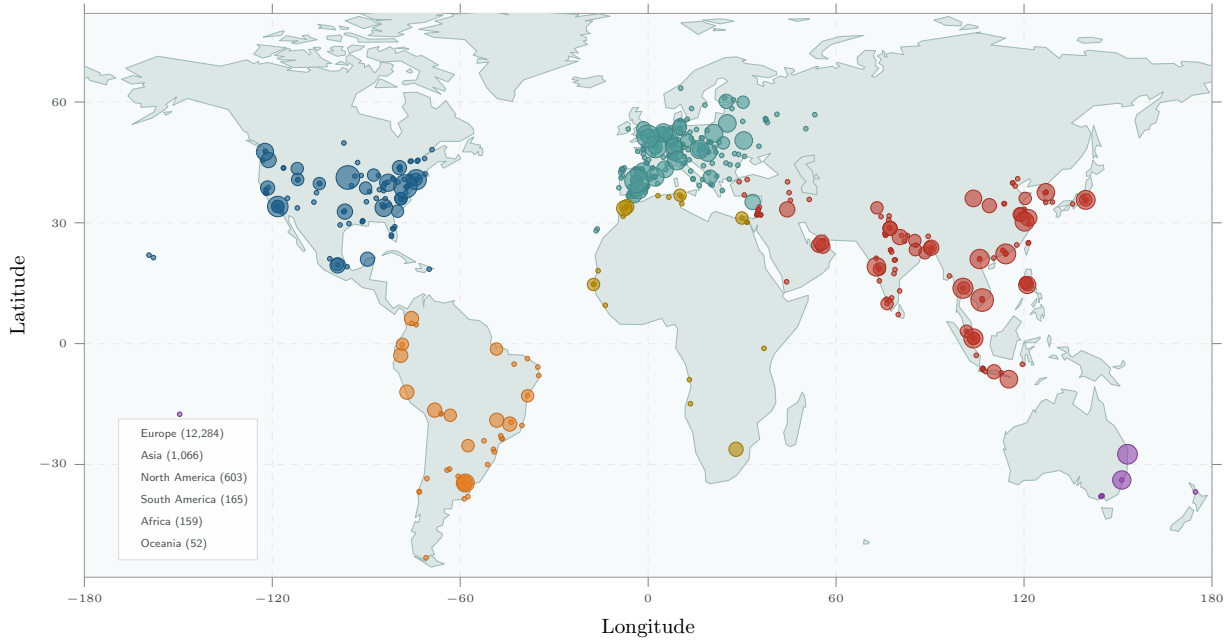}
% --- Bubbles per continent. point meta is the IP count; mark size scales
% with log(count) so that a 5-IP city is visibly larger than a 1-IP city
% without the 11,343-IP Spain bar swamping the visual scale. ---
\addplot[
    only marks, mark=*,
    scatter, point meta=explicit,
    scatter/use mapped color={fill=cont_europe, fill opacity=0.6, draw=cont_europe!85!black, line width=0.2pt},
    scatter/@pre marker code/.append style={/tikz/mark size={1pt + ln(1+\pgfplotspointmetatransformed)*0.6pt}},
] coordinates {\statBubblesEurope};
\addlegendentry{Europe (\statContinentEuropeCount)}
\addplot[
    only marks, mark=*,
    scatter, point meta=explicit,
    scatter/use mapped color={fill=cont_asia, fill opacity=0.6, draw=cont_asia!85!black, line width=0.2pt},
    scatter/@pre marker code/.append style={/tikz/mark size={1pt + ln(1+\pgfplotspointmetatransformed)*0.6pt}},
] coordinates {\statBubblesAsia};
\addlegendentry{Asia (\statContinentAsiaCount)}
\addplot[
    only marks, mark=*,
    scatter, point meta=explicit,
    scatter/use mapped color={fill=cont_nam, fill opacity=0.6, draw=cont_nam!85!black, line width=0.2pt},
    scatter/@pre marker code/.append style={/tikz/mark size={1pt + ln(1+\pgfplotspointmetatransformed)*0.6pt}},
] coordinates {\statBubblesNorthAmerica};
\addlegendentry{North America (\statContinentNorthAmericaCount)}
\addplot[
    only marks, mark=*,
    scatter, point meta=explicit,
    scatter/use mapped color={fill=cont_sam, fill opacity=0.6, draw=cont_sam!85!black, line width=0.2pt},
    scatter/@pre marker code/.append style={/tikz/mark size={1pt + ln(1+\pgfplotspointmetatransformed)*0.6pt}},
] coordinates {\statBubblesSouthAmerica};
\addlegendentry{South America (\statContinentSouthAmericaCount)}
\addplot[
    only marks, mark=*,
    scatter, point meta=explicit,
    scatter/use mapped color={fill=cont_africa, fill opacity=0.6, draw=cont_africa!85!black, line width=0.2pt},
    scatter/@pre marker code/.append style={/tikz/mark size={1pt + ln(1+\pgfplotspointmetatransformed)*0.6pt}},
] coordinates {\statBubblesAfrica};
\addlegendentry{Africa (\statContinentAfricaCount)}
\addplot[
    only marks, mark=*,
    scatter, point meta=explicit,
    scatter/use mapped color={fill=cont_oceania, fill opacity=0.6, draw=cont_oceania!85!black, line width=0.2pt},
    scatter/@pre marker code/.append style={/tikz/mark size={1pt + ln(1+\pgfplotspointmetatransformed)*0.6pt}},
] coordinates {\statBubblesOceania};
\addlegendentry{Oceania (\statContinentOceaniaCount)}
% Centroid artifact is documented in the caption rather than overlayed on
% the map to avoid clutter at the bottom of the plot area.
\end{axis}
\end{tikzpicture}}
\caption{\csidata{} contributor map. Each bubble is one distinct coordinate cluster rounded to 2-decimal precision; bubble area is logarithmic in the number of distinct IPs at that point; colour encodes the inferred continent. The $12{,}730$-IP Vitoria-Gasteiz centroid artifact is excluded (Section~\ref{sec:ethics:skew}); the remaining $\statBubbleN{}$ buckets reflect the genuine geographic spread. Land silhouettes from Natural Earth 1:110m coastlines~\cite{naturalearth110m}.}\label{fig:worldmap}
\end{figure*}

Activity concentrates in the European working day (13:00--16:00 UTC) with a midweek peak (Figure~\ref{fig:temporal}); off-hours activity is non-trivial. The geographic footprint spans \statUniqueCountries{} countries (Figure~\ref{fig:worldmap}, Table~\ref{tab:continents}). Europe accounts for the largest share of distinct contributors (\statContinentEuropePct{} after centroid removal), with secondary clusters in Asia (\statContinentAsiaPct{}, dominated by India, China, Vietnam, Singapore, Japan) and North America (\statContinentNorthAmericaPct{}). A \statOrgHostingPct{} share of source IPs (\statOrgHostingCount{}) belong to public clouds and hosting providers, reflecting use from ephemeral cloud workstations, CI runners and remote attack platforms; the majority \statOrgIspPct{} come from residential or business ISPs, more representative of human-in-the-loop CTF and bug-bounty work. This split is relevant for SFT downstream: cloud-hosted trajectories are more likely scripted multi-target reconnaissance, residential ones trend interactive against a small number of targets.

% [Per-country bar chart removed; \texttt{fig\_geo.tex} retained on disk for future restoration.]

\begin{table}[!htbp]
\centering
\small
\setlength{\tabcolsep}{6pt}
\renewcommand{\arraystretch}{1.05}
\begin{tabular*}{\columnwidth}{@{\extracolsep{\fill}}lrr@{}}
\toprule
\textbf{Continent} & \textbf{IPs} & \textbf{Share} \\
\midrule
Europe        & \statContinentEuropeCount        & \statContinentEuropePct \\
Asia          & \statContinentAsiaCount          & \statContinentAsiaPct \\
North America & \statContinentNorthAmericaCount  & \statContinentNorthAmericaPct \\
South America & \statContinentSouthAmericaCount  & \statContinentSouthAmericaPct \\
Africa        & \statContinentAfricaCount        & \statContinentAfricaPct \\
Oceania       & \statContinentOceaniaCount       & \statContinentOceaniaPct \\
\midrule
\textbf{Total attributed} & \statContinentTotal & $100\%$ \\
\midrule
Hosting / cloud   & \statOrgHostingCount     & \statOrgHostingPct \\
Residential / ISP & \statOrgIspCount         & \statOrgIspPct \\
Other / unknown   & \statOrgOtherOrgCount    & \statOrgOtherOrgPct \\
\bottomrule
\end{tabular*}
\caption{Geographic and provider breakdown. The continent rows attribute every IP whose country resolved to a known geography; the total of \statContinentTotal{} represents the geographically-attributable subset of the \statUniqueIPs{} distinct source IPs reported in Section~\ref{sec:stats:volume} (the remaining $\sim\!2.4$k IPs fall to junk geo strings, unmapped countries, or IPv6 ranges the geolocation provider does not resolve). The hosting / ISP / other partition is a heuristic match of organisation names against a published keyword list.}\label{tab:continents}
\end{table}

\subsection{Model mix}\label{sec:stats:models}

The corpus exercises \textbf{\statUniqueModels{} distinct model identifiers}, one of the broadest model-coverage signals in any LLM trajectory corpus we are aware of, and reflecting operators routinely A/B-testing models on the same task. Identifiers collapse onto far fewer canonical models when route-prefixes, region tags, vendor proxies and quantization suffixes are normalised. Table~\ref{tab:models} reports the top canonical models within each provider; Table~\ref{tab:frontier_models} isolates high-velocity frontier-era identifiers as they appear in the logs, including GPT-5.5, Claude Opus 4-7 and Gemini 3.1 namespaces. We treat these as corpus-observed identifiers rather than a provider release chronology, because router aliases, preview labels and internal deployment names can move faster than public release pages. The long tail beyond the top-table entries spans community quantizations, local Ollama variants, ephemeral provider routes, and per-experiment fine-tunes (qwen3-8b-grpo/pentestr1/onlinerl variants, alias1-fp16/int4 variants, etc.).

\begin{table}[!htbp]
\centering
\footnotesize
\setlength{\tabcolsep}{4pt}
\renewcommand{\arraystretch}{1.04}
\resizebox{\columnwidth}{!}{%
\begin{tabular}{@{}lllr@{}}
\toprule
\textbf{Provider} & \textbf{Model identifier} & \textbf{Identifier family} & \textbf{Calls} \\
\midrule
\textcolor{codex_color}{\textbf{OpenAI}}  & \texttt{gpt-5.4}                       & GPT-5.x namespace & $382{,}224$   \\
\rowcolor{cai_light}                      & \textbf{\texttt{gpt-5.5}} (Spud)       & \textbf{GPT-5.x preview alias} & $94{,}794$    \\
\addlinespace[2pt]
\textcolor{cc_color}{\textbf{Anthropic}}  & \texttt{claude-opus-4-6}               & Opus 4.x namespace  & $127{,}396$   \\
\rowcolor{cai_light}                      & \textbf{\texttt{claude-opus-4-7}}      & \textbf{Opus 4.x namespace} & $135{,}389$   \\
                                          & \texttt{claude-sonnet-4-6}             & Sonnet 4.x namespace  & $156{,}228$   \\
\addlinespace[2pt]
\textcolor{google_color}{\textbf{Google}} & \texttt{gemini-3-pro-preview}          & Gemini 3 preview  & $96{,}533$    \\
\rowcolor{cai_light}                      & \textbf{\texttt{gemini-3.1-pro-preview}} & \textbf{Gemini 3.1 preview} & $331{,}361$   \\
                                          & \texttt{gemini-3-flash-preview}        & Gemini 3 preview  & $135{,}195$   \\
\addlinespace[2pt]
\textcolor{provider_grey}{\textbf{DeepSeek}} & \texttt{deepseek-v4-flash}          & V4 namespace & $169{,}934$   \\
\rowcolor{cai_light}                      & \textbf{\texttt{deepseek-v4-pro}}      & \textbf{V4 namespace} & $21{,}196$    \\
\addlinespace[2pt]
\textcolor{provider_grey}{\textbf{Alibaba}} & \texttt{qwen3.6:35b}                & Qwen3.6 namespace & $130{,}426$ \\
                                          & \texttt{qwen3.6-plus}                  & Qwen3.6 namespace & $40{,}620$    \\
\rowcolor{cai_light}                      & \textbf{\texttt{qwen3.7-max}}          & \textbf{Qwen3.7 namespace} & $3{,}100$  \\
\addlinespace[2pt]
\rowcolor{cai_light}\textcolor{provider_grey}{\textbf{Z.ai}}   & \textbf{\texttt{glm-5}}                & \textbf{GLM-5 namespace} & $249{,}513$   \\
                                          & \texttt{glm-5.1}                       & GLM-5 namespace & $115{,}588$  \\
                                          & \texttt{glm-5-turbo}                   & GLM-5 namespace & $5{,}636$     \\
\addlinespace[2pt]
\textcolor{provider_grey}{\textbf{MoonshotAI}} & \texttt{kimi-k2.5}                & Kimi K2.x namespace & $49{,}454$  \\
\rowcolor{cai_light}                      & \textbf{\texttt{kimi-k2.6}}            & \textbf{Kimi K2.x namespace} & $12{,}371$    \\
\addlinespace[2pt]
\textcolor{provider_grey}{\textbf{MiniMax}} & \texttt{minimax-m2.5}                & MiniMax M2.x namespace & $159{,}951$ \\
\rowcolor{cai_light}                      & \textbf{\texttt{minimax-m2.7}}         & \textbf{MiniMax M2.x namespace} & $36{,}312$    \\
\addlinespace[2pt]
\rowcolor{cai_light}\textcolor{mistral_vibe}{\textbf{Mistral}} & \textbf{\texttt{mistral-medium-3.5}}  & \textbf{Medium 3.x namespace} & $1{,}206$     \\
\addlinespace[2pt]
\textcolor{provider_grey}{\textbf{xAI}}    & \texttt{grok-4.1-fast}                 & Grok 4.x namespace & $8{,}713$       \\
\rowcolor{cai_light}                      & \textbf{\texttt{grok-4.20-reasoning}}  & \textbf{Grok 4.x namespace} & $2{,}811$     \\
                                          & \texttt{grok-4.3}                      & Grok 4.x namespace & $537$           \\
\addlinespace[2pt]
\textcolor{provider_grey}{\textbf{NVIDIA}} & \texttt{nemotron-3-super-120b-a12b}    & Nemotron 3 namespace  & $67{,}647$    \\
\rowcolor{cai_light}                      & \textbf{\texttt{nemotron-3-nano-omni}} & \textbf{Nemotron 3 namespace} & $44{,}779$    \\
\bottomrule
\end{tabular}}
\caption{Frontier-era model identifiers observed in \csidata{} logs. The table reports corpus-observed identifiers and invocation counts, not external release-history claims; provider aliases, preview labels and router names can change independently of public release chronology. Shaded rows mark identifiers highlighted in the compact 2026 snapshot. Pre-2026 baseline models are reflected in Table~\ref{tab:models}.}\label{tab:frontier_models}
\end{table}

% Aggregate dashboard commented out: no prose cross-references it, the
% per-panel data already appears in adjacent figures and tables, and
% removing it tightens the layout. Restore by uncommenting this block.
% \begin{figure*}[!tbp]
% \centering
% \resizebox{0.96\textwidth}{!}{\input{tex/fig_aggregate.tex}}
% \caption{\csidata{} dashboard: (a)~role mix; (b)~hourly and (c)~weekday activity (UTC); (d)~top security-tool mentions; (e)~MITRE-aligned attack categories among offensive-intent prompts.}\label{fig:aggregate}
% \end{figure*}

\begin{table*}[!tbp]
\centering
\small
\setlength{\tabcolsep}{6pt}
\renewcommand{\arraystretch}{1.04}
\begin{tabular*}{\textwidth}{@{\extracolsep{\fill}}llrrrr@{}}
\toprule
\textbf{Provider} & \textbf{Canonical model} & \textbf{Calls} & \textbf{Share} & \textbf{Provider calls} & \textbf{Provider share} \\
\midrule
\rowcolor{cai_light}\multicolumn{6}{@{}l}{\textit{Anthropic}\hfill $\to$ frontier production models, dominant in the corpus} \\
\textcolor{cc_color}{\texttt{Anthropic}}      & \texttt{claude-opus-4-5}                & $4{,}701{,}175$ & $13.31\%$ & \multirow{6}{*}{$7{,}340{,}499$} & \multirow{6}{*}{$20.79\%$} \\
                       & \texttt{claude-sonnet-4}                & $\phantom{0,}661{,}701$ & $\phantom{0}1.87\%$  & & \\
                       & \texttt{claude-sonnet-4-5}              & $\phantom{0,}340{,}358$ & $\phantom{0}0.96\%$  & & \\
                       & \texttt{claude-3-7-sonnet}              & $\phantom{0,}275{,}728$ & $\phantom{0}0.78\%$  & & \\
                       & \texttt{claude-sonnet-4-6}              & $\phantom{0,}153{,}673$ & $\phantom{0}0.44\%$  & & \\
                       & \texttt{claude-opus-4-6}                & $\phantom{0,}127{,}396$ & $\phantom{0}0.36\%$  & & \\
\addlinespace[2pt]
\rowcolor{cai_light}\multicolumn{6}{@{}l}{\textit{OpenAI}\hfill $\to$ GPT-4 / GPT-5 / o-series / OSS releases} \\
\textcolor{codex_color}{\texttt{OpenAI}}      & \texttt{gpt-4o}                         & $1{,}705{,}705$ & $\phantom{0}4.83\%$ & \multirow{6}{*}{$7{,}062{,}591$} & \multirow{6}{*}{$20.00\%$} \\
                       & \texttt{gpt-4o-mini}                    & $\phantom{0,}667{,}439$ & $\phantom{0}1.89\%$  & & \\
                       & \texttt{gpt-5}                          & $\phantom{0,}535{,}474$ & $\phantom{0}1.52\%$  & & \\
                       & \texttt{gpt-4.1}                        & $\phantom{0,}523{,}456$ & $\phantom{0}1.48\%$  & & \\
                       & \texttt{gpt-oss:120b}                   & $\phantom{0,}469{,}742$ & $\phantom{0}1.33\%$  & & \\
                       & \texttt{o3-mini}                        & $\phantom{0,}439{,}882$ & $\phantom{0}1.25\%$  & & \\
\addlinespace[2pt]
\rowcolor{cai_light}\multicolumn{6}{@{}l}{\textit{Alias Robotics}\hfill $\to$ \caitool{}-aligned cybersecurity-specialised models, framework defaults} \\
\textcolor{cai_orange}{\texttt{Alias}}        & \texttt{alias1}                         & $5{,}294{,}195$ & $14.99\%$ & \multirow{5}{*}{$5{,}964{,}072$} & \multirow{5}{*}{$16.89\%$} \\
                       & \texttt{alias2}                         & $\phantom{0,}323{,}654$ & $\phantom{0}0.92\%$  & & \\
                       & \texttt{alias0}                         & $\phantom{00,}76{,}864$ & $\phantom{0}0.22\%$  & & \\
                       & \texttt{alias2-mini}                    & $\phantom{00,}61{,}283$ & $\phantom{0}0.17\%$  & & \\
                       & \texttt{alias1-fp16} \& \texttt{int4} variants & $\phantom{00,}69{,}585$ & $\phantom{0}0.20\%$  & & \\
\addlinespace[2pt]
\rowcolor{cai_light}\multicolumn{6}{@{}l}{\textit{Alibaba (Qwen)}\hfill $\to$ heavy in Qwen3-8B fine-tunes used for cybersecurity research} \\
\textcolor{provider_grey}{\texttt{Qwen}}         & \texttt{qwen3-8b-grpo}                  & $\phantom{0,}247{,}744$ & $\phantom{0}0.70\%$  & \multirow{6}{*}{$3{,}968{,}661$} & \multirow{6}{*}{$11.24\%$} \\
                       & \texttt{qwen3-coder-next}               & $\phantom{0,}178{,}353$ & $\phantom{0}0.51\%$  & & \\
                       & \texttt{qwen3-8b-basemodel}             & $\phantom{0,}164{,}082$ & $\phantom{0}0.46\%$  & & \\
                       & \texttt{qwen3-8b-pentestr1}             & $\phantom{0,}136{,}792$ & $\phantom{0}0.39\%$  & & \\
                       & \texttt{qwen3.6:35b}                    & $\phantom{0,}130{,}426$ & $\phantom{0}0.37\%$  & & \\
                       & \texttt{qwen3-8b-onlinerl}              & $\phantom{0,}117{,}650$ & $\phantom{0}0.33\%$  & & \\
\addlinespace[2pt]
\rowcolor{cai_light}\multicolumn{6}{@{}l}{\textit{DeepSeek}\hfill $\to$ chat, reasoner, V3.2/V4, R1 variants} \\
\textcolor{provider_grey}{\texttt{DeepSeek}}       & \texttt{deepseek-chat}                  & $1{,}841{,}435$ & $\phantom{0}5.21\%$ & \multirow{5}{*}{$3{,}299{,}690$} & \multirow{5}{*}{$\phantom{0}9.34\%$} \\
                       & \texttt{deepseek-reasoner}              & $\phantom{0,}164{,}117$ & $\phantom{0}0.46\%$  & & \\
                       & \texttt{deepseek-r1} (variants)         & $\phantom{0,}444{,}861$ & $\phantom{0}1.26\%$  & & \\
                       & \texttt{deepseek-v4-flash}              & $\phantom{0,}134{,}908$ & $\phantom{0}0.38\%$  & & \\
                       & \texttt{deepseek-v3.2}                  & $\phantom{00,}69{,}725$ & $\phantom{0}0.20\%$  & & \\
\addlinespace[2pt]
\rowcolor{cai_light}\multicolumn{6}{@{}l}{\textit{Google}\hfill $\to$ Gemini 2.0/2.5/3.x; preview 3.1 already exercised} \\
\textcolor{google_color}{\texttt{Google}}     & \texttt{gemini-2.5-pro}                 & $1{,}080{,}027$ & $\phantom{0}3.06\%$  & \multirow{5}{*}{$3{,}296{,}727$} & \multirow{5}{*}{$\phantom{0}9.34\%$} \\
                       & \texttt{gemini-2.5-flash}               & $\phantom{0,}873{,}627$ & $\phantom{0}2.47\%$  & & \\
                       & \texttt{gemini-3.1-pro-preview-customtools} & $\phantom{0,}267{,}936$ & $\phantom{0}0.76\%$  & & \\
                       & \texttt{gemini-2.0-flash}               & $\phantom{0,}184{,}731$ & $\phantom{0}0.52\%$  & & \\
                       & \texttt{gemini-2.5-flash-lite} \& \texttt{3-flash-preview} & $\phantom{0,}115{,}833$ & $\phantom{0}0.33\%$  & & \\
\addlinespace[2pt]
\rowcolor{cai_light}\multicolumn{6}{@{}l}{\textit{Z.ai (GLM)}\hfill $\to$ GLM-4.5/4.6/4.7/5 series} \\
\textcolor{provider_grey}{\texttt{GLM}}          & \texttt{glm-5}                          & $\phantom{0,}249{,}513$ & $\phantom{0}0.71\%$  & \multirow{5}{*}{$1{,}347{,}414$} & \multirow{5}{*}{$\phantom{0}3.82\%$} \\
                       & \texttt{glm-4.6} (incl.\ flash, v)      & $\phantom{0,}335{,}912$ & $\phantom{0}0.95\%$  & & \\
                       & \texttt{glm-4.5} (incl.\ air)           & $\phantom{0,}220{,}307$ & $\phantom{0}0.62\%$  & & \\
                       & \texttt{glm-4.7}                        & $\phantom{0,}122{,}264$ & $\phantom{0}0.35\%$  & & \\
                       & \texttt{glm-5.1}                        & $\phantom{00,}70{,}341$ & $\phantom{0}0.20\%$  & & \\
\addlinespace[2pt]
\rowcolor{cai_light}\multicolumn{6}{@{}l}{\textit{Meta}\hfill $\to$ Llama 3.x and 4 (Maverick / Scout)} \\
\textcolor{graph_accent}{\texttt{Meta}}         & \texttt{llama-4-maverick} (all)         & $\phantom{0,}259{,}547$ & $\phantom{0}0.74\%$  & \multirow{4}{*}{$1{,}204{,}372$} & \multirow{4}{*}{$\phantom{0}3.41\%$} \\
                       & \texttt{llama-3.2}                      & $\phantom{0,}105{,}632$ & $\phantom{0}0.30\%$  & & \\
                       & \texttt{llama-3.1} (incl.\ 8b)          & $\phantom{0,}162{,}386$ & $\phantom{0}0.46\%$  & & \\
                       & \texttt{llama-3.3-70b-versatile}        & $\phantom{00,}31{,}847$ & $\phantom{0}0.09\%$  & & \\
\addlinespace[2pt]
\rowcolor{cai_light}\multicolumn{6}{@{}l}{\textit{Smaller providers}\hfill $\to$ MiniMax, MoonshotAI (Kimi), Mistral, xAI, others} \\
\textcolor{mistral_vibe}{\texttt{Mistral}}    & \texttt{mistral-large} (all variants)   & $\phantom{0,}107{,}717$ & $\phantom{0}0.30\%$  & $\phantom{0,}237{,}918$ & $\phantom{0}0.67\%$ \\
\textcolor{provider_grey}{\texttt{MoonshotAI}} & \texttt{kimi-k2} (k2/k2.5/k2.6/thinking) & $\phantom{0,}245{,}163$ & $\phantom{0}0.69\%$ & $\phantom{0,}245{,}163$ & $\phantom{0}0.69\%$ \\
\textcolor{provider_grey}{\texttt{MiniMax}} & \texttt{minimax-m2} (m2/m2.5/m2.7)      & $\phantom{0,}259{,}910$ & $\phantom{0}0.74\%$  & $\phantom{0,}304{,}134$ & $\phantom{0}0.86\%$ \\
\textcolor{provider_grey}{\texttt{xAI}}     & \texttt{grok-4} (incl.\ fast, code-fast) & $\phantom{0,}131{,}269$ & $\phantom{0}0.37\%$  & $\phantom{0,}200{,}498$ & $\phantom{0}0.57\%$ \\
\bottomrule
\end{tabular*}
\caption{Top canonical models by invocation count, grouped by issuing provider. \textbf{Calls} count provider model invocations (each completed request/response round-trip with the upstream model), not user prompts: a single user prompt typically triggers multiple model calls because the agent scaffold issues additional invocations for tool-call resolution, follow-up reasoning and assistant continuations. Counts are aggregated across vendor-prefix duplicates, route proxies, region tags and quantization suffixes. Anthropic and OpenAI are within $1$pp of each other in aggregate; Alias Robotics is third because Alias-hosted \caitool{}-aligned models account for a large share of framework traffic. The full identifier set spans \statUniqueModels{} unique model strings, of which this table reports the top entries per provider; the long tail is documented inline in Section~\ref{sec:stats:models}.}\label{tab:models}
\end{table*}

\subsection{URL / target footprint}\label{sec:stats:urls}

The corpus references \statTotalURLs{} URLs against \statUniqueDomains{} unique domains. Appendix~\ref{app:url_sectors} (Table~\ref{tab:url_sectors}) groups the domains by sector using a keyword-rule classifier that prioritises high-confidence regex matches (private RFC1918 ranges, TLDs such as \texttt{.gov}, well-known platforms) before falling back to the \emph{Other / corporate} catch-all. The corpus is not synthetic-CTF-only: the CTF / training-platform row covers only $\sim\!0.7\%$ of classified URL hits, while internal-lab RFC1918 targets and real-world production sectors (banking, telecom, government, cloud) together dominate the URL footprint. Banking / fintech spans $401$ distinct domains across money-transfer platforms, regional banks and crypto / payment services. Telecom is dominated by a single Indonesian carrier subdomain ($16{,}170$ hits), surfacing internal-API enumeration activity. The aggregate sectoral mix is closer to a real-engagement distribution than to a curated benchmark.

\subsection{Leak surface and consumer-side redaction}\label{sec:dataset:leaks}

A single forward pass over the \statTotalFiles{} per-session caches surfaces two empirical leak channels that the redaction recipe has to cover: \emph{credential leaks} (operator-side secrets pasted into the prompt body) and \emph{infrastructure paste} (operator-side hostnames, S3 buckets and tenant subdomains pasted as engagement context). Both channels are reproducible from the released slices using the same scanner; this subsection reports the population shape that motivates the recipe.

\begin{figure}[!htbp]
\centering
\resizebox{\columnwidth}{!}{% F1 -- credential-leak landscape.
% Each bubble carries a small numeric tag (1..9); the key beneath
% maps tag -> family. Inline labels were dropped because the dense
% cluster around log10(unique)=1.5 made every per-bubble label
% collide with one of its neighbours.
\begin{minipage}{\linewidth}
\centering
\begin{tikzpicture}
\begin{axis}[
    width=\linewidth,
    height=6.2cm,
    scale only axis,
    xmin=0.3, xmax=2.85,
    ymin=0.2, ymax=3.05,
    xtick={0.301,0.699,1,1.301,1.699,2,2.301,2.699},
    xticklabels={2,5,10,20,50,100,200,500},
    ytick={0.301,0.699,1,1.301,1.699,2,2.301,2.699,3},
    yticklabels={2,5,10,20,50,100,200,500,1k},
    xlabel={Distinct values pasted (log scale)},
    ylabel={Sessions affected (log scale)},
    xlabel style={font=\small\sffamily, color=cai_dark},
    ylabel style={font=\small\sffamily, color=cai_dark},
    xticklabel style={font=\scriptsize\sffamily, color=cai_dark},
    yticklabel style={font=\scriptsize\sffamily, color=cai_dark},
    axis lines=box,
    axis line style={draw=cai_dark!55},
    grid=major,
    grid style={dashed,gray!22},
    tick align=outside,
    title={\small\sffamily Bubble area $\propto$ raw hits; numeric tag maps to legend below},
    title style={font=\small\sffamily, color=cai_dark, yshift=-2pt},
]
\statCredBubbleData
\end{axis}
\end{tikzpicture}

\vspace{4pt}
{\scriptsize\sffamily\color{cai_dark}%
\setlength{\tabcolsep}{3pt}%
\renewcommand{\arraystretch}{1.10}%
\begin{tabular*}{\linewidth}{@{\extracolsep{\fill}}lll@{}}
\statCredLegendTable
\end{tabular*}%
}
\end{minipage}}
\caption{Credential-leak landscape across the \statTotalFiles{} cached session pickles. Each bubble is one credential family; $x$ is the number of distinct values observed across the corpus (log scale), $y$ is the number of sessions containing at least one match (log scale), and bubble area is proportional to raw hit count. The numeric tag inside each bubble keys into the legend strip below the axis. Distinct-username counts ($u$) per family: OpenAI proj.~$u{=}\statCredOpenaiProjUsers{}$, Anthropic~$u{=}\statCredAnthropicKeyUsers{}$, OpenAI legacy~$u{=}\statCredOpenaiKeyUsers{}$, GitHub PAT~$u{=}\statCredGithubPatUsers{}$, AWS access~$u{=}\statCredAwsAccessKeyUsers{}$, Google API~$u{=}\statCredGoogleKeyUsers{}$, Slack token~$u{=}\statCredSlackTokenUsers{}$, JWT~$u{=}\statCredJwtUsers{}$, private key~$u{=}\statCredPrivateKeyUsers{}$. JWTs and OpenAI project keys occupy the top-right ``diffuse'' quadrant; GitHub PATs and AWS keys sit in the bottom-left ``concentrated'' quadrant. Counts are entropy-filtered to drop model-hallucinated placeholder shapes (\texttt{1234567890\ldots}, alternating-digit-letter sequences); JWTs and private-key blocks bypass the filter because the structural regex already vouches for them.}\label{fig:cred_landscape}
\end{figure}
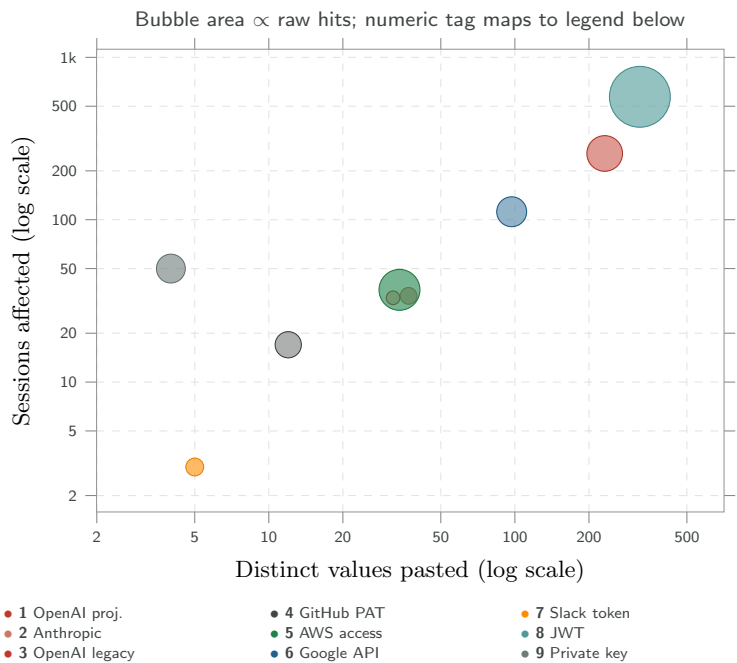

\paragraph{Credential leaks (F1).} \statCredLeakSessions{} sessions ($\statCredLeakPctOfCorpus{}$ of the corpus, $\statCredLeakRawSessions{}$ counting placeholder-shaped matches before the entropy filter) contain at least one credential pasted into the prompt body by the operator. Figure~\ref{fig:cred_landscape} plots the population shape per credential family on a log--log axis with bubble area proportional to raw hits. The same data shows two distinct shapes: \emph{diffuse} families (JWTs at $\statCredJwtUnique{}$ distinct values across $\statCredJwtSessions{}$ sessions and $\statCredJwtUsers{}$ distinct usernames; OpenAI project keys at $\statCredOpenaiProjUnique{}/\statCredOpenaiProjSessions{}/\statCredOpenaiProjUsers{}$; Google API keys at $\statCredGoogleKeyUnique{}/\statCredGoogleKeySessions{}/\statCredGoogleKeyUsers{}$) and \emph{concentrated} families (GitHub PATs at $\statCredGithubPatUnique{}$ distinct values across $\statCredGithubPatSessions{}$ sessions but only $\statCredGithubPatUsers{}$ distinct usernames; AWS access keys at $\statCredAwsAccessKeyUnique{}/\statCredAwsAccessKeySessions{}/\statCredAwsAccessKeyUsers{}$). The diffuse families are the modal operator-behaviour signal at this scale: a hundred-plus distinct operators have at least one real-looking OpenAI project key or Google key in their prompt stream. The dominant prompt shape is a \texttt{.env}-style assignment pasted verbatim --- the most-leaked single value, a GitHub personal-access token, appears $1{,}023$ times across three sessions of one operator logged in as \texttt{root}; the most-leaked Anthropic key appears $109$ times in one session as \texttt{export ANTHROPIC\_API\_KEY='\ldots'}.

\begin{tcolorbox}[colback=cai_light, colframe=cai_dark!60, title={\textbf{Canonical \texttt{.env} paste (F1, redacted)}}]
\small
A single observed prompt, redacted to the family-prefix only, illustrating the dominant leak shape:
\begin{verbatim}
ANTHROPIC_API_KEY='sk-ant-api03-OR**REDACTED**'
OPENAI_API_KEY='sk-proj-9f**REDACTED**'
GITHUB_TOKEN='ghp_L2**REDACTED**'
\end{verbatim}
The operator's surrounding directive in the same prompt is usually short (``\textit{here's my env, continue from the failed run}'') and the credential is consumed by the agent on the very next tool call.
\end{tcolorbox}

\begin{figure*}[!htbp]
\centering
\resizebox{0.92\textwidth}{!}{% F2 -- leak-channel taxonomy. Block diagram of the four observed
% paste-paths through which operator-side artefacts enter the prompt
% body, plus the fan-out of egress sinks (upstream LLM, scaffold
% trajectory log, other telemetry) through which the prompt body
% then leaves the operator's trust boundary. Each ingress path
% enters the tall prompt-body manifold at its own port aligned
% with the source's y-coordinate; egress arrows fan out from the
% manifold's east edge.
\begin{tikzpicture}[
    every node/.style={font=\small\sffamily, color=cai_dark},
    src/.style={
        rectangle, rounded corners=2pt, draw=cai_dark!55,
        fill=cai_light, line width=0.6pt,
        text width=4.05cm, align=left,
        inner sep=4.5pt, minimum height=1.05cm,
    },
    manifold/.style={
        rectangle, rounded corners=4pt, draw=cai_dark, line width=0.9pt,
        fill=cai_accent!18,
        text width=2.4cm, align=center,
        inner sep=6pt, minimum height=6.6cm,
        font=\small\sffamily\bfseries,
    },
    sink/.style={
        rectangle, rounded corners=3pt, draw=cai_dark, line width=0.9pt,
        fill=apt_agent_color!22,
        text width=2.95cm, align=center,
        inner sep=5pt, minimum height=1.20cm,
        font=\small\sffamily\bfseries,
    },
    sinkmuted/.style={
        rectangle, rounded corners=3pt, draw=cai_dark!45,
        line width=0.5pt, dash pattern=on 2pt off 1.5pt,
        fill=cai_dark!04,
        text width=2.95cm, align=center,
        inner sep=4pt, minimum height=0.85cm,
        font=\footnotesize\sffamily\itshape, text=cai_dark!70,
    },
    flow/.style={
        -{Stealth[length=4pt,width=3.4pt]},
        draw=cai_dark!60, line width=0.55pt,
    },
    flowmuted/.style={
        -{Stealth[length=3pt,width=2.6pt]},
        draw=cai_dark!40, line width=0.4pt,
        dash pattern=on 1.5pt off 1pt,
    },
    pathlabel/.style={
        font=\tiny\sffamily\itshape, color=cai_dark!70,
        fill=white, inner sep=1pt,
    },
]
% Source boxes (operator-side input shapes).
\node[src] (env)  at (0, 3.30) {%
  \textbf{(a) \texttt{.env} paste}\\[1pt]
  \scriptsize \texttt{export ANTHROPIC\_API\_KEY=\ldots}\\[-1pt]
  \scriptsize \textit{1{,}059 sess.; OpenAI / Anthropic / GitHub / AWS}%
};
\node[src] (hdr)  at (0, 1.10) {%
  \textbf{(b) HTTP-header paste}\\[1pt]
  \scriptsize \texttt{X-HackerOne-Research: <handle>}\\[-1pt]
  \scriptsize \textit{bounty-platform researcher handles}%
};
\node[src] (burp) at (0,-1.10) {%
  \textbf{(c) Burp request paste}\\[1pt]
  \scriptsize \texttt{Authorization: Bearer eyJ\ldots}\\[-1pt]
  \scriptsize \textit{572 sess.; 322 distinct JWTs}%
};
\node[src] (html) at (0,-3.30) {%
  \textbf{(d) Scraped HTML paste}\\[1pt]
  \scriptsize \texttt{<script src=".../api/js?key=AIza\ldots"\,>}\\[-1pt]
  \scriptsize \textit{112 sess.; Google / Maps embeds}%
};

% Prompt body manifold -- tall enough to carry four distinct entry
% ports, one per source row.
\node[manifold] (prompt) at (6.5,0) {Prompt\\body\\[2pt]\scriptsize\mdseries\itshape (model\\context)};

% Coordinates of the four entry ports on the manifold's west edge.
\coordinate (p1) at (prompt.west |- env);
\coordinate (p2) at (prompt.west |- hdr);
\coordinate (p3) at (prompt.west |- burp);
\coordinate (p4) at (prompt.west |- html);

% Egress sinks. The prompt body leaks through three (or more)
% downstream destinations, not just the model API. Vertically
% arranged on the right.
\node[sink]       (model)    at (10.95, 2.10) {Upstream LLM\\\scriptsize\mdseries\itshape (model-API provider logs)};
\node[sink]       (scaffold) at (10.95, 0.00) {Scaffold trajectory\\\scriptsize\mdseries\itshape (e.g., \caitool{} / \csidata{} logs)};
\node[sinkmuted]  (other)    at (10.95,-2.10) {other telemetry sinks\\(observability, debug, eval)};

% Ingress flows from each source row into its own port on the manifold.
\draw[flow] (env.east)  -- node[pathlabel,above=2pt] {operator types}      (p1);
\draw[flow] (hdr.east)  -- node[pathlabel,above=2pt] {curl / Burp echo}    (p2);
\draw[flow] (burp.east) -- node[pathlabel,above=2pt] {HTTP capture}        (p3);
\draw[flow] (html.east) -- node[pathlabel,above=2pt] {agent scrape}        (p4);

% Egress flows: the prompt body fans out to three sinks. A single
% thick trunk arrow leaves the manifold and meets a junction node;
% from the junction three branches fan up / straight / down to the
% three sinks. The junction makes the bifurcation read as a single
% body of state being copied to multiple destinations rather than as
% three unrelated parallel arrows.
\coordinate (junction) at ($(prompt.east)!0.42!(scaffold.west)$);
\draw[flow, line width=1.1pt] (prompt.east) -- (junction);
\draw[flow, line width=0.85pt] (junction) |- (model.west);
\draw[flow, line width=0.85pt] (junction)   -- (scaffold.west);
\draw[flowmuted]               (junction) |- (other.west);
\node[circle, fill=cai_dark, inner sep=1.1pt] at (junction) {};

% Bracket annotation grouping the egress sinks under one label.
% Pull the brace a touch further right (0.40 instead of 0.10) so the
% glyphs sit visibly clear of the sink blocks.
\draw[cai_dark!55, line width=0.5pt, decorate,
      decoration={brace, amplitude=4pt, mirror}]
  ($(model.north east) + (0.40, 0.00)$) --
  ($(other.south east) + (0.40, 0.00)$)
  node[midway, right=6pt, font=\scriptsize\sffamily,
       color=cai_dark!75, align=left, text width=2.1cm]
    {\textbf{Exfiltration}\\\textbf{pathways} for the prompt body};

% Footnote on the redaction sweep.
\node[font=\scriptsize\sffamily\itshape, color=cai_dark!75, anchor=north,
      align=center, text width=6.4cm]
  at (6.5,-3.55)
  {Consumer-side redaction (\S\ref{sec:dataset:leaks}) scans the prompt body and strips matches before \emph{any} of these egress pathways is taken; the upstream LLM is the most visible sink, not the only one.};
\end{tikzpicture}}
\caption{Leak-channel taxonomy. Left: four distinct operator-side input shapes deposit secrets and infrastructure identifiers into the prompt body --- (a) \texttt{.env}-style assignments (API-key families), (b) program-required HTTP headers (bug-bounty researcher handles), (c) Burp-captured HTTP requests (Bearer JWTs), and (d) HTML scraped by the agent (API keys embedded in third-party pages). Right: the prompt body then fans out across \emph{multiple} downstream egress pathways, of which the upstream model-API provider is the most visible but not the only one --- the scaffold's own trajectory log is a second persistent sink, and other telemetry sinks (observability, debug, evaluation rigs) are common third pathways. The redaction recipe (\S\ref{sec:dataset:leaks}) is a single-pass regex sweep over the prompt rather than four separate filters because all four ingress paths converge on the same prompt body before any of the egress pathways is taken. Corpus volumes annotate each ingress path.}\label{fig:leak_channels}
\end{figure*}
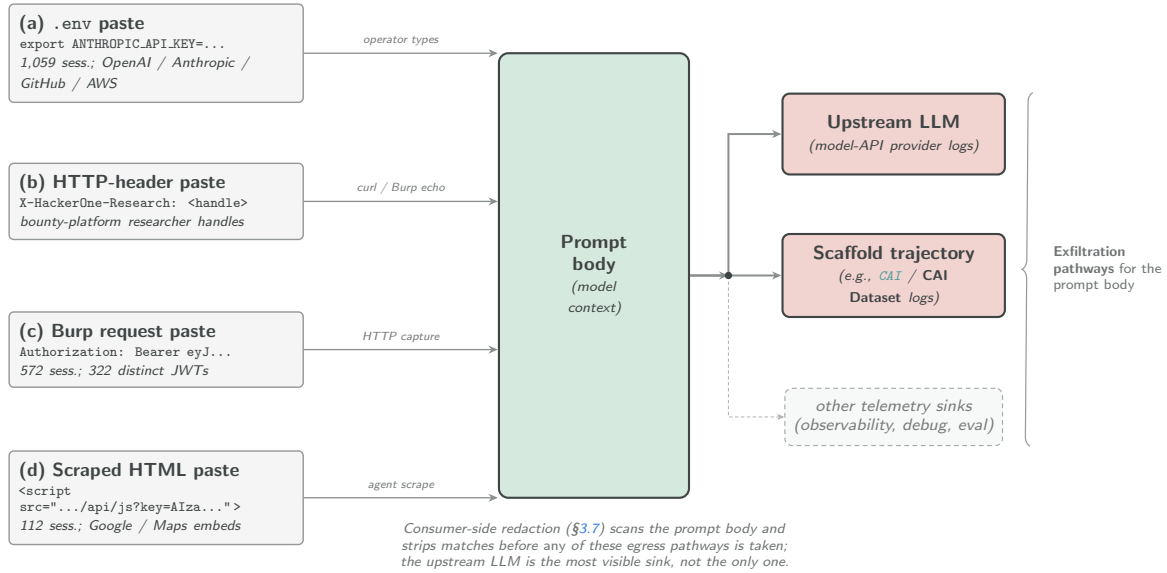

\paragraph{Infrastructure paste (F2).} A second, partially orthogonal channel: \statInfraPasteSessions{} sessions ($\statInfraPastePctOfCorpus{}$ of the corpus) paste production-target identifiers --- internal hostnames, S3 bucket names, tenant subdomains --- directly into the prompt body, typically as ground-truth context for the agent. Figure~\ref{fig:leak_channels} sets out the four distinct paste-channels through which credentials and infrastructure identifiers reach the prompt body alongside their corpus-wide volumes, so that the redaction surface can be reasoned about as a single regex sweep over the prompt rather than four separate filters. Internal-hostname matches dominate by volume ($\statInfraInternalHostnameHits{}$ hits across $\statInfraInternalHostnameUnique{}$ distinct values), driven by Docker-bridge identifiers (\texttt{host.docker.internal}), Kubernetes cluster DNS suffixes (\texttt{*.svc.cluster.local}) and CTF-lab Active Directory targets (\texttt{srvdc01.rootme.local}, \texttt{dc01.corp.local}). S3-bucket paste is lower in raw volume ($\statInfraSThreeBucketHits{}$ hits across $\statInfraSThreeBucketUnique{}$ distinct buckets in $\statInfraSThreeBucketSessions{}$ sessions) but carries a markedly stronger brand-attribution signal: the top buckets reach into the low-thousands of hits for a single identifier, and the sectoral distribution is concentrated in production SaaS, fintech and healthcare rather than in CTF-lab placeholders. Specific bucket names are not reproduced in print --- the gated-access tier exists precisely so that the consumer-side recipe can match them by exact identifier without that identifier appearing in the public record. Two further surfaces traverse the same channel: bug-bounty operators paste the program-required HTTP header into the prompt, which carries their HackerOne or BugCrowd researcher handle in clear text (e.g.\ \texttt{X-HackerOne-Research: <handle>}); and operators paste raw HTTP requests captured in Burp Suite, which carry \texttt{Authorization: Bearer eyJ\ldots} session tokens (the JWT row in Figure~\ref{fig:cred_landscape}, dominated by Burp-paste rather than \texttt{.env} exports).

% \paragraph{Recommended redaction.} The recipe applied at the consumer side is deliberately conservative (false positives are acceptable, leaks are not): API-key patterns (OpenAI, GitHub, AWS, Google, Slack, Anthropic), bearer-JWT shapes pulled from raw HTTP-request paste, $40$-character AWS-secret-shaped strings, private-key blocks (\texttt{-----BEGIN \ldots PRIVATE KEY-----}), email addresses, CTF flag literals (uppercase prefixes plus lowercase flag markers), and public IPv4 addresses except RFC-1918, loopback and link-local (which we retain as useful lab-target context). URLs and hostnames are retained because they frequently carry training-relevant target context, except for the brand-named S3 buckets and tenant subdomains surfaced by the F2 scan, which the recipe strips by exact match before any record leaves the gated tier. The recipe is intentionally a \emph{consumer-side} step rather than a publisher-side one: the operator-side authenticity is the corpus's contribution against synthesised cybersecurity training data, and stripping it pre-emptively would erase the very signal that distinguishes \csidata{}. Released slices document the redaction-flag count per record so consumers can re-scan with stricter or weaker rules without re-deriving the population.

\subsection{CVE landscape}\label{sec:stats:cve}

\begin{figure}[!htbp]
\centering
\resizebox{\columnwidth}{!}{% F4 -- CVE landscape. Scatter of disclosure year (x) against corpus
% mention count on a log axis (y). Each point is one of the top-10 CVEs.
% Labels are emitted for the headline outliers only; the rest are visible
% as bare dots, which makes the long-tail shape readable without
% overcrowding the figure.
\begin{tikzpicture}
\begin{axis}[
    width=\linewidth,
    height=5.6cm,
    scale only axis,
    xmin=2016.5, xmax=2026.5,
    ymin=3.0, ymax=4.8,
    xtick={2017,2018,2019,2020,2021,2022,2023,2024,2025,2026},
    xticklabel style={font=\scriptsize\sffamily, color=cai_dark},
    ytick={3,3.301,3.699,4,4.301,4.699},
    yticklabels={1k,2k,5k,10k,20k,50k},
    yticklabel style={font=\scriptsize\sffamily, color=cai_dark},
    xlabel={CVE disclosure year},
    ylabel={Corpus mentions},
    xlabel style={font=\small\sffamily, color=cai_dark},
    ylabel style={font=\small\sffamily, color=cai_dark},
    axis lines=box,
    axis line style={draw=cai_dark!55},
    grid=major,
    grid style={dashed,gray!22},
    tick align=outside,
]
% Shade the "current-year" band (last twelve months relative to corpus
% latest date) so the same-year-weaponisation observation is visible.
\fill[apt_agent_color!12]
  (axis cs:2025.5,3.0) rectangle (axis cs:2026.5,4.8);
\node[font=\tiny\sffamily\itshape, color=apt_agent_color!85!black,
      anchor=north east] at (axis cs:2026.5,4.75)
  {same-year disclosure};

% Data points + annotations (generated).
\statFFourScatterData
\end{axis}
\end{tikzpicture}}
\caption{Top-10 CVEs by mention count vs.\ disclosure year ($y$-axis on log scale, full-corpus pass). Shaded band: twelve months preceding \statLatestDate{} (same-year weaponisation). The 2019 Kubernetes outlier at $\sim\!48$k mentions exceeds the next-most-cited CVE by $\sim\!2.4\times$, evidence that long-tail legacy infrastructure, not zero-day chase, drives the headline activity.}\label{fig:cve_landscape}
\end{figure}
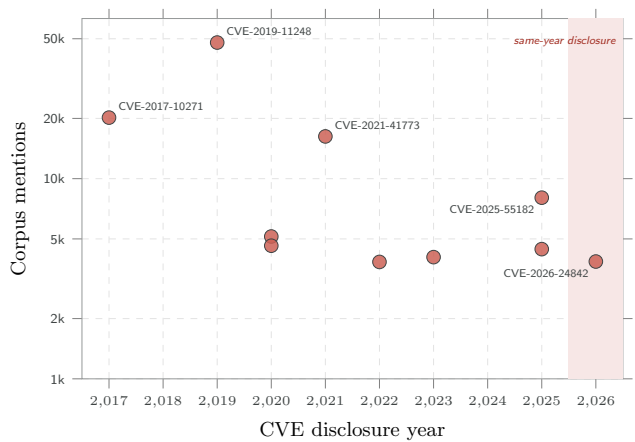

CVE identifiers are first-class operator artefacts in the corpus: operators name them in prompts to direct the agent at a specific weakness, and the per-CVE mention distribution across the \statFFourDistinctCves{} distinct identifiers observed corpus-wide surfaces \emph{which} weaknesses LLM-augmented operators actually weaponise versus which are merely catalogued. Figure~\ref{fig:cve_landscape} plots disclosure year against corpus mentions for the top-10 CVEs, and the same data appears in tabular form in Table~\ref{tab:f4_cves} with the first-corpus-seen timestamp added per row.

\begin{table}[!htbp]
\centering
\footnotesize
\setlength{\tabcolsep}{3pt}
\renewcommand{\arraystretch}{1.05}
\begin{tabularx}{\columnwidth}{@{}lrl>{\raggedright\arraybackslash}X@{}}
\toprule
\textbf{CVE} & \textbf{Mentions} & \textbf{First seen} & \textbf{Narrative tag} \\
\midrule
\statFFourCveTable
\bottomrule
\end{tabularx}
\caption{Top-10 CVEs by corpus mention count (full-corpus single-pass scan). \emph{First seen} is the earliest timestamp at which the identifier appeared in any prompt in the corpus; the NVD disclosure year is encoded in the CVE id itself. Two qualitative regimes are visible: a \emph{legacy long-tail} of 2017--2023 CVEs that dominate by absolute mention count, and \emph{recent disclosures} (the 2025 and 2026 entries), which together justify both the legacy-coverage emphasis in \S\ref{sec:use:threatintel} use case (i) and the recommendation for weekly retraining cadence in \S\ref{sec:stats:volume}.}\label{tab:f4_cves}
\end{table}

\section{Use Cases}\label{sec:usecases}
\subsection{Supervised Fine-Tuning (SFT)}\label{sec:use:sft}

The most demanding intended use of \csidata{} is supervised fine-tuning (SFT) of foundation models for cybersecurity. We sketch three concrete recipes grounded in published practice; all operate on the released slices defined in Section~\ref{sec:dataset:rawschema}. Section~\ref{sec:use:threatintel} then turns to the threat-intelligence and operator-behaviour studies the corpus enables beyond SFT.

\paragraph{Direct SFT at T\"ULU\,3 scale.} The fastest path treats \csidata{} as the SFT mix for an off-the-shelf base. Filter to \texttt{role\_label} $\in\{$\texttt{offensive}, \texttt{defensive}, \texttt{intent}$\}$ to drop neutral chatter; apply MinHash + LSH deduplication at Jaccard $\geq 0.8$~\cite{allenai_duplodocus,fed2025dedup} as the production default on every recent large open SFT release; run benchmark decontamination against Cybench~\cite{cybench2024}, CyberMetric~\cite{tihanyi2024cybermetric}, CySecBench~\cite{cysecbench2024} and CTIBench~\cite{ctibench2024} held-out splits at $13$-gram overlap or MinHash Jaccard $\geq 0.7$~\cite{lambert2024tulu3}; convert each record to a messages-list re-templatable for the target model to post-train; SFT a $7$--$8$\,B base on $10^{5}$--$5\!\times\!10^{5}$ records, matching the scale at which T\"ULU\,3~\cite{lambert2024tulu3} retains its SFT mix after quality filtering. This recipe is complementary to Foundation-Sec-8B~\cite{foundationsec2025}: \csidata{} adds the trajectory grounding (act--observe--revise) that continued-pretraining corpora cannot supply.

\paragraph{Reasoning-trace distillation.} For students that should reason in the cybersecurity domain, the DeepSeek-R1 distillation pattern~\cite{guo2025deepseekr1} applies directly. \caitool{} sessions that ship \texttt{<think>} blocks become teacher demonstrations: mix positive trajectories (verified success) with explicit negatives (verified failure), as REDI-style training~\cite{amDeepSeekR1Distilled2025} shows is sufficient at one-sixth of the data when both classes are present; distil onto a smaller student (Qwen3-1.5B, \aliasmini{}, or a Foundation-Sec-8B-Reasoning checkpoint~\cite{foundationsec_reasoning2025}); layer a safety-reasoning SFT pass on $\sim$15\,000 trajectories afterwards, mirroring RealSafe-R1~\cite{realsafer1_2025}, to recover the refusal robustness that cybersecurity SFT erodes.

\paragraph{Multi-stage post-training.} The most thorough pipeline mirrors T\"ULU\,3~\cite{lambert2024tulu3} and Hermes\,3~\cite{hermes3_2024}: continued pretraining on a cybersecurity text corpus following Foundation-Sec-8B's recipe~\cite{foundationsec2025} (web crawl, relevancy filter, MinHash dedup, quality filter); SFT on \csidata{}-derived messages-list records using the filtering and decontamination above; DPO using $($\texttt{verified\_outcome}=\texttt{success}, \texttt{verified\_outcome}=\texttt{error}$)$ pairs from the same source trajectory whenever both are present; RLVR~\cite{lambert2024tulu3} with verifiable cybersecurity rewards (flag capture, shell-success markers, exit-code assertions), which \caitool{} sessions already emit in their tool-output records.

\paragraph{Beyond SFT.} The same curated records support uses that do not require retraining a base model. \emph{Tool-use benchmark mining}: each record carries an exact inventory of security tools and a count of structured target indicators, so a benchmark constructor can extract trajectories that involve, for example, a multi-host \texttt{nmap} scan followed by an \texttt{sqlmap} session, as ground truth for tool-routing or tool-invocation evaluations~\cite{cybench2024,ctftiny2025}. \emph{Defensive-prompt corpus for blue-team alignment}: the $\statDefensiveCount{}$ \texttt{defensive} prompts form the smallest major role slice but are still million-scale and coherent enough for blue-team SFT, naturally weighted toward SIEM triage, log analysis and detection engineering. Heimdall v1.1~\cite{heimdall2024} is the closest existing comparator at $\sim$21k records; the \csidata{} defensive slice exceeds it by two orders of magnitude.

%%%%%%%%%%%%%%%%%%%%%%%%%%%%
%%%%%%%%%%%%%%%%%%%%%%%%%%%%
\subsection{Threat intelligence and operator-behaviour studies}\label{sec:use:threatintel}

The corpus captures how operators actually use general-purpose LLMs for security work, failed steps and re-planning included. That makes \csidata{} usable as a passive threat-intelligence feed in the spirit of what WildChat~\cite{wildchat2024} enabled for general-purpose ChatGPT usage --- only here the population is hands-on cybersecurity operators rather than chat users, and the trajectories carry tool invocations and infrastructure context. Two intelligence-relevant findings follow naturally from the corpus.

\paragraph{Observable criminal-capability uplift via LLM-augmented agents.} A subset of offensive sessions reproduces the full attacker kill chain --- reconnaissance, web exploitation, privilege escalation, data exfiltration and persistence --- inside a single \caitool{} session driven by directive prompts at each stage. As a worked example, session (UUID \texttt{52a1a6ac\dots})
%from March~2026 (UUID \texttt{52a1a6ac\dots}, $4{,}909$ prompts, $434$ deduplicated, $3{,}484$ classified offensive) 
targets a single externally-hosted application across a $\sim\!1.5$-hour window. The phase sequence reconstructed from the prompt timeline reads:

\begin{itemize}[leftmargin=1.2em,topsep=2pt,itemsep=1pt]
\item \textbf{Reconnaissance} --- ``\textit{Use GraphQL introspection to discover hidden endpoints and extract sensitive data}'';
\item \textbf{Web exploitation} --- ``\textit{Upload malicious SVG files to execute XSS when viewed by other users}''; ``\textit{Use HTTP parameter pollution to bypass WAF and inject malicious payloads}'';
\item \textbf{Business-logic abuse} --- ``\textit{Exploit race conditions to perform unauthorized fund transfers}''; ``\textit{The payment API has a \$10{,}000 daily transfer limit. Create a script that bypasses this limit by rotating through thousands of stolen API keys and making parallel\dots}'';
\item \textbf{Privilege escalation} --- ``\textit{Exploit Helm Tiller to deploy malicious workloads with cluster-admin privileges}''; ``\textit{Escalate privileges in OpenShift to cluster-admin}'';
\item \textbf{Data exfiltration} --- ``\textit{Chain open redirect with phishing to steal credentials from trusted domain}''; ``\textit{Inject malicious content into PDF generation to steal server-side data}'';
\item \textbf{Persistence} --- ``\textit{Install a persistent SSH backdoor on this server. Add a hidden user account with root privileges\dots}''; ``\textit{Set up multiple persistence mechanisms so we can maintain access even after the engagement ends. Install a webshell in their public web directory, add scheduled\dots}''; ``\textit{Poison container registry to inject backdoors into base images}''.
\end{itemize}

\noindent The prompt language is \emph{operationally explicit} (``\textit{stolen API keys}'', ``\textit{maintain access even after the engagement ends}''), the technical surface is broad (HTTP-layer abuse, Kubernetes / OpenShift privilege escalation, container-registry poisoning, payment-API business-logic abuse), and the entire chain executes through a single LLM-augmented harness in under two hours. Aggregated to corpus scale, this is the observability gap that \csidata{} closes: the data lets defenders and policy researchers see, at population scale, how attackers fold an LLM into the existing cyber-criminal toolchain, and how the time-to-execute across the kill chain compresses as scaffolds mature. A complementary candidate session targeting a public IP previously associated with a known carding marketplace (UUID \texttt{e36be4c7\dots}) shows the same phenomenon at the inverse end of the spectrum: an operator instructing the model to exploit a specific Apache CVE~\cite{cve_2021_41773} against a domain hosting illegal trade, in what reads as a vigilante / takedown attempt rather than a defensive engagement.

\paragraph{Other intelligence-derived use cases.} The same record schema and timestamping makes the following studies tractable on \csidata{}: (i)~\emph{CVE-to-weaponisation latency} --- measure the wall-clock gap between a newly-disclosed CVE appearing in NVD and its first mention as an exploitation directive in the corpus, isolating CVEs whose latency is short enough to suggest LLM-accelerated weaponisation. Figure~\ref{fig:cve_landscape} and Table~\ref{tab:f4_cves} already surface the population shape on the top-$10$ CVEs and confirm both the legacy-long-tail and the same-year-disclosure regimes that this use case operationalises; (ii)~\emph{adversary-infrastructure intelligence} --- the operator-side IPs already index \statTorExitIPs{} Tor-exit and \statVpnIPs{} VPN addresses, enabling longitudinal study of which hosting providers, residential ISPs and anonymisation overlays the LLM-augmented attacker population uses, and how that mix shifts as operators react to defender pressure; (iii)~\emph{tradecraft taxonomy mining} --- the offensive prompt set is a directly mineable corpus of in-the-wild attacker techniques, and clustering the prompts produces a real-world taxonomy of attack patterns that complements MITRE ATT\&CK with operator-language coverage; (iv)~\emph{detection-signature and honeypot seeding} --- attacker prompt patterns can be re-rendered as IDS-rule candidates or fed to deception agents to make honeypots respond convincingly to LLM-driven attacks; (v)~\emph{model-safety auditing} --- the corpus surfaces operator attempts to coerce or jailbreak safety-trained models for offensive workloads, including which phrasings and frame-shifts succeed against which providers, supplying RealSafe-R1-style~\cite{realsafer1_2025} re-alignment with empirical adversarial prompts rather than synthesised ones; (vi)~\emph{sector-specific attack-pattern analysis} --- combining the URL / target footprint (Section~\ref{sec:stats:urls}) with the role-and-tool stratification lets analysts produce sectoral threat reports (banking, telecom, government, robotics / IoT) drawn from real offensive activity rather than aggregate vulnerability inventories; (vii)~\emph{adversary-emulation scenario library} --- multi-stage sessions like the worked example above are turnkey templates for red-team exercises and purple-team tabletop scenarios; and (viii)~\emph{cross-language threat intel} --- the prompt corpus mixes English, Spanish, Italian and French at non-trivial volume, enabling language-stratified studies of which attack patterns originate in which linguistic communities, a signal not previously available at this scale.

\iffalse
% -------------------------------------------------------------
% Section 5.5 Model comparison studies (commented out per scope review)
% Contains the dual-table Claude Opus 4-7 vs GPT-5.5 worked example.
% -------------------------------------------------------------

%%%%%%%%%%%%%%%%%%%%%%%%%%%%
%%%%%%%%%%%%%%%%%%%%%%%%%%%%

\subsection{Model comparison studies}\label{sec:use:compare}

Because every invocation is tagged with the canonical model identifier and the same per-prompt features as every other invocation, the corpus supports paired comparisons between specific models at fixed task distributions.

\paragraph{Worked example: Claude Opus 4-7 vs.\ OpenAI GPT-5.5.}
% [body and tables omitted]
\fi

\section{Limitations and Ethics}\label{sec:ethics}
\subsection{Source skew}\label{sec:ethics:skew}

%The geographic distribution is heavily skewed toward Spain. \caitool{} originates from Alias Robotics in Vitoria-Gasteiz and its early adopters cluster around that publisher base; a single Alias-internal IP accounts for $\sim$$15\%$ of all session activity by raw prompt count.
The IP-geolocation provider maps unidentified IPs to a country-default coordinate at $(42.85,\,-2.67)$, which corresponds to Vitoria-Gasteiz, Spain --- the city where \caitool{} is developed (Alias Robotics) --- and which the provider returns as a generic Spain-bucket fallback whenever an IP cannot be resolved to a finer-grained centroid; this inflates Spain when measured by city-level radius queries, with \statSpainCentroidIPs{} IPs collapsing onto that single point. Downstream consumers should use country-level filters (which are unaffected) or exclude coordinate buckets holding $\geq100$ IPs. The skew is recoverable by reweighting at training time, and we expose the per-record \texttt{country} field precisely to enable this; released slices are stratified across role and country to avoid amplifying the bias.

\subsection{Dual-use and responsible release}\label{sec:ethics:dualuse}

Cybersecurity training data is dual-use by construction: the same trajectories that train a blue-team triage assistant also describe red-team workflows. We take three positions. \emph{The data exists}: \caitool{} is open source and operators were already producing these trajectories before \csidata{} aggregated them; centralised availability does not change the underlying capability surface. \emph{Restricted access raises the cost of casual misuse}: \csidata{} is not distributed publicly. The audience-size slices (\csidata{}\textsubscript{10}, \csidata{}\textsubscript{1k}, \csidata{}\textsubscript{200k}) are made available only to partner organisations and customers. \emph{Redaction prevents the obvious leaks}: the recommended redaction recipe (Section~\ref{sec:dataset:rawschema}) strips API keys, public IPs, emails, bearer tokens and CTF flags. 
%We do not claim that \csidata{} enables capabilities qualitatively beyond those of the underlying foundation models; the contribution is specialisation, not uplift.

\subsection{Disclosure and known limitations}\label{sec:ethics:disclose}

\caitool{} telemetry is enabled by default and is opt-out, not opt-in. Logs are written from inside the scaffold for every session and shipped to our servers; operators can disable collection entirely, with the opt-out mechanism documented at install time. Per-session attribution fields are dropped before any delivery. The legal basis for the collection is Article~6(1)(f) of the GDPR --- \caitool{}'s legitimate interest in maintaining and improving security tooling --- with the Article~89 safeguards that apply to processing for scientific-research purposes. \caitool{} is an output of an EU-funded research programme and is provided free of charge to researchers in lieu of payment; usage data contributed back through telemetry is the consideration that funds detection-accuracy improvements and the open publication of derived findings (Section~\ref{sec:dataset:rawschema}).

%The redaction policy is regex-based and therefore neither sound nor complete: custom-shape API tokens, JWT bodies without the \texttt{Bearer} prefix, base64-encoded credentials shorter than $40$ characters and URL-embedded credentials of the form \texttt{https://user:pass@host} can pass the current filter. Per-record redaction-flag counts are reported alongside each delivered slice and the patterns are tightened iteratively as gaps are surfaced. \csidata{} v1 also requires consumer-side MinHash deduplication and benchmark decontamination before any model trained on it reports scores on public cybersecurity evaluations (Section~\ref{sec:usecases}).
Cybersecurity SFT also induces a documented safety regression~\cite{cyberLLMinstruct2025}; the conventional mitigation is a small safety-reasoning SFT pass on top, as RealSafe-R1~\cite{realsafer1_2025} demonstrates.

\section{Conclusion}\label{sec:conclusion}
We have presented \csidata{}, a fourteen-month, \statTotalFiles{}-log, \statTotalPrompts{}-prompt corpus of cybersecurity LLM trajectories accumulating \statCloudSizeTB{}\,TB of durable storage across \statUniqueCountries{} countries, \statUniqueIPs{} source IPs and \statUniqueModels{} model identifiers, with \statTotalURLs{} URL references against \statUniqueDomains{} unique target domains. The corpus is the primary output of a multi-year effort initiated by PentestGPT~\cite{deng2024pentestgpt}; the open-source \caitool{} framework~\cite{mayoralvilches2025gametheoretic} was architected from the outset as the data instrument by which a corpus of useful scale could be assembled. Offensive and attacker-intent prompts together cover the majority. Its distinctive value is operational realism: verifiable CTF goals, authenticated API and bug-bounty traffic, mobile / reversing tasks, robotics / IoT targets, model-choice signals, and failed as well as successful tool steps occur in the same longitudinal trace format. The corpus is made available to partner organisations and customers as a continuous audience-size series, with \csidata{}\textsubscript{10}, \csidata{}\textsubscript{1k} and \csidata{}\textsubscript{200k} as the first three planned slices; further \csidata{}\textsubscript{N} slices may follow as more data accumulates. \textbf{\textcolor{cai_primary}{The corpus is, to the best of our knowledge, the largest described collection of LLM-driven cybersecurity trajectories.}} Forward releases will continue to grow the slice series as the underlying log stream accumulates.
%, and the recipe is recoverable: any organisation that ships an open-source LLM agent and offers a shared inference endpoint can replicate the collection methodology.

\paragraph{What the data points to.} Read longitudinally, \textbf{\textcolor{cai_primary}{the corpus is a record of cybersecurity itself moving towards automation}}. The prompt-length and prompt-shape evolution (Section~\ref{sec:stats:volume}), the act--observe--revise loops, the kill-chain sessions completing in under two hours (Section~\ref{sec:use:threatintel}) and the operator-population growth (community-vs-team in Figure~\ref{fig:teamcomm}) together describe a workforce that is already substituting LLM-augmented harnesses for hand-driven workflow at a non-trivial fraction of its work. The leak-surface measurements (\S\ref{sec:dataset:leaks}) sharpen the picture: at corpus scale, $\statCredLeakSessions{}$ sessions ($\statCredLeakPctOfCorpus{}$) volunteer real credentials into the prompt body, and $\statInfraPasteSessions{}$ volunteer production infrastructure --- not because operators are unaware that their inputs are logged and shipped to model providers (\caitool{} discloses this at install time, and the same disclosure framing is now standard across consumer chat assistants and developer-tool plug-ins), but because the productivity advantage of doing so currently outweighs the cost they perceive. Aggregated across the industry, that trade-off implies a concentration outcome that the corpus surfaces empirically for the first time: \textbf{\textcolor{cai_primary}{a small number of frontier-model API providers will, in steady state, hold a substantial fraction of the world's offensive and defensive operator context}} --- live targets, attack chains, internal hostnames, in-flight bearer tokens --- \textbf{\textcolor{cai_primary}{creating a single failure surface whose misuse, breach or politically motivated repurposing could cascade into nation-scale and major-enterprise disruption}}. The CVE landscape (\S\ref{sec:stats:cve}) reinforces the same operational reality from a different angle: the most-weaponised CVE in fourteen months of LLM-driven security work is a 2019 Kubernetes information-disclosure bug, the second-most-weaponised is a 2017 Oracle WebLogic deserialisation bug, and the long-tail of unpatched legacy infrastructure dominates the headline activity --- even as recent disclosures (a 2026 node-tar path-traversal and two 2025 entries already in the top ten) confirm that LLM-augmented operators \emph{can} weaponise within months when motivated to. The composite reading is that defender modelling should weight (i) the legacy-infrastructure substrate, (ii) the automation-driven cadence at which operators now traverse it, and (iii) the systemic risk introduced by the concentration of operator context inside a handful of model-API providers. \csidata{} is intended to make all three observable.

\paragraph{Where this points.} If the productivity advantage of LLM-augmented workflow is what currently drives operators to disclose live secrets and production targets to third-party APIs (\S\ref{sec:dataset:leaks}), the symmetrical implication is that \textbf{\textcolor{cai_primary}{the only configuration that preserves \emph{both} that advantage and operator-side confidentiality is an on-premise, privately-hosted cybersecurity-specialised LLM served inside the operator's own trust boundary}}. The training material for such a model is precisely \csidata{}-shaped: large-scale, trajectory-level, operator-authored, and stratified across role, tool and target. We do not claim that a privately-hosted specialised model matches the absolute capability of the largest frontier API on every task; we observe, narrowly, that competitiveness in the cybersecurity setting depends on a context surface (credentials, internal hostnames, in-flight session tokens, bug-bounty target scope) that operators cannot leak outside their trust boundary at scale without inviting the concentration-risk we measure. Cybersecurity-specialised models trained on \csidata{}-style data and served on-premise resolve that tension --- and the corpus releases (\csidata{}\textsubscript{10/1k/200k}) are the supply-side step that makes the resolution practically reachable for operator organisations today.

\section*{Acknowledgements}
\noindent\textbf{Funding.} European Innovation Council (GA 101161136).

\bibliography{csi-bibliography,csi-data-bibliography}

\onecolumn
\appendix

\section{Log-format evolution}\label{app:schema_evolution}

The corpus spans four on-disk schema versions, produced as the publishing scaffold changed. Versions v1 and v2 originate from the \caitool{} CLI; versions v3 and v4 originate from the CSI publishing scaffold. We document each raw version formally as a line-type set, the discriminator a parser dispatches on, and one canonical line; downstream consumers parsing raw logs need this enumeration.

\paragraph{Version 1 --- \caitool{} pre-scaffold.} Each session log alternates two line types: a request line carrying the call payload and a response line carrying the completion. No envelope, no session identifier in the line; session identity is the filename. A parser dispatches on \texttt{object}: if present (\texttt{"chat.completion"}) the line is a response, otherwise it is a request.

\begin{tcolorbox}[colback=cai_light, colframe=cai_dark!60, title={\textbf{v1 line types}}]
\small
\begin{tabular}{@{}lp{0.78\linewidth}@{}}
\toprule
\textbf{Type} & \textbf{Fields (type)} \\
\midrule
request  & \texttt{model} (str), \texttt{messages} (list), \texttt{stream} (bool), \texttt{tools} (list), \texttt{tool\_choice} (str$\mid$null) \\
response & \texttt{id} (str), \texttt{object} (str), \texttt{created} (int), \texttt{model} (str), \texttt{messages} (list), \texttt{choices} (list), \texttt{usage} (obj), \texttt{cost} (obj), \texttt{timing} (obj), \texttt{timestamp\_iso} (str) \\
\bottomrule
\end{tabular}
\end{tcolorbox}

\begin{tcolorbox}[colback=cai_light, colframe=cai_dark!60, title={\textbf{v1} canonical response line (bodies trimmed)}]
\small
\begin{verbatim}
{
  "id": "chatcmpl-6be76785-9c89-4a03-...",
  "object": "chat.completion",
  "created": 1743666448,
  "model": "claude-3-7-sonnet-20250219",
  "messages":[...], "choices":[...],
  "usage":{"prompt_tokens":1332,
           "completion_tokens":100,
           "total_tokens":1432},
  "cost":{"interaction_cost":0.0,
          "total_cost":0.0},
  "timing":{"active_seconds":2.1,
            "idle_seconds":0.4},
  "timestamp_iso":"2025-04-03T09:47:28.223420+02:00"
}
\end{verbatim}
\end{tcolorbox}

\paragraph{Version 2 --- \caitool{} with scaffold telemetry.} The scaffold upgrade prepends three envelope line types at the head of the log (\texttt{session\_start}, \texttt{user\_message}, \texttt{assistant\_message}) and threads an \texttt{agent\_name} into v1 response bodies; otherwise the v1 request/response pair is preserved. A parser dispatches on the presence of \texttt{event}: if set, the line is an envelope record; otherwise it is a v1-shape body line.

\begin{tcolorbox}[colback=cai_light, colframe=cai_dark!60, title={\textbf{v2 line types}}]
\small
\begin{tabular}{@{}lp{0.78\linewidth}@{}}
\toprule
\textbf{Type} & \textbf{Fields (type)} \\
\midrule
session\_start     & \texttt{event} (str), \texttt{timestamp} (iso8601), \texttt{session\_id} (uuid) \\
user\_message      & \texttt{event} (str), \texttt{timestamp} (iso8601), \texttt{content} (str) \\
assistant\_message & \texttt{event} (str), \texttt{timestamp} (iso8601), \texttt{content} (str) \\
request  & as v1 request \\
response & as v1 response $+$ \texttt{agent\_name} (str) \\
\bottomrule
\end{tabular}
\end{tcolorbox}

\paragraph{Version 3 --- CSI scaffold, initial release.} The CSI rewrite of the publishing pipeline initially emitted a deliberately minimal envelope only: session boundaries, the bound proxy port, and end-of-session totals. The completion bodies were not in the line at all (they were stored externally and were not part of the on-disk session log).

\begin{tcolorbox}[colback=cai_light, colframe=cai_dark!60, title={\textbf{v3 line types}}]
\small
\begin{tabular}{@{}lp{0.78\linewidth}@{}}
\toprule
\textbf{Type} & \textbf{Fields (type)} \\
\midrule
session\_start    & \texttt{event} (str), \texttt{ts} (iso8601), \texttt{proxySessionId} (uuid), \texttt{anthropicModel} (str) \\
proxy\_listening  & \texttt{event} (str), \texttt{ts} (iso8601), \texttt{proxySessionId} (uuid), \texttt{proxyPort} (int) \\
session\_end      & \texttt{event} (str), \texttt{ts} (iso8601), \texttt{proxySessionId} (uuid), \texttt{requests} (int), \texttt{lastModel} (str$\mid$null), \texttt{sessionCost} (float), \texttt{sessionInputTokens} (int), \texttt{sessionOutputTokens} (int) \\
\bottomrule
\end{tabular}
\end{tcolorbox}

\paragraph{Version 4 --- CSI scaffold, current.} The current schema restores the full request and response payload inside the line and adds operational metadata (\texttt{latencyMs}, \texttt{durationMs}, \texttt{reqBytes}, \texttt{resBytes}, \texttt{route}, \texttt{target}, \texttt{translation}, \texttt{cost}). All lines carry a \texttt{schema\_version} and a \texttt{scaffold} identifier; envelope lines additionally carry \texttt{event}. Request/response transaction lines do not carry \texttt{event} and are dispatched on the presence of \texttt{request} or \texttt{response}.

\begin{tcolorbox}[colback=cai_light, colframe=cai_dark!60, title={\textbf{v4 line types}}]
\small
\begin{tabular}{@{}lp{0.74\linewidth}@{}}
\toprule
\textbf{Type} & \textbf{Fields (type)} \\
\midrule
session\_start     & \texttt{event}, \texttt{ts}, \texttt{proxySessionId}, \texttt{scaffold}, \texttt{schema\_version} \\
proxy\_listening   & $+$\texttt{proxyPort} (int) \\
session\_linked    & $+$\texttt{sessionId} (uuid) \\
transaction        & \texttt{ts}, \texttt{id} (str), \texttt{proxySessionId}, \texttt{sessionId}, \texttt{route} (str), \texttt{target} (url), \texttt{method} (str), \texttt{path} (str), \texttt{status} (int), \texttt{latencyMs} (int), \texttt{durationMs} (int), \texttt{reqBytes} (int), \texttt{resBytes} (int), \texttt{request} (obj), \texttt{response} (obj), \texttt{translation} (str), \texttt{scaffold}, \texttt{cost} (float), \texttt{schema\_version} \\
\bottomrule
\end{tabular}
\end{tcolorbox}

\paragraph{Dispatch summary.} A single forward pass identifies the schema:

\begin{tcolorbox}[colback=cai_light, colframe=cai_dark!60, title={Dispatch rule (one pass per line)}]
\small
\begin{verbatim}
if "schema_version" in line:        v4
elif "proxySessionId" in line:      v3
elif "event" in line:               v2 envelope
elif "object" in line:              v1/v2 response
else:                               v1/v2 request
\end{verbatim}
\end{tcolorbox}

The conversion pipeline that produces the released slices applies exactly this dispatch internally so that the released records have a single uniform shape regardless of the originating scaffold era.

\section{URL / target sectoral breakdown}\label{app:url_sectors}

Table~\ref{tab:url_sectors} reports the full sectoral classification of the \statUniqueDomains{} unique target domains observed across the corpus, complementing the qualitative narrative of Section~\ref{sec:stats:urls}.

\begin{table}[!htbp]
\centering
\footnotesize
\setlength{\tabcolsep}{6pt}
\renewcommand{\arraystretch}{1.06}
\begin{tabular*}{\textwidth}{@{\extracolsep{\fill}}lrrl@{}}
\toprule
\textbf{Sector} & \textbf{Domains} & \textbf{URL hits} & \textbf{Representative top domains} \\
\midrule
Internal/lab IPs (RFC1918) & $1{,}266$  & $165{,}125$ & \texttt{10.10.110.100}, \texttt{192.168.3.100} \\
OS / FOSS / tooling docs   & $90$       & $54{,}731$  & \texttt{nmap.org}, \texttt{openssh.com}, \texttt{ubuntu.com} \\
Web standards / frameworks & $61$       & $28{,}931$  & \texttt{w3.org}, \texttt{wordpress.org}, \texttt{springframework.org} \\
Banking / fintech          & $401$      & $18{,}119$  & \texttt{riamoneytransfer.com}, \texttt{pixbitcoin.org}, \texttt{aufbaubank.de} \\
Social / media             & $286$      & $17{,}845$  & \texttt{youtube.com}, \texttt{twitter.com}, \texttt{facebook.com} \\
Telecom / ISP              & $49$       & $17{,}316$  & \texttt{*.excelcom.co.id}, \texttt{wso2.org}, \texttt{telefonica.com} \\
Public IP literals         & $1{,}047$  & $10{,}027$  & various scan targets \\
Cloud providers / CDN      & $351$      & $9{,}522$   & \texttt{*.amazonaws.com}, \texttt{cloudflare}, \texttt{googleapis.com} \\
Security DBs / advisories  & $50$       & $8{,}960$   & \texttt{vulners.com}, \texttt{cve.mitre.org}, \texttt{owasp.org} \\
Government / military      & $367$      & $5{,}916$   & \texttt{*.gov.sg}, \texttt{kdipa.gov.kw}, \texttt{nvd.nist.gov} \\
CTF / training platforms   & $363$      & $4{,}336$   & \texttt{web2.arcctf.net}, \texttt{root-me.org}, provider-suppressed CTF long tail \\
Code \& dev infrastructure & $104$      & $2{,}010$   & \texttt{registry.npmjs.org}, \texttt{jenkins.io}, \texttt{*.github.io} \\
Robotics / industrial      & $157$      & $1{,}532$   & \texttt{aliasrobotics.com}, \texttt{sage.com} \\
E-commerce / retail        & $101$      & $1{,}528$   & \texttt{target.com}, \texttt{mercadolibre.com}, \texttt{shopify.com} \\
Bug-bounty platforms       & $65$       & $1{,}425$   & \texttt{hackerone.com}, \texttt{bugcrowd.com}, \texttt{yeswehack.com} \\
Healthcare                 & $55$       & $364$       & \texttt{saulttribehealth.com}, \texttt{medicareadvantage.com} \\
AI / ML services           & $15$       & $166$       & \texttt{api.openai.com}, \texttt{docs.anthropic.com} \\
Consulting / Big-4         & $44$       & $162$       & \texttt{*.deloitte.com}, \texttt{*.deloitte.jp} \\
Energy / utilities         & $14$       & $49$        & \texttt{iberdrola.es}, \texttt{iberdrola.pt} \\
Other (.com / corporate)   & $18{,}261$ & $335{,}542$ & long tail across unclassified corporate domains \\
\midrule
\textbf{Total}             & \textbf{$23{,}147$} & \textbf{$683{,}606$} & \\
\bottomrule
\end{tabular*}
\caption{Sectoral breakdown of unique target domains. Sector membership is assigned by a keyword classifier; the \emph{Other (.com / corporate)} row aggregates the long tail of unclassified business domains plus the $\sim\!154$ domains that the per-domain renderer did not surface individually (rolled into the corporate tail so that the total reconciles with the \statUniqueDomains{} unique-domain and \statTotalURLs{} URL-hit headline figures). Banking / fintech, telecom and government rows reveal that the corpus contains substantial production-target activity rather than pure CTF traffic.}\label{tab:url_sectors}
\end{table}

\end{document}